\documentclass[twocolumn]{aastex62}
\usepackage{lineno}
\usepackage{tabularx}
\usepackage{amsmath,amssymb}
\usepackage{color}
\usepackage{xcolor}
\usepackage{multirow}
\usepackage{CJK}

\definecolor{iceberg}{rgb}{0.44, 0.65, 0.82}
\definecolor{lavenderblue}{rgb}{0.8, 0.8, 1.0}
\definecolor{lavenderpink}{rgb}{0.98, 0.68, 0.82}
\usepackage{subfigure}

\submitjournal{ApJ}
\graphicspath{{./}{figures/}}
\usepackage{color, colortbl}
\usepackage{romannum}
\begin{document}
\begin{CJK*}{UTF8}{gbsn}

\title{GRB-SN Association within the Binary-Driven Hypernova Model}

\author[0000-0001-5717-6523]{Y.~Aimuratov}
\affiliation{Fesenkov Astrophysical Institute, Observatory 23, 050020 Almaty, Kazakhstan}
\affiliation{ICRANet, Piazza della Repubblica 10, I-65122 Pescara, Italy}

\author[0000-0002-3262-5545]{L.~M.~Becerra}
\affiliation{Escuela de F\'isica, Universidad Industrial de Santander, A.A.678, Bucaramanga, 680002, Colombia }
\affiliation{ICRANet, Piazza della Repubblica 10, I-65122 Pescara, Italy}

\author[0000-0001-7749-4078]{C.L.~Bianco}
\affiliation{ICRANet, Piazza della Repubblica 10, I-65122 Pescara, Italy}
\affiliation{ICRA, Dipartamento di Fisica, Sapienza Universit\`a  di Roma, Piazzale Aldo Moro 5, I-00185 Rome, Italy}
\affiliation{Universit\'e de Nice Sophia-Antipolis, Grand Ch\^ateau Parc Valrose, Nice, CEDEX 2, France}
\affiliation{INAF, Istituto di Astrofisica e Planetologia Spaziali, Via Fosso del Cavaliere 100, 00133 Rome, Italy}

\author[0000-0002-0542-5601]{C.~Cherubini}
\affiliation{ICRANet, Piazza della Repubblica 10, I-65122 Pescara, Italy}
\affiliation{ICRA, University Campus Bio-Medico of Rome, Via Alvaro del Portillo 21, I-00128 Rome, Italy}
\affiliation{Department of Science and Technology for Sustainable Development and One Health, Universit\'a Campus Bio-Medico di Roma, Via Alvaro del Portillo, 21 - I-00128 Rome, Italy}

\author[0000-0003-3142-5020]{M.~Della~Valle}
\affiliation{INAF - Osservatorio Astronomico di Capodimonte, Salita Moiariello 16, I-80131, Napoli, Italy}
\affiliation{ICRANet, Piazza della Repubblica 10, I-65122 Pescara, Italy}

\author[0000-0003-1952-0834]{S.~Filippi}
\affiliation{ICRANet, Piazza della Repubblica 10, I-65122 Pescara, Italy}
\affiliation{ICRA, University Campus Bio-Medico of Rome, Via Alvaro del Portillo 21, I-00128 Rome, Italy}
\affiliation{Department of Engineering, Universit\'a Campus Bio-Medico di Roma, Via Alvaro del Portillo, 21 - I-00128 Rome, Italy}

\author[0000-0002-1343-3089]{Liang~Li (李亮)}
\affiliation{ICRANet, Piazza della Repubblica 10, I-65122 Pescara, Italy}
\affiliation{ICRA, Dipartamento di Fisica, Sapienza Universit\`a  di Roma, Piazzale Aldo Moro 5, I-00185 Rome, Italy}
\affiliation{INAF, Osservatorio Astronomico d'Abruzzo, Via M. Maggini snc, I-64100, Teramo, Italy}

\author[0000-0002-2516-5894]{R.~Moradi}
\affiliation{Key Laboratory of Particle Astrophysics, Institute of High Energy Physics, Chinese Academy of Sciences, Beijing 100049, People’s Republic of China}
\affiliation{ICRA, Dipartamento di Fisica, Sapienza Universit\`a  di Roma, Piazzale Aldo Moro 5, I-00185 Rome, Italy}
\affiliation{INAF, Osservatorio Astronomico d'Abruzzo, Via M. Maggini snc, I-64100, Teramo, Italy}

\author{F.~Rastegarnia}
\affiliation{ICRANet, Piazza della Repubblica 10, I-65122 Pescara, Italy}
\affiliation{Department of Physics, Faculty of Physics and Chemistry, Alzahra University, Tehran, Iran}

\author[0000-0003-4904-0014]{J.~A.~Rueda}
\affiliation{ICRANet, Piazza della Repubblica 10, I-65122 Pescara, Italy}
\affiliation{ICRA, Dipartamento di Fisica, Sapienza Universit\`a  di Roma, Piazzale Aldo Moro 5, I-00185 Rome, Italy}
\affiliation{ICRANet-Ferrara, Dip. di Fisica e Scienze della Terra, Universit\`a degli Studi di Ferrara, Via Saragat 1, I-44122 Ferrara, Italy}
\affiliation{Dipartamento di Fisica e Scienze della Terra, Universit\`a degli Studi di Ferrara, Via Saragat 1, I-44122 Ferrara, Italy}
\affiliation{INAF, Istituto di Astrofisica e Planetologia Spaziali, Via Fosso del Cavaliere 100, 00133 Rome, Italy}

\author[0000-0003-0829-8318]{R.~Ruffini}
\affiliation{ICRANet, Piazza della Repubblica 10, I-65122 Pescara, Italy}
\affiliation{ICRA, Dipartamento di Fisica, Sapienza Universit\`a  di Roma, Piazzale Aldo Moro 5, I-00185 Rome, Italy}
\affiliation{Universit\'e de Nice Sophia-Antipolis, Grand Ch\^ateau Parc Valrose, Nice, CEDEX 2, France}
\affiliation{INAF, Viale del Parco Mellini 84, 00136 Rome, Italy}

\author[0000-0003-2011-2731]{N.~Sahakyan}
\affiliation{ICRANet, Piazza della Repubblica 10, I-65122 Pescara, Italy}
\affiliation{ICRANet-Armenia, Marshall Baghramian Avenue 24a, Yerevan 0019, Republic of Armenia}

\author[0000-0001-7959-3387]{Y.~Wang (王瑜)}
\affiliation{ICRANet, Piazza della Repubblica 10, I-65122 Pescara, Italy}
\affiliation{ICRA, Dipartamento di Fisica, Sapienza Universit\`a  di Roma, Piazzale Aldo Moro 5, I-00185 Rome, Italy}
\affiliation{INAF, Osservatorio Astronomico d'Abruzzo, Via M. Maggini snc, I-64100, Teramo, Italy} 

\author{S.~R.~Zhang (张书瑞)}
\affiliation{ICRANet, Piazza della Repubblica 10, I-65122 Pescara, Italy}
\affiliation{ICRANet-Ferrara, Dip. di Fisica e Scienze della Terra, Universit\`a degli Studi di Ferrara, Via Saragat 1, I-44122 Ferrara, Italy}
\affiliation{School of Astronomy and Space Science, University of Science and Technology of China, Hefei 230026, China }
\affiliation{CAS Key Laboratory for Research in Galaxies and Cosmology, Department of Astronomy, University of Science and Technology of China, Hefei 230026, China}

\email{jorge.rueda@icra.it, ruffini@icra.it, rahim.moradi@icranet.org}

\date{Received date /Accepted date }

\begin{abstract}
The observations of supernovae (SNe) Ic occurring after the prompt emission of long gamma-ray bursts (GRBs) are addressed within the binary-driven hypernova (BdHN) model where GRBs originate from a binary composed of a $\sim10M_\odot$ carbon-oxygen (CO) star and a neutron star (NS). The CO core collapse gives the trigger, leading to a hypernova with a fast-spinning newborn NS ($\nu$NS) at its center. The evolution depends strongly on the binary period, $P_{\rm bin}$. For $P_{\rm bin}\sim5$min, BdHNe I occur with energies $10^{52}$--$10^{54}$erg. The accretion of SN ejecta onto the NS leads to its collapse, forming a black hole (BH) originating the MeV/GeV radiation. For $P_{\rm bin}\sim 10$min, BdHNe II occur with energies $10^{50}$--$10^{52}$erg and for $P_{\rm bin}\sim$hours, BdHN III occurs with energies below $10^{50}$erg. {In BdHNe II and III,} no BH is formed. The $1$--$1000$ms $\nu$NS originates, in all BdHNe, the X-ray-optical-radio afterglows by synchrotron emission. The hypernova follows an independent evolution, becoming an SN Ic, powered by nickel decay, observable after the GRB prompt emission. We report $24$ SNe Ic associated with BdHNe. Their optical peak luminosity and time of occurrence are similar and independent of the associated GRBs. {From previously identified $380$ BdHN I comprising redshifts up to $z=8.2$, we analyze} four examples with their associated hypernovae. By multiwavelength extragalactic observations, we identify seven new Episodes, theoretically explained, fortunately not yet detected in galactic sources, opening new research areas. Refinement of population synthesis simulations is needed to map the progenitors of such short-lived binary systems inside our galaxy.
\end{abstract}


\section{Introduction}\label{sec:1}

The pioneering work of the BeppoSAX telescope, linking for the first time the success of gamma-ray astronomy with the discovery of gamma-ray bursts (GRBs) \citep{1973ApJ...182L..85K} and the CGRO/BATSE era \citep{1982AIPC...77..443F} to the X-ray astronomy of binary X-ray sources \citep{GiacconiRuffini1978}, led to the discovery of the GRB X-ray afterglow \citep{Costa1997} and the determination of the GRB cosmological nature \citep{1997Natur.387..878M}. Following these successes, we have returned to address the fundamental issue of the observational coincidence of GRBs with Ic supernovae (SNe):

1) Our theoretical framework started with the induced gravitational collapse scenario \citep{2012ApJ...758L...7R} introduced to {originate stellar} mass black holes (BHs) powering long GRBs {associated with type Ic SNe}. It was soon followed by the binary-driven hypernovae (BdHNe) model \citep{2014A26A...565L..10R}, which assumes a binary system composed of a carbon-oxygen (CO) star of $\lesssim 10 M_\odot$ and a companion neutron star (NS) as the GRB progenitor. The GRB trigger occurs when the CO core collapses, originating a newborn NS ($\nu$NS) and an SN Ic. The SN {ejecta accretes onto the NS companion and the $\nu$NS because of matter fallback \citep{2019ApJ...871...14B, 2022PhRvD.106h3002B}.} 

2) The first evidence for such BdHN was presented by analyzing two sources: GRB 090618 at $z=0.54$ \citep{2012A&A...543A..10I, 2012A&A...548L...5I} and GRB 090423 \citep{2009Natur.461.1258S,2009Natur.461.1254T,2014A&A...569A..39R}. The extraordinary result of GRB 090423 was that it was observed at $z=8.2$, which was and still is the farthest GRB in our Universe with a spectroscopic confirmation. We are currently examining GRB 090429B, with a photometric redshift $z=9.4$ \citep{2011ApJ...736....7C}, within the BdHN model (Ruffini et al., in preparation). In the meantime, the existence of $380$ BdHNe has been presented   \citep{2021MNRAS.504.5301R}. Their distribution ranges from the above $z=8.2$ to close extra-galactic GRBs in the local Universe. Their enormous energies range between $10^{49}$ erg and a nearly $10^{55}$ erg of GRB 220101A (and GRB {221009A}). A crucial point is that the compact CO-NS systems of the BdHN model are the final stage of {a peculiar} binary evolution, short-lived and rare, as GRBs are. The probability of their occurrence in our Galaxy is extremely low. Since the progenitors are short-lived, their frequency of occurrence essentially mimics the evolution of the cosmic star formation rate with redshift, peaking at $z\sim 2$--$2.5$ (e.g., \citealp{2014ARA&A..52..415M}{; see also \citealp{2008ApJ...683L...5Y, 2012MNRAS.423.3049G, 2016ApJ...823..154G, 2017ApJ...834..170G}}). Based on the low rate of long-duration GRBs in the current cosmic epoch in our Galaxy \citep{2007ApJ...657L..73G}, which is {$\sim 3$ orders of magnitude lower than the observed core-collapse SN rate \citep{2017PASP..129e4201S}}, the potential GRB progenitors currently \textit{ready to explode} in the Milky Way are, in the most optimistic view, a handful of objects. The observed density rate of BdHN I is $\sim 1$ Gpc$^{-3}$ yr$^{-1}$ \citep{2016ApJ...832..136R, 2018ApJ...859...30R}. Therefore, it is not surprising that we can acknowledge the existence of such compact binary progenitors only through their cataclysmic fate leading to GRBs thanks to their extragalactic, cosmological nature. Interestingly, the above feature could not be fortuitous since an energetic GRB inside our Galaxy might represent a catastrophe for life on Earth \citep[see, e.g.,][]{2015ARep...59..469C}.

3) The crucial topic of extreme interest has been the byproducts of the GRB observations: a) the discovery of supernovae of characteristic energy of $10^{49}$ erg associated with all different classes of BdHN, this article is dedicated to this topic; b) the discovery of seven different episodes characterizing the most general GRB and presenting new physical processes in ultrarelativistic regimes impossible to discover within our Galaxy; c) the fundamental knowledge developed in decades of observations in Earth-based accelerators pointing to vacuum polarization processes \citep[see][and references therein]{2010PhR...487....1R} are here discovered in ultrarelativistic regimes and overcritical quantum electrodynamical processes. These processes, when occurring outside our Galaxy, give the unique opportunity to extend the knowledge reached on our planet, but, at the same time, they indicate the danger of the occurrence of these events for the survival of life if they should occur in our Galaxy. An unexpected additional result has been the possibility to apply, in the comprehension of BdHNe, the still untested configuration of rapidly rotating self-gravitating systems that have attracted the attention of the greatest scientists in world history: from {Isaac Newton \citep[\emph{Principia}, Book III, Propositions XVIII-XX][]{1687pnpm.book.....N} to Colin Maclaurin \citep{maclaurin1742treatise}, Carl Gustav Jacob Jacobi \citep{1834AnP...109..229J}, George Darwin \citep{1886RSPS...41..319D}, James Hopwood Jeans \citep{BOOK_Jeans}, and more recently Subrahmanyan Chandrasekhar \citep{1969efe..book.....C}; see Section \ref{sec:10}, for details.}

A new era for relativistic astrophysics started, grounded on the classical results obtained on compact stellar X-ray sources originating from binary massive systems derived on Galactic observations \citep{2006csxs.book..623T}, as well as on the concepts of BHs expressed by the mathematical equations of Roy Kerr \citep{1963PhRvL..11..237K} and by the mass-energy formula of Christodoulou-Ruffini \citep{1970PhRvL..25.1596C, 1971PhRvD...4.3552C} and Hawking \citep{1971PhRvL..26.1344H} finally here reaching confirmations in extragalactic sources. It opens to the fundamental issues of understanding the role of GRBs and their intriguing possible interaction with the birth and the end of life in the Universe.

\begin{figure*}
    \centering
    \includegraphics[width=0.9\hsize,clip]{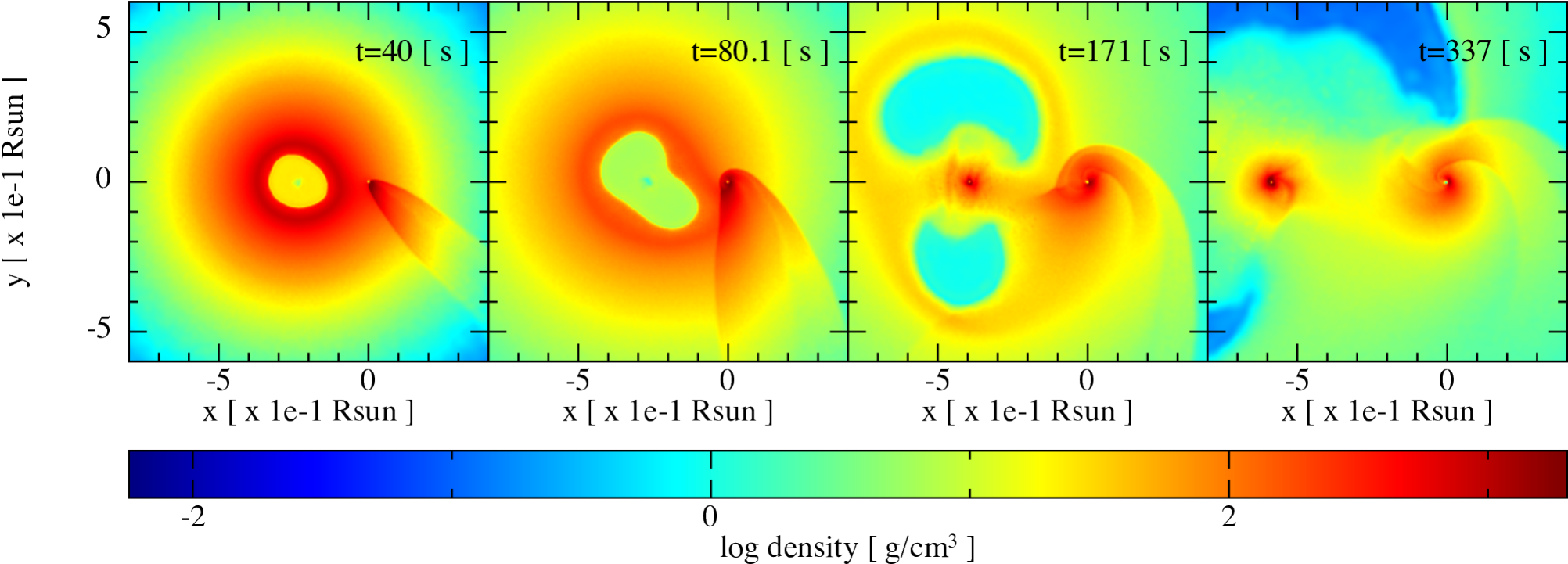}\\
    \includegraphics[width=0.9\hsize,clip]{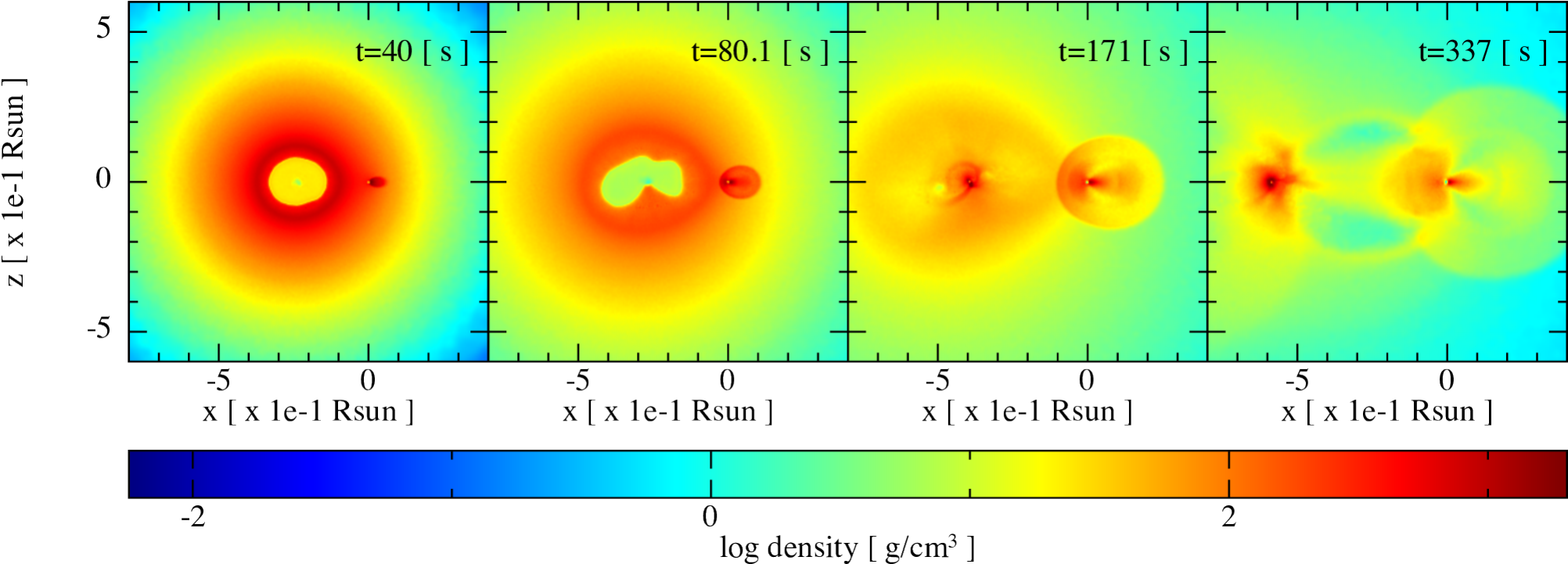}
    \caption{SPH simulation of a BdHN I: model ``30m1p1eb'' of Table 2 in \citet{2019ApJ...871...14B}. The binary progenitor comprises a CO star of $\approx 9 M_\odot$  (produced by a ZAMS star of $30 M_\odot$) and a $2 M_\odot$ NS companion. The orbital period is $\approx 6$ min. From left to right, each snapshot corresponds to selected increasing times where $t = 0$ s refers to the SN shock breakout. The upper and lower panel shows the mass density on the equatorial plane and the plane orthogonal to the latter. The reference system is rotated and translated to align the x-axis with the line joining the binary components. The origin of the reference system is located at the NS companion position. In the first snapshot at $t = 40$ s, particles in the NS gravitational capture region form a tail behind the NS companion. These particles then circularize around the NS, forming a thick disk visible in the second snapshot at $t = 80$ s. Part of the ejecta produces a fallback accretion process onto the $\nu$NS visible in the third snapshot at $t=171$ s. At $t = 337$ s (about one orbital period), a disk structure is visible around the $\nu$NS and the NS companion.
    This figure has been produced with the SNsplash visualization program \citep{2011ascl.soft03004P}. The figure highlights the $\nu$NS is coeval with the SN explosion and remains at the SN center while the ejecta expands. It also shows that the timescale of the physical phenomena leading to the transient activity of the GRB (e.g., the hypercritical accretion and its consequences) is shorter than the timescale of changes in the orbital properties, e.g., an orbital widening or eventual binary disruption owing to mass loss \citep{1961BAN....15..265B, 1972NPhS..239...67V, 2015PhRvL.115w1102F}.}
    \label{fig:3DSPH}
\end{figure*}

Since their discovery, the enormous energetics led to the idea that GRBs are associated with massive stars' gravitational collapse, leading to NSs or BHs. The community widely accepts the seminal proposal that mergers of NS-NS or NS-BH binaries are the progenitors of short GRBs \citep{1986ApJ...308L..47G, 1986ApJ...308L..43P, 1989Natur.340..126E, 1991ApJ...379L..17N}. For long GRBs, our sources of interest here, the traditional model is based on a \textit{collapsar}, the core-collapse of a single massive star leading to a BH (or a magnetar) surrounded by an accretion disk \citep{1993ApJ...405..273W}. We refer the reader to \citet{2002ARA26A..40..137M, 2004RvMP...76.1143P}, for comprehensive reviews.

In the GRB traditional model, the prompt emission originates in the dynamics, expansion, and transparency of a \textit{fireball}, an optically thick electron-positron ($e^-e^+$)-photon plasma in equilibrium with baryons \citep{1978MNRAS.183..359C, 1986ApJ...308L..43P, 1986ApJ...308L..47G, 1991ApJ...379L..17N, 1992ApJ...395L..83N}. The fireball expands in a collimated relativistic jet with Lorentz factor $\Gamma \sim 10^2$--$10^3$ \citep{1990ApJ...365L..55S, 1992MNRAS.258P..41R, 1993MNRAS.263..861P, 1993ApJ...415..181M, 1994ApJ...424L.131M}. In this picture, the interaction of internal and external shocks with the surrounding and interstellar medium is responsible for the prompt emission and the afterglow, including the very-high-energy (VHE) emission by synchrotron self-Compton radiation \citep{2002ARA26A..40..137M, 2004RvMP...76.1143P, 2019Natur.575..455M, 2019Natur.575..448Z}. We refer to \citet{2018pgrb.book.....Z} for the latest developments of the GRB traditional model.

From the energetics, dynamics, and radiation efficiency, two difficulties arise in the traditional model. 1) Only a small fraction of the energy of the ultra-relativistic jet is radiated by the synchrotron emission, so much of the kinetic energy remains in the jet. 2) The radiation from the jet implies the absence of afterglow in some long GRBs, while it is clear that the afterglow is present in all GRBs.

We now turn to one of this article's main topics, the GRB-SN connection. The follow-up of the optical afterglow, extended by the Neil Gehrels Swift Observatory \citep{2005SSRv..120..143B,2005SSRv..120..165B,2005SSRv..120...95R}, led to the discovery of the association of long GRBs with type Ic supernovae (SNe), first marked with the temporal and spatial coincidence of GRB 980425 and SN 1998bw \citep{1998Natur.395..670G}. Since then, further observations have confirmed the GRB-SN connection \citep{2006ARA&A..44..507W, 2011IJMPD..20.1745D, 2012grb..book..169H, 2017AdAst2017E...5C}. The association of GRBs with SNe Ic is possibly one of the most relevant observational clues for theoretical models. Several theoretical and observational consequences from the GRB-SN connection constrain models of GRBs and the associated SNe Ic:

\textit{(i) Long GRBs and SNe have different energetics}. SNe radiate energies {$\sim 10^{49}$--$10^{52}$~erg}, while GRBs show energies in the much wider range {$\sim 10^{49}$--$10^{55}$~erg}. The energy release of energetic GRBs is associated with the gravitational collapse to a BH, while SNe originate in the core-collapse of a massive star to an NS.

{\textit{(ii) Most (if not all) long-duration GRBs originate from binary stars.} a) In recent decades, growing evidence has shown that long-duration GRBs are associated with the explosions of massive stars. This fact has been well established both on a statistical basis, e.g., \citet{2006Natur.441..463F, 2008ApJ...689..358R, 2008ApJ...687.1201K}, and from stellar evolution which, even if constraining the zero-age main-sequence (ZAMS) mass of the SN progenitor is highly model-dependent, points undoubtedly to massive stellar progenitors from the modeling of the photometric and spectroscopic follow-up of SNe-Ibc associated with GRBs, e.g., SN 1998bw, $25$--$40 M_\odot$ \citep{2006ARA&A..44..507W, 2006ApJ...640..854M}; SN 2003dh $35$--$40 M_\odot$ \citep{2003omeg.conf..223N, 2003ApJ...599L..95M}; SN 2003lw $25 M_\odot$ \citep{2006ApJ...645.1323M}; 2008D $30 M_\odot$ \citep{2009ApJ...700.1680T}; 2010bh $25 M_\odot$ \citep{Bufano...2012ApJ...753...67B};  2016jca $35 M_\odot$ \citep{2019MNRAS.487.5824A}. b) It is well known that a significant fraction of massive stars is in binaries (about $70\%$, e.g., \citealp{2007ApJ...670..747K} and \citealp{2012Sci...337..444S}). c) In addition, although stellar evolution models predict the direct formation of a BH from the gravitational collapse for progenitor stars $\gtrsim 25 M_\odot$ \citep{2003ApJ...591..288H}, two observational facts pose serious challenges to GRB-SN models in which both a BH and SN originate from a single-star: 1) the direct gravitational collapse of a massive star to a BH should occur without an SN emission; 2) observed pre-SN progenitors have masses $\lesssim 18~M_\odot$ \citep[see][for details]{2009ARA&A..47...63S, 2015PASA...32...16S}}. Therefore, it is unlikely that the GRB and the SN can originate from the very same single star. Indeed, it is an extreme request for the gravitational collapse of a massive star to form a collapsar, a jetted fireball, and an SN explosion. Some models attempt to supply (partial) solutions to these issues, like an efficient neutrino emission from the accretion disk \citep[e.g.][]{1999ApJ...524..262M} or the presence of an outflow/wind where the nucleosynthesis of the nickel for the optical SN can occur \citep[see, e.g.,][]{2005ApJ...629..341K, 2012ApJ...744..103M, 2012ApJ...750..163L}. The direct conclusion from the abovementioned points is that most long-duration GRBs occur in binaries. Indeed, \citet{2007A&A...465L..29C} tested the idea of producing rapidly rotating Wolf-Rayet (WR) stars in massive close binaries as possible progenitors of collapsars. The above facts also motivated our development of a model for long-duration GRBs that fully exploit the binary nature of progenitors.

\textit{(iii) The SNe associated with GRBs are of type Ic}. The lack of hydrogen (H) and helium (He) in the spectra of type Ic SNe has the leading explanation that they originate in bare He, CO, or WR  stars that lose the outermost hydrogen and helium layers during their evolution \citep[see, e.g.,][]{2011MNRAS.415..773S, 2020MNRAS.492.4369T}. Numerical simulations indicate that the most natural mechanism for He/CO/WR stars to get rid of their H/He envelope is from interactions with a compact-star companion (e.g., NS) through multiple mass-transfer and common-envelope phases \citep[see, e.g.,][]{1988PhR...163...13N, 1994ApJ...437L.115I, 2007PASP..119.1211F, 2010ApJ...725..940Y, 2011MNRAS.415..773S, 2015PASA...32...15Y, 2015ApJ...809..131K}. 

Although the above is not a complete list of possible drawbacks of the single-star scenario, it is already clear that considering alternatives is natural. In their pioneering work, \citet{1999ApJ...526..152F} show that various binary stellar evolution channels can lead to diverse GRB events. This alternative binary approach has contributed, as mentioned above, in the study of short GRBs \citep[see, e.g.,][]{2016ApJ...831..178R,2017ApJ...844...83A}, as well as an enigmatic long-lasting {GRB 060614} without SN \citep{2006Natur.444.1050D} interpreted as a white dwarf (WD)-NS merger \citep{2009A&A...498..501C,2018JCAP...10..006R} and the weakest GRBs from WD-WD mergers \citep[see, e.g.,][]{2019JCAP...03..044R, 2022IJMPD..3130013R}.

We specialize in the BdHN model of long GRBs based on the IGC scenario \citep{2012ApJ...758L...7R}. Following the evolution of stripped-envelope binaries, the BdHN model proposes as GRB progenitor a CO-NS binary at the end of the thermonuclear life of the CO star, i.e., the second core-collapse SN event in the binary lifetime. The first SN formed the NS companion of the CO star. The CO nature of the exploding star explains why the SNe associated with GRBs are type Ic. {This SN explosion in the CO-NS binary triggers the physical processes that explain} the seven episodes observed in the GRB \citep{2012ApJ...758L...7R, 2012A&A...548L...5I, 2014ApJ...793L..36F, 2015PhRvL.115w1102F, 2015ApJ...812..100B, 2016ApJ...833..107B, 2019ApJ...871...14B}. Figure \ref{fig:3DSPH} shows an example of numerical simulation performed by \citet{2019ApJ...871...14B} of the explosion of a CO star leading to a newborn NS ($\nu$NS) and the SN Ic, in the presence of an NS companion. These simulations, which include hydrodynamics, neutrino emission, and general relativistic effects, show a variety of outcomes of the system, leading to a variety of GRB events, a BdHN classification, which we discuss below. One of the most relevant results is that, among the possible fates, the NS companion can reach the point of gravitational collapse, forming a rotating, newborn Kerr BH. As recalled, the BdHN progenitors have not been simulated in population synthesis or binary stellar evolution models. Thus, in our numerical simulations, we have to use pre-SN stars resulting from the stellar evolution of single stars and assume the presence of the NS companion. Therefore, the binary evolution leading to the compact BdHN system could start with a different ZAMS mass than the one we are currently considering. Namely, single and binary evolutionary paths can lead to different ZAMS masses starting from a given pre-SN star mass. The latter scenario can lead to a less massive ZAMS progenitor than the former {\citep[see, e.g.,][for the case of binary progenitors of type II SNe]{2019A&A...631A...5Z}}. For the early phases of the BdHN model, we have scrutinized the simulations derived by \citet{2006csxs.book..623T} (e.g., Figs. $16.12$ and $16.15$). Such simulations are based on X-ray observations of stellar evolution in our Galaxy. We generally confirm the applicability of these models up to the common-envelope phase. {Following that phase, the explanation of the multiwavelength observations (from X-rays to GeV and ultrahigh energy) of long GRBs within the BdHN model predicts the existence of CO-NS binaries with orbital periods from hours to days (BdHN II and III) to minutes (BdHN I), taking into due account the relevant role of the angular momentum (see Section \ref{sec:2}, and references therein). In view of the low occurrence rate of GRBs in a single galaxy, the necessity of forming CO-NS binaries has been evidenced only by extragalactic observations, whose comprehension has been made possible under the complementary information gained from galactic systems \citep[e.g.,][]{2006csxs.book..623T}.}

In Section~\ref{sec:2}, we recall the basics of the BdHN model and address how the interplay between the SN, $\nu$NS, and NS companion leads to the variety of long GRBs.

Section~\ref{sec:3} recalls a relativistic formulation's framework in the source's cosmological rest frame, including the $k$-correction. 

In Section~\ref{sec:4}, we analyze $24$ SNe associated with GRBs. We show that the SN bolometric peak luminosity and its time of occurrence in the source cosmological rest-frame are nearly the same for all sources (see Figs.~\ref{fig:redlpeak} and \ref{fig:redtpeak}). We also present the prompt gamma-ray energy ($E_{\rm iso}$) of the associated GRB. We show that $E_{\rm iso}$ spans over six orders of magnitude, while the SN bolometric peak luminosity and the time of occurrence of the peak remain relatively constant; see Figs.~\ref{fig:iso-lpeak} and \ref{fig:iso-tpeak}. These results constrain GRB models and will be explained within the BdHN model in the following sections.

Section \ref{sec:5} describes the physical phenomena in the different BdHN types and relates them to specific GRB observables, namely the seven episodes of BdHNe; see Table \ref{tab:observables} for details. 

In Section \ref{sec:6}, after recalling the observations that made possible the identification of GRB 180720B as BdHN I \citep{GCN23019}, we address the seven episodes characterizing the source as a BdHN I.

Section~\ref{sec:7} investigates the second BdHN I fully understood in the BdHN model: GRB 190114C \citep{2021MNRAS.504.5301R, 2021PhRvD.104f3043M}. We recall the observations that identified this source as BdHN I \citep{2019GCN.23715....1R} and discuss its corresponding seven episodes, following an analogous presentation for GRB 180720B in Section \ref{sec:6}. 

In Section~\ref{sec:8}, we turn to the case of a BdHN II, GRB 190829A. 

Since in BdHN II, the BH is not formed, the number of episodes in this GRB reduces from seven to three, which we address in detail. 

In Section~\ref{sec:9}, we analyze the only example analyzed to date of a BdHN III: GRB 1711205A. Similar to BdHN II, in BdHN III, the BH is not formed. 

The number of episodes in this GRB reduces from seven to two, which we present in detail. 

Section \ref{sec:10} summarizes new physical phenomena triggered by the SN occurrence in BdHNe, not previously studied in the GRB physics literature.

Finally, we outline conclusions in Section~\ref{sec:11}.

\section{BdHN classification}\label{sec:2}

The BdHN model assumes that some long GRB progenitors are binaries composed of a CO star of mass of $\sim 10 M_\odot$ and a companion NS of $ \sim 2.0 M_\odot$. It also assumes that the gravitational collapse of the CO star generates an SN explosion and creates a newborn NS ($\nu$NS) at its center. The $\nu$NS with a mass of $1.5 M_\odot$ is assumed to spin with a period of $\sim 1$--$100$~ms. It further assumes that $\sim 7$--$8 M_\odot$ are ejected during the SN explosion. The theoretical motivations and the observation constraints leading to these assumptions are given in Sections~\ref{sec:5}--\ref{sec:10}, and implications are presented in the Conclusions (Section~\ref{sec:11}). The SN ejecta drives an accretion process onto the NS companion and a fallback accretion onto the $\nu$NS. The accretion rates proceed at hypercritical rates (i.e., highly super-Eddington) due to the efficient neutrino emission \citep{2014ApJ...793L..36F, 2016ApJ...833..107B, 2018ApJ...852..120B}. We differentiate three types of BdHN: I, II, and III, as a function of their overall energetics. A dependence of these energetics from the total initial angular momentum of the Co star-NS binary is evidenced. The shorter the binary period, the higher the BdHN total radiated energy. 

\subsection{BdHN I}\label{sec:BdHNI}

We indicate by BdHN I the most energetic class of long GRBs with energies in the range of $10^{52} \,{\rm erg} \lesssim E_{\rm iso} \lesssim 10^{54}$~erg. Their orbital period is of the order of $\gtrsim 5$ min, which implies an orbital separation of $\sim 10^{10}$ cm, just bigger than the CO star radii \citep[see, e.g.,][]{2014ApJ...793L..36F, 2016ApJ...833..107B, 2019ApJ...871...14B}.  The hypercritical accretion of the SN ejecta onto the companion NS leads it to reach the critical mass, consequently forming a Kerr BH. Simulations show that the peak accretion rate onto the NS companion can reach $\Dot{M}_{\rm peak} \sim 10^{-3}$--$10^{-2} M_\odot$ s$^{-1}$, which implies accreting $0.5$--$1 M_\odot$ in about one orbital period time \citep{2016ApJ...833..107B, 2019ApJ...871...14B}. The NS gains a large angular momentum, $\Delta J \sim G M_{\rm NS} \Delta M_{\rm acc}/c \sim 10^{49}$ g cm$^2$ s$^{-1}$, hence it reaches the critical mass at millisecond rotation rates. The accretion energy gain when bringing the NS to the critical mass and the energy involved in the BH formation process set a lower edge of $\sim 10^{52}$ erg of energy released in a BdHN I. Therefore, BdHNe I explain the long GRBs with energies $E_{\rm iso} \gtrsim 10^{52}$ erg \citep[see][for details]{2018ApJ...859...30R}. The fallback accretion onto the $\nu$NS also proceeds at hypercritical rates, and the presence of the NS companion generates a double-peak accretion (\citealp{2019ApJ...871...14B}; see also \citealp{2022PhRvD.106h3002B} for recent simulations and implications). The first peak of accretion is of a few $10^{-3} M_\odot$ s$^{-1}$ and lasts for about one-tenth of the orbit \citep{2019ApJ...871...14B}. The $\nu$NS reaches a high rotation period of $0.5$ ms, near the mass-shedding limit \citep{2015PhRvD..92b3007C}. The fast-spinning $\nu$NS gives origin to the GRB afterglow as explained in Section~\ref{sec:5}. Examples of BdHNe I are GRB 180720B (see Section~\ref{sec:6}) , GRB 190114C (Section~\ref{sec:7}), and GRB 130427A \citep{2019ApJ...886...82R, 2021MNRAS.504.5301R}.

In \citet{2012ApJ...758L...7R, 2015PhRvL.115w1102F, 2016ApJ...832..136R}, we have advanced that the CO-NS compact binaries leading to BdHN I could form in an evolution path similar to the one leading to the so-called ultra-stripped binaries {(see, e.g., \citealp{2015MNRAS.451.2123T, 2017ApJ...846..170T}, as well as, e.g., \citealp{2006MNRAS.368.1742D, 2020A&A...642A.106D}, for alternative stellar evolution scenarios)}. However, population synthesis simulations of those systems lead to binaries with orbital periods longer than the ones of the BdHN systems \citep[see, e.g., Fig. 16.15 in][]{2006csxs.book..623T}. We currently consider with great interest scrutinizing the possibility that the evolution following the common-envelope phases (the last evolution stages of the binary) can have {a relevant role of the angular momentum of the stellar components, as suggested by the BdHN modeling of long GRBs,} branching off a formation channel of BdHN systems. {Certainly, BdHN progenitors can form in our own Galaxy, and likely some currently observed binary X-ray sources could, in due time, lead to a BdHN. However, it is observationally established that the probability of occurrence of a GRB in a single galaxy is extremely low, e.g., for the Milky Way, the observed GRB rate suggests one source every million years or so. The GRB detection rate in the Earth originates from extragalactic sources, which, given the GRB's enormous energetics, allow us to sample an enormous volume containing billions of galaxies, leading to nearly daily detections.} We recall that the observed density rate of BdHN I is $\sim 1$ Gpc$^{-3}$ yr$^{-1}$ \citep{2016ApJ...832..136R, 2018ApJ...859...30R}, so a small subpopulation of $\approx 0.01\%$--$0.1\%$ of ultra-stripped binaries following such a particular evolution branch might be sufficient to explain the BdHN I population \citep[see][for details]{2015PhRvL.115w1102F, 2016ApJ...832..136R}, given that ultra-stripped binaries comprise $0.1\%$-–$1\%$ of the total SNe \citep{2015MNRAS.451.2123T}; see also Section \ref{sec:11}.

\subsection{BdHN II}

These binaries are characterized by longer orbital periods of $\sim 20$--$40$ min, so binary separations of a few $10^{10}$ cm. Numerical simulations show that in these binaries, the accretion rate onto the NS companion occurs at lower rates, $\dot{M}_{\rm peak} \sim 10^{-5}$--$10^{-4} M_\odot$ s$^{-1}$. The NS does not reach the critical mass in these systems, so it does not form a BH. The above range of accretion rates implies that the BdHN II subclass can explain long GRBs with energies $E_{\rm iso}\sim 10^{50}$--$10^{52}$ erg \citep[see, e.g.,][]{2016ApJ...832..136R, 2018ApJ...859...30R}. 

Regarding the $\nu$NS, although the first peak of fallback accretion is similar to that of BdHN I, the second peak is considerably lower, so in the end, the fallback accretion leads the $\nu$NS to a slower rotation than its BdHN I counterpart. Still, the $\nu$NS in BdHN II reaches rotation periods of a $\sim 10$ ms, sufficient to explain the afterglow by the associated synchrotron radiation; see Section~\ref{sec:5}. Examples of BdHN II are GRB 180728A \citep{2019ApJ...874...39W} and GRB 190829A; see Section~\ref{sec:8}. 

\subsection{BdHN III}

There are CO-NS binaries with orbital periods that can be even hours, corresponding to the binary separation of the order of a few $10^{11}$ cm. The accretion rate onto the NS companion is negligible, and the SN explosion likely disrupts the binary. In these cases, the fallback accretion onto the $\nu$NS and its interaction with the SN ejecta are the only ones responsible for the long GRB emission. This BdHN III system explains low-luminous GRBs with an energy release of $E_{\rm iso}\sim 10^{49}$--$10^{50}$ erg, and the $\nu NS$ reaches the period of $\sim 50$--$100$ ms, which are sufficient to explain the afterglow by the associated synchrotron emission; see Section~\ref{sec:5}. An example of BdHN III is GRB 171205A, for which we refer the reader to the recent and detailed analysis and simulations presented in \citet{2022ApJ...936..190W} and Section~\ref{sec:9}. 

From all the above, all BdHNe types are endowed with an X-ray afterglow that can be explained by synchrotron radiation powered by the fast-spinning $\nu$NS. 

{If the binary is not disrupted by the mass loss in the SN explosion (see \citealp{2015PhRvL.115w1102F} for details), BdHNe I produce NS-BH binaries and BdHN II NS-NS binaries. In BdHN III, the SN is expected to disrupt the system. For a few minutes binary, the merger time is of the order of $10^4$ yr, when they will lead to short GRBs. Given the short time to merge, the survived newborn compact-object binaries will not travel far from the long GRB site, which implies a direct link between long and short GRBs \citep{2015PhRvL.115w1102F, 2018ApJ...859...30R}. Interestingly, the recent analysis of the population of long and short GRBs by \citet{2023arXiv230605855B} supports the above long-short GRB connection which is a unique prediction of the BdHN model.
}

We now turn to the observational data of $24$ long GRBs and associated Ic SNe and proceed to a selected sample of two BdHN I, one BdHN II, and one BdHN III and their associated HNe.

\begin{deluxetable*}{lllL|LCL|LLLl}
\rotate
\tablewidth{0pt}
\tabletypesize{\scriptsize}
\tablecaption{GRB-SN spectroscopically confirmed sample \label{tab:24-grb-sn-spectroscopic}}
\tablehead{
\colhead{} 
& \colhead{} 
& \colhead{} 
& \colhead{} 
& \multicolumn{3}{c}{This Study} 
& \multicolumn{4}{c}{Literature} \\[-6pt]
\colhead{GRB} 
& \colhead{SN} 
& \colhead{SN} 
& \colhead{$z$} 
& \colhead{$E_{\rm iso,\gamma}$} 
& \colhead{$L_{\rm p, SN}$} 
& \colhead{$t_{\rm p, SN}$} 
& \colhead{$E_{\rm iso,\gamma}$} 
& \colhead{$L_{\rm p, SN}$} 
& \colhead{$t_{\rm p, SN}$} 
&\colhead{Data source}\\[-6pt]
\colhead{Name} 
& \colhead{Name} 
& \colhead{Type} 
& \colhead{Redshift} 
& \colhead{(erg)} 
& \colhead{($\times 10^{42}$~erg/s)} 
& \colhead{(days)} 
& \colhead{($\times 10^{52}$~erg)} 
& \colhead{($\times 10^{42}$~erg/s)} 
& \colhead{($\times 10^{6}$~s)} 
& \colhead{References}
}
\startdata
980425  & 1998bw  & Ic-BL  &  0.0085 & (8.6\pm0.2) \times 10^{47}   & 7.33    & 15.16 & 0.000086
& 14.5 
& 1.30464 & (1)-(5) \\
011121  & 2001ke  & Ic     &  0.362  & (7.8\pm2.1) \times 10^{52}   & 5.90       & 17 & 7.8
& \sim5.9, 13.7 & 1.4688 & (3), (6) \\
021211  & 2002lt  & Ic     &  1.006  & (1.12\pm0.13) \times 10^{52} & 7.20       & 14.00 & 0.828
& - & 2.16 & (3), (7)--(11) \\
030329  & 2003dh  & Ic     &  0.1687 & (1.5\pm0.3) \times 10^{52}   & 10.1       & 12.75 & 1,515
& 10.1 & 1.1016 & (3), (7) \\
031203  & 2003lw  & Ic     &  0.1055 & (8.6\pm4.0) \times 10^{49}   & 12.6       & 17.33 & 0.0098 
& 12.6 & 1.497312 & (3), (7)\\
050525  & 2005nc  & Ic     &  0.606  & (2.5\pm0.43) \times 10^{52}  & 4.47       & 13.10 & 2.945
& - & - & (3), (12) \\
060218  & 2006aj  & Ic-BL  &  0.0334 & (5.3\pm0.3) \times 10^{49}   & 6.47       & 10.42 & 0.0053
& 6.47 & 0.90029 & (3) \\
081007A & 2008hw  & Ic     &  0.5295 & (1.5\pm0.4) \times 10^{51}   & 14.0       & 12.00 & 0.15
& \sim14 & \sim1.0368 & (3)\\
091127  & 2009nz  & Ic     &  0.4904 & (1.5\pm0.2) \times 10^{52}   & 12.0       & 15.00 & 1.5
& \sim12& \sim1.296 & (3)\\
100316D & 2010bh  & Ic-BL  &  0.0592 & >5.9 \times 10^{49}          & 5.67       & 8.76  & >0.0059
& 5.67 & 0.756864 & (3)\\
101219B & 2010ma  & Ic     &  0.5519 & (4.2\pm0.5) \times 10^{51}   & 15.0       & 11.80 & 0.42
& 15  & 1.01952 & (3)\\
111209A & 2011kl  & SLSN-I &  0.677  & (5.82\pm0.73) \times 10^{53} & 29.1       & 14.80 & 58.2
& 29.1 & 1.27872 & (3) \\
120422A & 2012bz  & Ib/c   &  0.2825 & (2.4\pm0.8) \times 10^{50}   & 14.8       & 14.45 & 0.024
& 14.8 & 1.24848 & (3)\\
120714B & 2012eb  & Ib/c   &  0.3984 & (5.94\pm1.95) \times 10^{50} & 6.20       & 13.60 & 0.3174195
& - & 13.6 \pm 0.7
& (3), (13)\\
130215A & 2013ez  & Ic     &  0.597  &  &  &  & 3.1 
& - & - & (3)\\
130427A & 2013cq  & Ic     &  0.3399 & (8.1\pm0.8) \times 10^{53}   & 9.12       & 12.68 & 89
& - & - & (3), (14)--(17) \\
130702A & 2013dx  & Ic-BL  &  0.145  & (6.4\pm1.3) \times 10^{50}   & 10.8       & 12.94 & 0.064
& 10.8, 19.2 & 1.118016 & (3), (6) \\
130831A & 2013fu  & Ib/c   &  0.4791 & (4.6\pm0.2) \times 10^{51}   & 6.90       & 11.90 & 0.59221795
& - & 1.60704\pm 0.05789 
& (3), (13), (18) \\
161219B & 2016jca & Ic-BL  &  0.1475 & (8.50\pm8.46) \times 10^{49} & 4.90       & 10.70 & 0.0858
& 10.4 & 0.92448 & (6), (19), (20)\\
171010A & 2017htp & Ic-BL  &  0.33   & (1.80\pm0.30) \times 10^{53} & 8.4        & 12.80 & 18, 22 & 21 \pm 9 & - & (21)--(23)\\
171205A & 2017iuk & Ic-BL  &  0.0368 & (5.72\pm0.80) \times 10^{49} & 6.5        & 15.08 & 0.00218 & - & 1.09728 
& (24) \\
180728A & 2018fip & Ic-BL  &  0.117  & (2.30\pm0.10) \times 10^{51} & 5.8        & 12.70 & 0.2545
& - & 1.27008 \pm 0.25056 & (25)--(27) \\
190114C & 2019jrj & Ic     &  0.4245 & (3.0\pm0.5) \times 10^{53}   & 6.0        & 10.50 & 30
& - & 1.62432 \pm 0.31968 & (28)--(30) \\
190829A & 2019oyw & Ic-BL  &  0.0785 & (2.0\pm0.3) \times 10^{50}   & 6.27       & 18.00 & 0.018
& - & 0.794016
\pm 0.0216 & (31)--(32) \\
\enddata
\tablecomments{\textbf{Information on SN type} is retrieved from Transient Name Server (\url{www.wis-tns.org}) and SIMBAD Astronomical Database (\url{http://simbad.cds.unistra.fr/simbad/}), except for the following events: SN 2001ke: \citet{Bloom...2002ApJ...572L...45B}; SN 2009nz: \citet{Berger...2011ApJ...743...204B}; SN 2011kl: \citet{Greiner...2015Natur...523...189G}; SN 2019jrj: \citet{2022AandA...659A..39M}; SN 2019oyw: \citet{2021AandA...646A..50H}. 
\textbf{References for} $z$: 980425: \citet{1998Natur.395..670G}; 011121: \citet{Infante...2001GCN...1152...1I}; 021211: \citet{Vreeswijk...2003GCN...1785...1V}; 030329: \citet{Thone...2007ApJ...671...628T}; 031203: \citet{Prochaska...2003GCN...2482...1P}; 050525: \citet{DellaValle...2006ApJ...642L...103D}; 060218: \citet{Pian...2006Natur...442...1011P}; 081007A: \citet{Berger...2008GCN...8335...1B}; 091127: \citet{Vergani...2011AA...535A..127V}; 100316D: \citet{Bufano...2012ApJ...753...67B}; 101219B: \citet{Sparre...2011ApJ...735L...24S}; 111209A: \citet{Vreeswijk...2011GCN...12648...1V}; 120422A: \citet{Schulze...2014AA...566A.102S}; 120714B: \citet{Fynbo...2012GCN...13477...1F}; 130215A: \citet{Cucchiara...2013GCN...14207...1C}; 130427A: \citet{Flores...2013GCN...14491...1F}; 130702A: \citet{Mulchaey...2013ATel...5191...1M}; 130831A: \citet{Cucchiara...2013GCN...15144...1C}; 161219B: \citet{Tanvir...2016GCN...20321...1T}; 171010A: \citet{Kankare...2017GCN...22002...1K}; 171205A: \citet{2017GCN.22180....1I}; 180728A: \citet{Rossi...2018GCN...23055...1R}; 190114C: \citet{2019GCN.23695....1S}; 190829A: \citet{Valeev...2019GCN...25565...1V}; 
\textbf{References for data sources}: 
(1) \citet{1998astro.ph..7345H}, (2) \citet{1999hea..work..123I}, (3) \citet{2017AdAst2017E...5C}, (4) \citet{Yamazaki...2003ApJ...594L..79Y}, (5) \citet{2016MNRAS.457..328L}, (6) \citet{Lian...2022ApJ...931...90L}, (7) \citet{Ulanov...2005NCimC...28...351U}, (8) \citet{Ghirlanda...2004ApJ...616..331G}, (9) \citet{Fox...2003ApJ...586L...5F}, (10)\citet{Pandey...2003AA...408L..21P}, (11) \citet{Della...2004AIPC...727...403D}, (12) \citet{Amati...2006MNRAS...372...233A}, (13) \citet{Klose...2019AA...622A...138K}, (14) \citet{Golenetskii...2013GCN...14487...1G}, (15) \citet{2019ApJ...886...82R}, (16) \citet{Levan...2014ApJ...792..115L}, (17) \citet{Vurm...2014ApJ...789L..37V}, (18) \citet{Cano...2014AA...568A...19C}, (19) \citet{Minaev...2019...arxiv...mnras}, (20) \citet{Frederiks...2016GCN.20323....1F}, (21) \citet{Frederiks...2017GCN.22003...1F}, (22) \citet{Kumar...2022NewA...9701889K}, (23) \citet{Bright...2019MNRAS...486...2721B}, (24) \citet{2018AandA...619A..66D}, (25) \citet{Frederiks...2018GCN.23061...1F}, (26) \citet{2021MNRAS.504.5301R}, (27) \citet{2018GCN.23066....1R}, (28) \citet{2019GCN.23707....1H}, (29) \citet{2019GCN.23715....1R}, (30) \citet{Jordana-Mitjans...2020ApJ...892...97J}, (31) \citet{Tsvetkova...2019GCN.25660...1T}, (32) \citet{2021AandA...646A..50H}.
}
\end{deluxetable*}


\begin{figure*}
\centering
\includegraphics[width=0.850\hsize,clip]{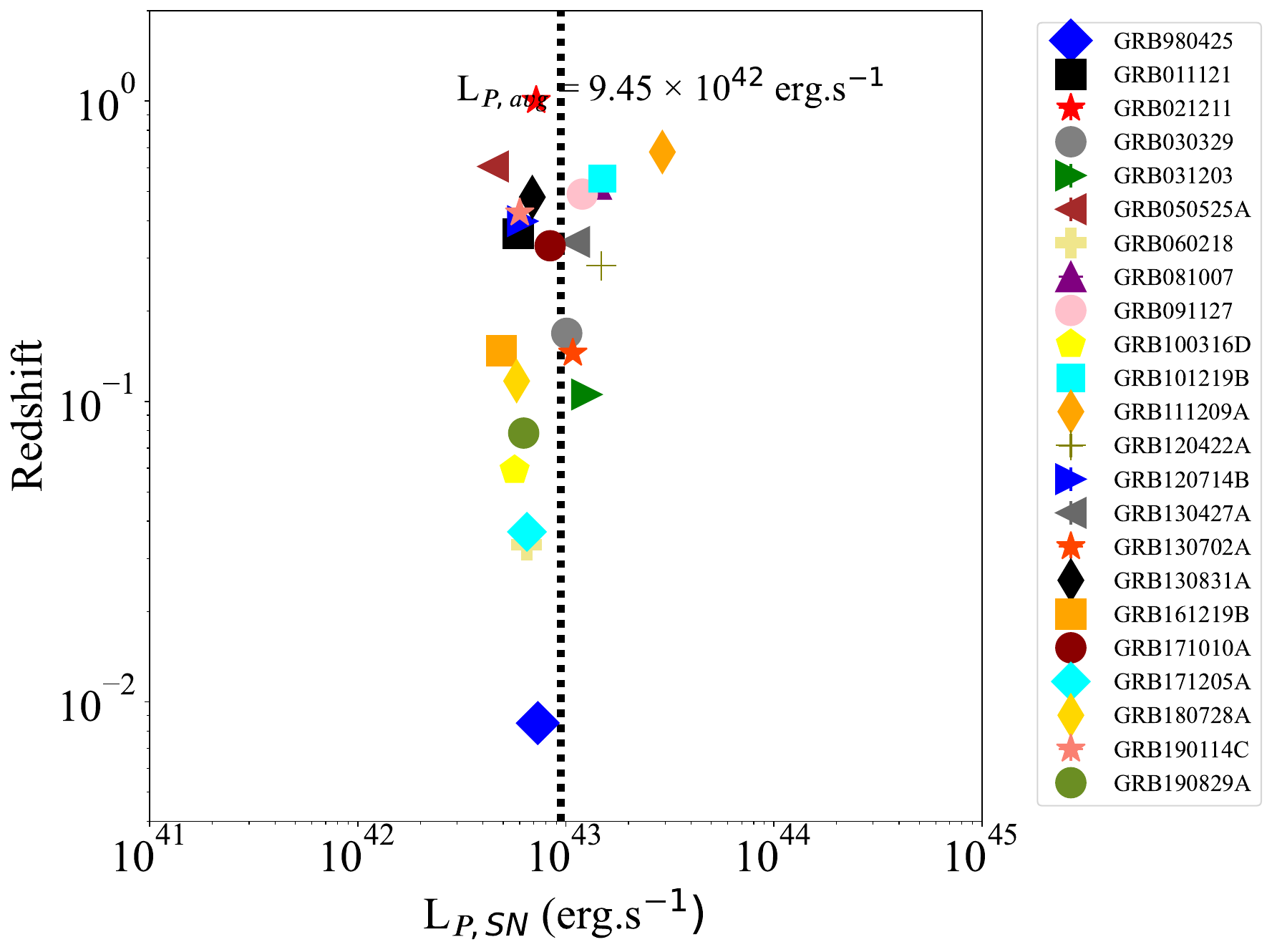} 
\caption{GRB redshifts ($z$) versus the values of peak luminosity of the bolometric light curve of the associated SN ($L_{\rm P, SN}$). The plot shows the spread in data points and the lack of correlation between these two quantities.}\label{fig:redlpeak} 
\end{figure*}

\begin{figure*}
\centering
\includegraphics[width=0.850\hsize,clip]{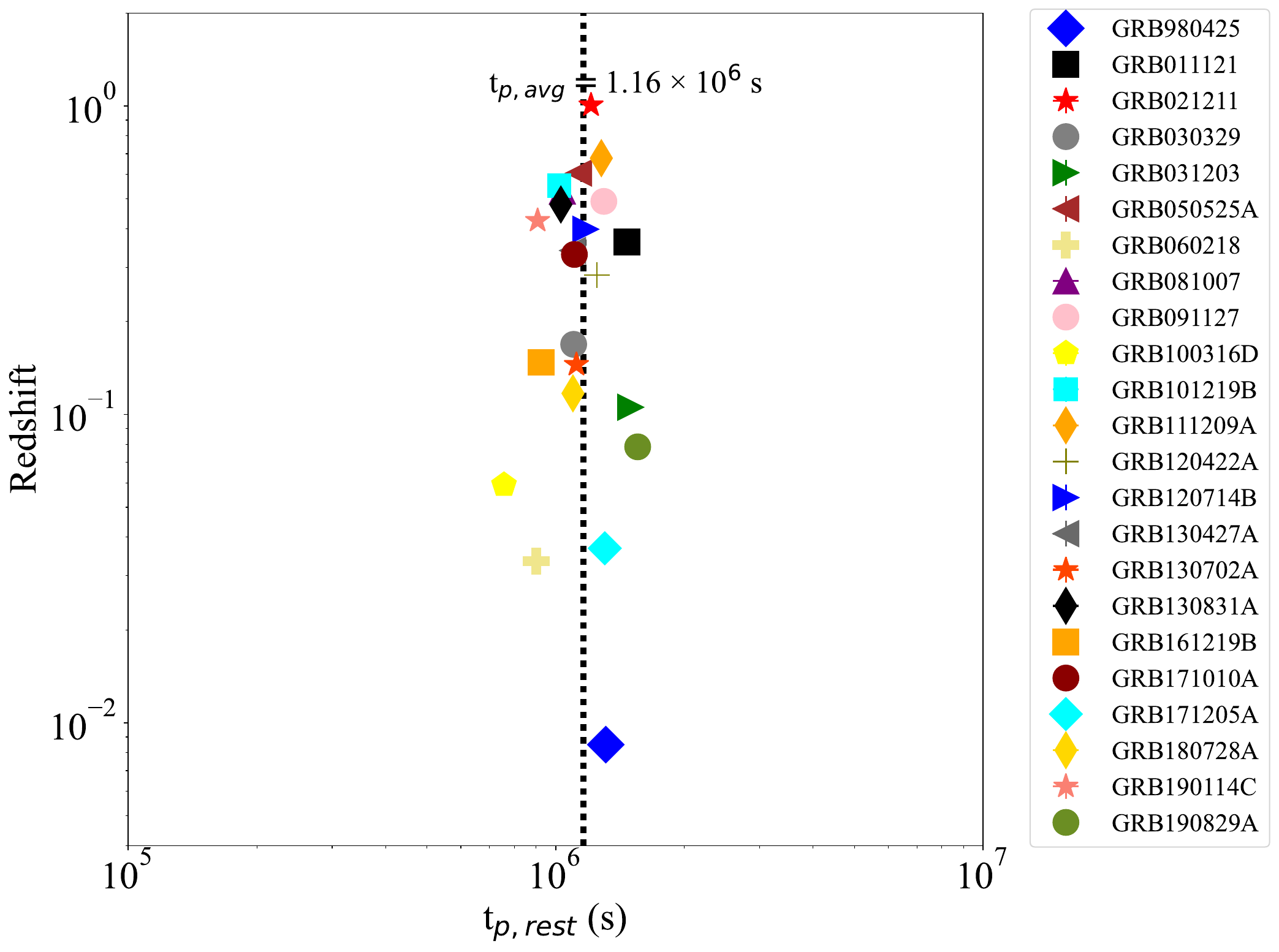} 
\caption{GRB redshifts ($z$) versus the peak time of luminosity of the bolometric light curve of the associated SN ($t_{\rm P, SN}$). The plot shows the lack of correlation between these two quantities.}\label{fig:redtpeak}
\end{figure*} 

\begin{figure*}
\centering
\includegraphics[width=0.850\hsize,clip]{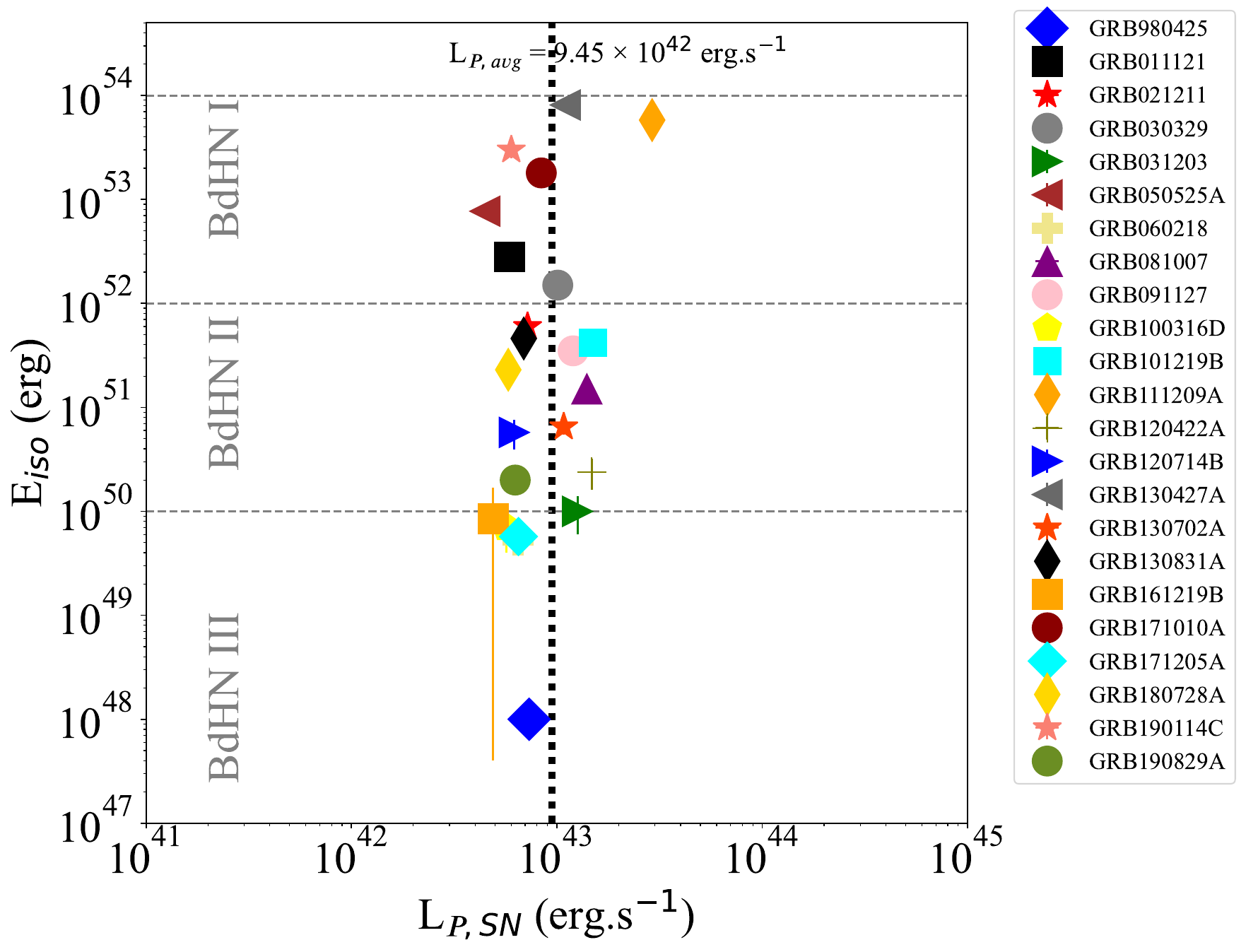} 
\caption{Isotropic-equivalent energy ($E_{\rm \gamma, iso}$) of GRB versus the peak luminosity of the bolometric light curve of the associated SN ($L_{\rm P, SN}$). The plot shows the lack of correlation: the SN luminosities stay within an order of magnitude spread, while the GRB energy spans $\sim 6$ orders of magnitude.}\label{fig:iso-lpeak} 
\end{figure*} 

\begin{figure*}
\centering
\includegraphics[width=0.850\hsize,clip]{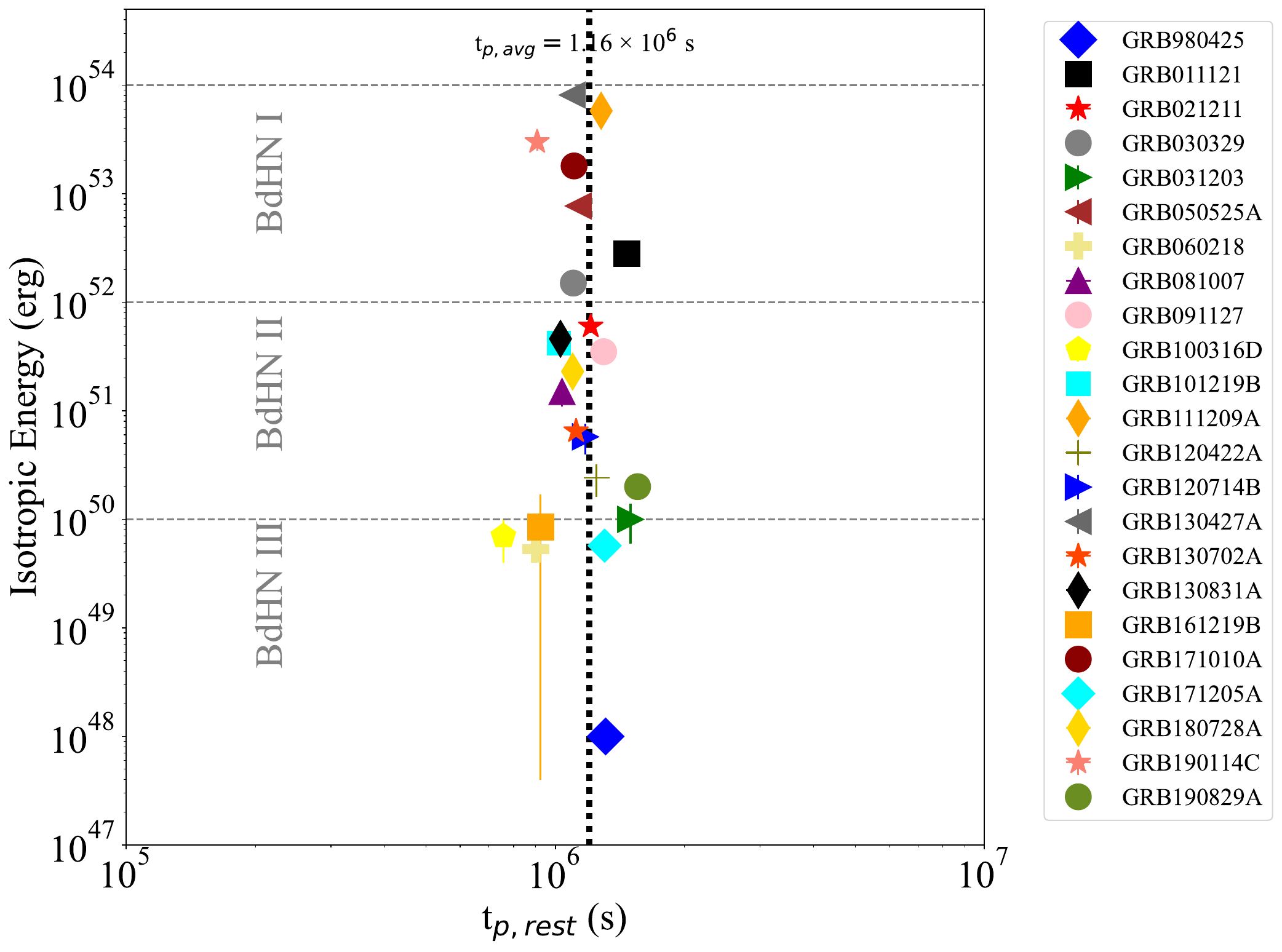} 
\caption{Isotropic-equivalent energy ($E_{\rm \gamma, iso}$) of GRB versus the peak time of luminosity of the bolometric light curve of the associated SN ($t_{\rm P, SN}$). The plot shows the lack of correlation: the SN peaking times (in the rest-frame) stay within an order of magnitude spread, while the GRB energy spans $\sim$6 orders of magnitude.}\label{fig:iso-tpeak}
\end{figure*} 

\begin{table*}
    \centering
    {
    \caption{Physical phenomena occurring in BdHN I, II, and III, and their associated observations in the GRB data. References in the table: $^a$\citet{2019ApJ...874...39W, 2022ApJ...936..190W, PhysRevD.106.083004},$^b$\citet{ 2014ApJ...793L..36F, 2016ApJ...833..107B, PhysRevD.106.083002, PhysRevD.106.083004, 2022ApJ...936..190W}, $^c$\citet{2019ApJ...886...82R, 2021A&A...649A..75M, 2021PhRvD.104f3043M},     $^d$\citet{2001A&A...368..377B, 2021PhRvD.104f3043M, 2022EPJC...82..778R}, $^e$\citet{2019ApJ...886...82R, 2020EPJC...80..300R, 2021A&A...649A..75M, 2022ApJ...929...56R}, $^f$\citet{2018ApJ...852...53R}, $^g$\citet{2018ApJ...869..101R, 2019ApJ...874...39W, 2020ApJ...893..148R}, $^h$ \citet{2017AdAst2017E...5C} and this paper. UPE stands for ultrarelativistic prompt emission, SXFs for soft X-ray flares, HXFs for hard X-ray flares, CED for classical electrodynamics, QED for quantum electrodynamics, SN for supernova, and HN for hypernova.}
    \small\addtolength{\tabcolsep}{-3pt}
    \scriptsize{
   \begin{tabular}{|l|c|c|c|c|c|c|ccc|ccc|c|}
     \hline
     \multirow{4}{*}{\parbox{2.0cm}{Physical\\phenomenon}} & \multirow{4}{*}{\parbox{0.7cm}{\centering BdHN type}} & \multicolumn{12}{c}{GRB Episodes} \vline\\
     \cline{3-14}
      & & 0 & \multicolumn{1}{c}{ I} \vline & II & \multicolumn{1}{c}{III}\vline  & IV & \multicolumn{3}{c}{V}\vline  & \multicolumn{3}{c}{VI} \vline& \multicolumn{1}{c}{VII}\vline\\
     &&(SN-rise) &\multicolumn{1}{c}{ ($\nu$NS-rise)} \vline& (NS-rise) &\parbox{1.5cm}{\centering (BH-rise overcritical)}  & \parbox{1.7cm}{\centering (BH-rise undercritical)} & \multicolumn{3}{c}{(BH echoes)}\vline    & \multicolumn{3}{c}{(Afterglows)} \vline& (SN Ic \& HN)\\
     \cline{3-14}
     & & SN-rise & $\nu$NS-rise  & NS-rise & UPE  & \parbox{1.3cm}{\centering Jetted emission} & Cavity & HXF & SXF & \,\,X\, & Opt. & Rad. & Opt. SN \& HN\\
     & &(X-$\gamma$)&(X-$\gamma$)  & (X-$\gamma$) & (X-$\gamma$) & (GeV) &(X-$\gamma$) &(X-$\gamma$) &(X)  & & & &\\
    
     \hline
    CO core-collapse $^a$ & I,II,III  & $\bigotimes$&  &  & & & & & & & & &  \\
    \cline{1-2}
    $\nu$NS accretion$^b$ & I,II,III & & $\bigotimes$ & & & & & & & & & & \\
    \cline{1-2}
    NS accretion$^b$  & I,II & & &$\bigotimes$ &  & & & & & & & &\\
    \cline{1-2}
    \parbox{2.0cm}{BH QED$^d$} & I & & & &$\bigotimes$& & & & & & & &\\
     \cline{1-2}
     \parbox{2.0cm}{BH CED$^e$} & I & & & & & $\bigotimes$ & & & & & & &\\
    \cline{1-2}
    BH disk accretion$^f$ & I & & & & & & $\bigotimes$& $\bigotimes$ & $\bigotimes$ & & & &\\
     \cline{1-2}
    \parbox{2.2cm}{$\nu$NS synchr. $+$
    pulsar emission$^g$}  & I,II,III & & & & & & & & &$\bigotimes$ &$\bigotimes$ &$\bigotimes$ &\\
     \cline{1-2}
    \parbox{2.0cm}{Nickel decay $+$ ejecta kinetic energy$^h$} & I,II,II & & & & & & & & & & & &$\bigotimes$\\
    \hline
    & & Section~\ref{sec:SNrise} & \multicolumn{1}{c}{Section~\ref{sec:nuns} }\vline & Section~\ref{sec:nuns} & \multicolumn{1}{c}{ Section~\ref{sec:episodesUPE}} \vline & Section~\ref{sec:episodesGEV} & \multicolumn{3}{c}{Section~\ref{sec:episodesCav}}\vline   & \multicolumn{3}{c}{Section~\ref{sec:3afterglows}}  \vline &Section~\ref{sec:episodesclassicSN}\\
    \hline
    \end{tabular}
    }
    }
    \label{tab:observables}
\end{table*}

\section{Cosmological rest-frame time and $k$-correction}\label{sec:3}

We here introduce the conversion factor adopted in deriving a luminosity and time both in the cosmological rest-frame of the source \citep[see][]{2018ApJ...852...53R}. This conversion, known as $k$-correction, has been often neglected in the literature \citep{2007ApJ...671.1903C,2007ApJ...671.1921F,2010MNRAS.406.2149M}.

The observation time ($t_{\rm obs}$) of the source is related to the time measured in the cosmological rest-frame ($t_{\rm rf}$) on the earth by $t_{\rm obs}=(1+z)t_{\rm rf}$. The observed flux $f_{obs}$, namely the energy per unit area and time in a fixed detector energy bandwidth $[\epsilon_{obs,1};\epsilon_{obs,2}]$, is
\begin{equation}
    f_{obs,[\epsilon_{obs,1};\epsilon_{obs,2}]} = \int_{\epsilon_{obs,1}}^{\epsilon_{obs,2}} \epsilon~n_{obs}(\epsilon) d \epsilon\, ,
\label{eq:fluxIntegration}
\end{equation}
where $n_{obs}$ is the photon spectrum, i.e., the number of observed photons per unit energy, area, and time.

The total energy emitted in the$[\epsilon_{obs,1};\epsilon_{obs,2}]$ bandwidth per unit time, which by definition is in the source cosmological rest-frame, is
\begin{equation}
    L_{[\epsilon_{obs,1}(1+z);\epsilon_{obs,2}(1+z)]} = 4 \pi D_L^2(z) f_{obs,[\epsilon_{obs,1};\epsilon_{obs,2}]}\, ,
\label{eq:CorrectedLuminosity}
\end{equation}
where $D_L(z)$ is the source luminosity distance. 

To express the luminosity $L$ in the cosmological rest-frame energy band, $[E_1;E_2]$, common to all sources, we rewrite Eq.(\ref{eq:CorrectedLuminosity}) as
\begin{align}
   L_{[E_1;E_2]}  &= 4 \pi D_L^2 f_{obs,\left[\frac{E_1}{1+z};\frac{E_2}{1+z}\right]} =\\ 
   &4 \pi D_L^2 k[\epsilon_{obs,1};\epsilon_{obs,2};E_1;E_2;z] f_{obs,[\epsilon_{obs,1};\epsilon_{obs,2}]}\, 
\label{eq:kCorrectedLuminosity}
\end{align}
where the $k$-correction factor is defined as
\begin{align}
   k[\epsilon_{obs,1};\epsilon_{obs,2};E_1;E_2;z] =
   & \frac{f_{obs,\left[\frac{E_1}{1+z};\frac{E_2}{1+z}\right]}}{f_{obs,[\epsilon_{obs,1};\epsilon_{obs,2}]}}\\& =\frac{\int_{E_1/(1+z)}^{E_2/(1+z)} \epsilon~n_{obs}(\epsilon) d \epsilon}{\int_{\epsilon_{obs,1}}^{\epsilon_{obs,2}} \epsilon~n_{obs}(\epsilon) d \epsilon}\, .
\label{eq:kCorrectionFactor}
\end{align}

Throughout this article, we use a $\Lambda$CDM cosmology with H$_0 ~\rm =69.6 kms^{-1} ~ Mpc^{-1}$, $\Omega_{\rm M}=0.286$, $\Omega_{\rm \Lambda}=0.714$ for performing the $k$-correction related to the cosmological-rest frame of sources. 

\section{Type Ic Supernovae associated with BdHN I, BdHN II, and BdHN III}\label{sec:4}

We address the observations of a sample of $24$ spectroscopically well-identified SNe associated with long GRBs (GRB-SN). In Table~\ref{tab:24-grb-sn-spectroscopic}, we give the name of the SN, the SN type, the cosmological redshift, our best estimate of the $E_{\rm iso}$ of the associated long GRB, the peak luminosity of the SN ($L_{p, \rm SN}$), and the time of occurrence of the peak ($t_{p, \rm SN}$). We also give the analogous information from the literature in the following three columns. 

The optical observations are performed during the long-lived multiwavelength afterglow of each GRB. As pointed out by \citet[][and references therein]{2017AdAst2017E...5C}, the spectroscopic analysis of the light curve close to their maxima, through the identified presence of strong absorption/emission lines \citep{Cappellaro...2022NCimR...45...549C}, allows classifying the type of the SN, e.g., Ib/c or Ic-BL. The photometric observation also indicates the evidence for an emerging SN by a characteristic rise in the optical afterglow at around $7$--$20$~days after the main GRB trigger. The rise in apparent magnitude points to the energy deposited in the expanding outflow by the decay of radioactive nickel mass synthesized during the SN explosion (see Section~\ref{sec:5}).

Since the first evidence of the GRB-SN association, GRB 980425-SN 1998bw in 2018 \citep{1998Natur.395..670G}, to the end of 2019, there have been detected about $60$~GRB-SN events. We collected the data from literature and catalogs \citep{Poolakkil...2021ApJ...913...60P, Lien...2016ApJ...829...7L}, Gamma-ray Coordinates Network (GCN),\footnote{\url{https://gcn.gsfc.nasa.gov}} online tables\footnote{\url{https://www.mpe.mpg.de/~jcg/grbgen.html}}$^,$\footnote{\url{https://user-web.icecube.wisc.edu/~grbweb\_public/index.html}} and databases.\footnote{\url{https://www.wis-tns.org}}$^,$\footnote{\url{http://simbad.cds.unistra.fr/simbad/}} Among these associations, there are $24$~SNe identified spectroscopically and $26$~SNe showing only a prominent ``bump'' in the late optical afterglow and any obtained spectra.\footnote{It was noted by \citet{Cappellaro...2022NCimR...45...549C} that due to recent emergence of transient surveys, the current SN discovery rate is counting to about a thousand event per year. Thus, only a small fraction of them receives a spectroscopic confirmation.} Interestingly, half of the sample occurred within \textit{Fermi} space observatory operational era, thus extending information to the high-energy counterpart of the accompanying GRBs \citep{Ajello...2019ApJ...878...52A}; see Table~\ref{tab:24-grb-sn-spectroscopic}.
 
Due to incomplete data in some of the observed GRB-SN, we cannot use the entire population. Therefore, we further focus on the $24$ spectroscopically confirmed SNe associated with GRBs to the end of $2019$. 

The peak luminosity integrated over the optical bands is similar in all observed SNe associated with GRBs independent of their redshift; see Fig.~\ref{fig:redlpeak}. The same applies to the time of occurrence of the peak measured since the GRB trigger and is independent of the redshift of the SN; see Fig.~\ref{fig:redtpeak}. As we will point out in Section~\ref{sec:5}, the determination of the trigger time strongly depends on the luminosity of the GRB and the instrument with the indeterminacy of $\sim 10^{4}$~s. The average peak bolometric luminosity is $L_{\rm p,avg}=(9.45\pm 3.8)\times 10^{42}$~erg s$^{-1}$ and the average peaking time in the rest-frame is $t_{\rm p,avg}=(1.16\pm 0.24)\times 10^{6}$~s.

Quite apart from this universality, it follows from Figs.~\ref{fig:iso-lpeak} and \ref{fig:iso-tpeak} that the peak luminosity of the associated SN Ic and its time of occurrence are not correlated to the $E_{\rm iso}$ of the BdHN I, II, and III.

As recalled in the Introduction, we assume that the progenitors of the SN Ic associated with long GRBs are composed of a $\sim 10 M_\odot$ CO star and a $\sim2 M_\odot$ companion NS. As recalled in Section \ref{sec:2}, the same progenitors also characterize the BdHNe. In both cases, the trigger is marked by the collapse of the CO core. From the results presented above, a new problem arises: how can the thermonuclear evolution of the SN Ic, characterized by a standard energy of $\sim 10^{49}$ erg, be unaffected by the presence of BdHN I, II, and III with energies in the range of $\sim 10^{49}$--$10^{54}$~erg. To answer this fundamental question and the above energetic difference, we proceed in Section~\ref{sec:5} to illustrate the physical processes in the seven fundamental episodes characterizing a most general BdHN and their spectral properties. In Sections~\ref{sec:6}--\ref{sec:9}, we provide BdHN I, II, and III examples.

\section{BdHN emission episodes}\label{sec:5}

The advantage of introducing the BdHN model may be to bring a certain amount of clarity in a field in which a great deal of confusion exists even in interpreting the specific spectral data \citep[see, e.g.,][]{2022arXiv221112187L}.

We have recalled in the Introduction the differences in addressing the fundamental question of what is considered a long GRB: in the traditional literature, the GRB is described by a single event originating from a ``collapsar'' and manifesting itself by an ultra-relativistic jetted emission. A much more scientifically complex and vaster picture starts from a binary progenitor.

We have also recalled how the large observational support and the equally profound theoretical comprehension following the breakthrough of the BeppoSAX promoted the unification of traditional gamma-ray astronomy to X-ray astronomy. This led to an expansion to additional multi-wavelength observations. The leading conceptual progress has emerged from explaining the spatial and temporal coincidence of two very different astrophysical events: the occurrence of SN Ic and the occurrence of long GRBs.

The BdHN model is rooted in the explanation of this coincidence, as explained in this article: we soon realized that both systems have a common origin in a progenitor composed of a CO core and a binary NS companion (see Section \ref{sec:1}). Their evolution leads to an SN explosion which, in addition to a large amount ($7$--$8 M_\odot$) of ejecta, gives origin to a millisecond pulsar at its center. We have indicated in Section~\ref{sec:2} the crucial role of the initial large angular momentum of the CO-NS binary systems due to the short initial binary period $P_{\rm bin}$. Three different BdHN types originate from very different energies: BdHN I with $P_{\rm bin}$  of $\sim 4$--$5$ min and energies ranging in $10^{52}$--$10^{54}$~erg, BdHN II with $P_{\rm bin}\sim$ 20 min and energies ranging $10^{50}$--$10^{52}$~erg, BdHN III with $P_{\rm bin}$ up to a few hours and energies below $10^{50}$~erg. Equally remarkable is the fact that the same progenitors, as shown in Table~\ref{tab:observables} and Figs.~\ref{fig:redlpeak}--\ref{fig:iso-tpeak}, lead to SNe Ic of a standard energy of $10^{49}$~erg and an HN with the kinetic energy of $\sim 10^{52}$ erg. This result points to a thermonuclear evolution of the SN Ic largely independent of the associated GRB.

The present effort is dedicated to addressing the physics and evolution of GRBs and SN Ic with quantum and classical field theories, which are currently full of conceptual holes. Within the BdHN model, we address the explanation of the above observational facts and justify the assumptions we have made. We have identified seven basic Episodes in the most general BdHN. Each Episode has been characterized through a specific new physical process, partly an extension to new extreme regimes of previously known processes or new processes introduced here for the first time. This has been made possible by observations in the extragalactic of phenomena never observed in our local Universe. Each Episode has been duly scrutinized, and the new physical laws introduced for their explanation have been validated by a time-resolved spectral analysis. The importance of these Episodes can hardly be overestimated since they offer the most reliable guide we have in classifying and interpreting the rapidly growing and already very complex observational picture. After some general considerations, we refer in the following sections to the seven specific Episodes, the related observable, and the BdHN type in which they are present. We then proceed in the following sections to specific examples; two BdHNe I in Section~\ref{sec:6} on GRB 180720B (in Table~\ref{tab:observables} we identify the physical phenomena), on GRB 190114C in Section~\ref{sec:7}, GRB 190829A as a BdHN II in Section~\ref{sec:8}, and GRB 171205A as BdHN III in Section~\ref{sec:9}.

{In the following, we refer to as``MeV'' emission the radiation in the $100$ keV--$10$ MeV energy range typical, e.g., of Fermi-GBM; ``GeV'' emission the radiation in the $100$ MeV--$10$ GeV energy range, typical of Fermi-LAT, and ``TeV'' emission the radiation at higher energies, above $100$ GeV, e.g., typical of H.E.S.S. and MAGIC.}

\setcounter{subsection}{-1}
\subsection{The SN-rise}\label{sec:SNrise}

As mentioned in Section \ref{sec:4}, the BdHN process, which includes the formation of an SN Ic and the associated GRB, is triggered by the gravitational collapse of the CO core. The early detection of this event, namely the first appearance of the SN related to the CO core collapse (SN-rise), is quite rare. It depends on various factors, including the GRB energy, the distance of the source, and especially the operation of the  multi-wavelength detectors at the unpredictable moment of the occurrence of the gravitational collapse. The possible examples in BdHN I {are GRB 160625B \citep{2021MNRAS.504.5301R}, GRB 221009A and GRB 220101A (Ruffini et al, in preparation)}. We are progressing in determining this episode's spectral signature, which is essential to identify the underlying physical processes originating the SN explosion. 
These observational features constrain SN explosion models, which still need theoretical developments to provide successful explosions in the presence of a CO core with substantial rotation and match the GRB-SN features. Although we have mentioned the difficulties in the observational identification of this Episode, we have recently identified it in a handful of GRBs (Ruffini et al., in preparation).

Subsequently to the SN-rise, the hypercritical accretion of the $7$--$8 M_\odot$ onto the $\nu$NS and the NS companion shows up as Episodes of the GRB  prompt emission \citep{2016ApJ...833..107B, 2019ApJ...874...39W, 2022ApJ...936..190W, 2022PhRvD.106h3002B, 2022ApJ...936..190W}.

\subsection{The $\nu$NS-rise}\label{sec:nuns}

The prompt GRB emission starts with the transfer of energy and angular momentum due to the accretion of the SN ejecta both on a very rapidly spinning $\nu$NS and the slower rotating companion NS. The period of the $\nu$NS ranges from $1$ ms in the case of a BdHN I to $\sim 100$ ms periods in the case of a BdHN III. We have indicated as $\nu$NS-rise this first BdHN Episode. This process occurs in all three BdHNe types, with a characteristic CPL spectrum \citep[see, e.g.,][]{2022ApJ...939...62R}. In parallel to the $\nu$NS emission, the SN ejecta accretion that occurs on the companion NS is energetically much weaker. However, in the case of BdHN I, the hypercritical accretion onto the NS companion, a few seconds after the trigger given by the $\nu$NS-rise, leads to the formation of the BH and the new Episode of the ultra-relativistic prompt emission (UPE)  occurs, with a clear CPL + thermal emission (see Section~\ref{sec:episodesUPE}). Initially, the UPE and the $\nu$NS-rise emissions have comparable luminosities. In the case of GRB 180720B, a first $\nu$NS-rise I Episode, lasting $4.84$ s, is followed by a prominent UPE I Episode lasting $1.21$ s, both identifiable by their different spectral properties. Soon after, the $\nu$NS-rise II Episode starts, lasting for $3.02$ s, followed by the UPE II Episode for $1.82$ s; see details in Table.~\ref{tab:Summary}. What is fascinating and identifiable is the non-interference of the emission process from the $\nu$NS-rise and the UPE. A similar behavior is present in GRB 190114C; see details in Table~\ref{tab:Summary1}.

In both cases of GRB 180720B and GRB 190114C, the millisecond rotation of $\nu$NS has given the possibility of examining the equilibrium configurations of a triaxial Jacobi ellipsoid soon evolving into a Maclaurin spheroid with possible emission of gravitational waves \citep{2022PhRvD.106h3004R}. Such possibility, theoretically indicated as necessary in the early evolution of the crab nebula pulsar \citep{1969ApJ...158L..71F}, can now be submitted to direct observations in BdHN I.

Following the $\nu$NS-rise, which again we recall exists in all BdHN types, the synchrotron radiation emitted by the rapidly spinning $\nu$NS, in the wavelengths ranging from X-rays to optical to radio gives origin to the afterglows. It is satisfactory that the afterglows are identically present in all BdHN types; see Section~\ref{sec:3afterglows}.

Numerical simulations show that the accretion process can be observed as a double-peak emission, where the relative time and intensity of the peaks depend on the orbital period and the angular momentum of the NS at the beginning of the accretion process \citep[see][for details]{2019ApJ...871...14B, 2022PhRvD.106h3002B}. The NS companion can reach the critical mass for BH formation before the second peak of fallback accretion onto the $\nu$NS \citep[see][for recent simulations]{2019ApJ...871...14B, 2022PhRvD.106h3002B}. Since the accretion process and associated $\nu$NS-rise is not exclusive of binaries forming a BH, the above double--peak emission from the accretion can appear as the prompt emission in a BdHN II, as in the case of GRB 190829A \citep{2022ApJ...936..190W}. The prompt emission appears without a double-peak structure in BdHN III, like in GRB 171205A \citep{2022arXiv220802725W}; see Section~\ref{sec:9}.

We refer to Section \ref{sec:6} for details on the $\nu$NS-rise in GRB 180720B, Section \ref{sec:7} for GRB 190114C, Section \ref{sec:8} for GRB 190829A, and Section \ref{sec:9} for GRB 171205A.

\subsection{The UPE phase}\label{sec:episodesUPE}

The UPE phase is the first new process that has made possible the extrapolation of the well-known quantum electrodynamics (QED) process of vacuum polarization, which for a long time approached in earth-bound experiments without reaching observational support, and now observing the new regime of overcritical fields in extragalactic astrophysics sources \citep[see][and references therein]{2010PhR...487....1R}.

These processes were pioneered by decades of theoretical works in the 1930s by Paul Dirac \citep{ 1930PCPS...26..361D}, Gregory Breit and John Archibald Wheeler \citep{1934PhRv...46.1087B} and by Fritz Sauter \citep{1931ZPhy...69..742S,1931ZPhy...73..547S}, Werner Heisenberg and Hans Euler \citep{1936AnP...418..398E,1936ZPhy...98..714H}, and later in the 1940s by Julian Schwinger \citep{1948PhRv...74.1439S,1949PhRv...75..651S,1949PhRv...76..790S}, and Richard Feynmann  \citep{1948RvMP...20..367F,1949PhRv...76..749F,1949PhRv...76..769F}; see e. g. \citet{2009PhRvD..79l4002C,2010PhR...487....1R}. Despite many efforts, the inverse of the Breit-Wheeler process, namely pair creation by two photons, was never observed in Earth-bound experiments neither in the past at DESY and SLAC, nor in the present in Brookhaven and Darmstadt, nor at ELI \url{https://eli-laser.eu/} or XFEL \url{https://www.xfel.eu}. It is today clear that these processes are routinely observed in GRBs on the vastest possible energy scales up to $10^{54}$ erg/s, on the shortest time intervals up to $10^{-9}$ s, and highest energies up to $\sim 10^{18}$ eV. 

A novel \textit{hierarchical} (\textit{self-similar}) structure has been evidenced in the UPE spectra of GRB 190114C and GRB 180720B, composed of a black body (BB) plus a cutoff power-law (CPL) model; see sections~\ref{sec:6} and \ref{sec:7}. Namely, the spectra of the UPE, rebinned in time intervals up to a fraction of a second, are all fitted by analogous BB+CPL models. This feature implies a microscopic phenomenon at work on ever shorter timescales. The explanation of the UPE phase of these BdHN I require the interplay of general relativity, QED, and plasma physics in an overcritical regime, which has been observed for the first time.

In BdHN I, ionized matter and the magnetic field inherited from the collapsed NS surround the newborn Kerr BH. These three components comprise the \textit{inner engine} that drives the GRB radiation above MeV energies, i.e., the prompt and the GeV emission \citep{2019ApJ...886...82R, 2020EPJC...80..300R, 2021A&A...649A..75M, 2021MNRAS.504.5301R, 2021PhRvD.104f3043M}.

The QED process at work in the UPE originates in the vacuum polarization of the BH vicinity by the electric field, $E$, induced by the gravitomagnetic interaction of the Kerr BH and the magnetic field, $B_0$. At the BH horizon, $r=r_H =(1+\sqrt{1-\alpha^2})G M/c^2$, the electric field is approximately given by \citep[see, e.g.,][]{2019ApJ...886...82R, 2020EPJC...80..300R}
\begin{equation}\label{eq:Efield}
    E(r_H)\sim \frac{v_H}{c} B_0\sim \frac{\Omega_H r_H}{c} B_0 = \frac{\alpha\,B_0}{2}\approx \frac{Q_{\rm eff}}{r_H^2},
\end{equation}
where $M$, $J$, $\alpha = c J/(G M^2)$, and $\Omega_H= c\,\alpha/(2\,r_H)$ are, respectively, the BH mass, angular momentum, dimensionless spin parameter, and angular velocity. The last expression introduces the \textit{effective charge} (the BH has zero net charge), defined by $Q_{\rm eff} = (G/c^3) 2 B_0 J$ \citep[see][for details]{2019ApJ...886...82R, 2020EPJC...80..300R, 2021A&A...649A..75M}. 

For a magnetic field strength $B_0 > 2 B_c/\alpha_0$, or conversely, for an initial BH spin parameter $\alpha_0 \geq 2 B_c/B_0$, the induced electric field is initially overcritical, i.e., $E(r_H) \geq E_c = m_e^2 c^3/(e \hbar)\approx 1.32\times 10^{16}$ V cm$^{-1}$. Therefore, in a short time-scale of the order of the Compton time, $\sim \hbar/(m_e c^2) \approx 10^{-21}$ s, the approximate vacuum around the BH is rapidly filled with electron-positron pairs ($e^+e^-$), forming an optically thick plasma. The $e^+e^-$ pairs self-accelerate and engulf baryons from the low-density medium around the BH. The plasma reaches transparency at large distances from the BH (e.g., $R_{\rm tr}\sim 10^9$ cm), with large Lorentz factor (e.g., $\Gamma\sim 10^2$; see \citealp{2021PhRvD.104f3043M}). There is no single transparency event but a train of transparencies that continues when the electric field reaches the critical value. This occurs when the spin parameter has been reduced from its initial value, $\alpha_0$, to $\alpha \sim 2 B_c/B_0$.

The $e^+e^-$ plasma energy comes from the electric energy stored in the electric field induced by the interaction of the external magnetic field and the gravitomagnetic field of the Kerr BH. Thus, the ultimate energy reservoir is the BH extractable energy, $E_{\rm ext} = (M-M_{\rm irr})c^2$, where $M_{\rm irr}$ is the BH irreducible mass. The latter is related to the other BH parameters by the mass-energy formula \citep{1970PhRvL..25.1596C,1971PhRvD...4.3552C,1971PhRvL..26.1344H}
\begin{equation}
\label{aone}
M^2 = \frac{c^2 J^2}{4 G^2 M^2_{\rm irr}}+M_{\rm irr}^2.
\end{equation}
As shown in \citep{2021PhRvD.104f3043M, 2022EPJC...82..778R}, each transparency process reduces the BH angular momentum by a small fractional amount $\Delta J/J \sim 10^{-9}$, leading to a slightly smaller angular momentum $J^* = J- \Delta J$. The BH mass changes by $\Delta M \approx \Omega_H \Delta J/c^2$ (keeping the BH irreducible mass approximately constant in the process), so $\Delta M/M \sim \Delta J/J$. Therefore, the system starts a new process with the same magnetic field $B_0$, kept constant, and a new effective charge of $Q^*_{\rm eff}= Q_{\rm eff}-\Delta Q_{\rm eff}$, with $\Delta Q_{\rm eff}/Q_{\rm eff} = \Delta J/J$.
 
We refer to Section \ref{sec:6} (and \citealp{2022EPJC...82..778R}) for details on the UPE phase in GRB 180720B, and Section \ref{sec:7} (and \citealp{2021PhRvD.104f3043M}) for GRB 190114C.
 
The UPE structure has been found as well in GRB 160625B ($z=1.406$), extending from $t_{\rm rf}=77.72$~s to $t_{\rm rf}=87.70$~s, and GBR 160509A ($z=1.17$), spanning from $t_{\rm rf}=4.84$~s to $t_{\rm rf}=8.53$~s \cite[see][for more details]{2023ApJ...945...10L}. There, the detailed time-resolved spectral analysis of the UPE phase of GRB 160625B has been given in Table 2, Fig. 4, as well as the luminosity and the temperature of the thermal components as a function of the rest-frame time in Fig. 5. The same analysis has been carried out for the UPE phase of GRB 160509A, presented in Table 4, Fig. 8, as well as Fig. 9 of \citet{2023ApJ...945...10L}. Although the UPE has been successfully analyzed in both sources, we are verifying the remaining six Episodes.

Thus, the UPE is expected to be present only in the prompt emission of BdHN I. The $\ nu$NS-rise instead dominates the prompt emission of BdHN II. We advance the possibility that a UPE-like emission could also occur under some conditions around a highly magnetic, fast-rotating NS, and the differences between the two cases could be checked through the prompt emission of BdHNe I and II.

\subsection{High-energy jetted (GeV) emission}\label{sec:episodesGEV}

{The UPE ends when} the strength of the induced electric field {becomes} lower than the critical field's. {Hence,} the vacuum polarization's QED process is no longer active. Yet, the {induced electric field is} sufficiently large to power the GeV emission by {the following} classical electrodynamics (CED) process. The electric field accelerates charged particles that move along and spiral around the magnetic field lines given the magnetic dominance, i.e., $\rm \bf{B}^2-\bf{E}^2 > 0$, leading to radiation {by acceleration, e.g., synchrotron emission}. In particular, for a magnetic field aligned and parallel to the BH spin, electrons move outward in the polar region around the BH rotation axis ($\theta=0$) comprised at angles $-60^\circ \lesssim \theta \lesssim 60^\circ$ in the northern hemisphere, and the analogous region in the southern hemisphere because of the reflection symmetry of the Kerr BH spacetime. For the involved pitch angles \citep[see, e.g.,][for details]{2021A&A...649A..75M}, those electrons emit most of the synchrotron radiation at GeV energies with a luminosity that explains the observed GeV radiation in (some, see below) long GRBs \citep{2019ApJ...886...82R, 2020EPJC...80..300R, 2021A&A...649A..75M}. We refer the reader to \citet{2022ApJ...929...56R} for a fully general relativistic treatment of the above process. As for the UPE phase, the BH extractable energy powers the GeV emission, which decreases with time following a power-law with an index of $\alpha_{\rm GeV} = -1.19 \pm 0.04$. Thus, the mass and angular momentum of the BH keeps decreasing with time. In this case, each process of emission extracts a fraction of the BH mass-energy $\Delta M/M\sim 10^{-18}$ and angular momentum $\Delta J/J \sim 10^{-16}$ \citep[see, e.g.,][]{2021A&A...649A..75M, 2022ApJ...929...56R}.

Unlike the isotropic afterglow emission, which originates from the $\nu$NS and is present in \textit{all} types of BdHN, the GeV radiation occurs only in BdHN I since the Kerr BH power it and is anisotropic, occurring in a double-cone of semi-aperture angle $\approx 60^\circ$, centered on the BH rotation axis. Therefore, it is not observable in every BdHN I, which explains the absence of observed GeV emission in a fraction of them \citep[see][for details]{2021MNRAS.504.5301R}.

We refer to Section \ref{sec:6} for details on the GeV emission in GRB 180720B (see also \citealp{2019ApJ...886...82R}), and Section \ref{sec:7} for GRB 190114C (see also \citealp{2020EPJC...80..300R, 2021A&A...649A..75M}).

\subsection{The BH echoes}\label{sec:episodesCav}

The hypercritical accretion onto the NS companion and the consequent BH formation in BdHN I decrease the matter density around the BH \citep{2019ApJ...871...14B}. Numerical simulations show that the expanding $e^+e^-$ plasma causes a further decrease of the density from $10^{-7}$ g cm$^{-3}$ to a value as low as $10^{-14}$ g cm$^{-3}$. The collision and partial reflection of the expanding $e^+e^-$ plasma with the cavity walls generates emission, known as \textit{cavity}, characterized by a spectrum similar to a Comptonized blackbody with a peak energy of a few hundreds of keV  \citep{2019ApJ...883..191R}.


The density of the matter surrounding the newborn BH site is highly asymmetric (see Fig. \ref{fig:3DSPH}). Consequently, the number of baryons that the $e^+e^-$ plasma loads during its expansion have an angular dependence. The transparency of the plasma in regions with ${\cal B}\lesssim 10^{-2}$ explains the radiation of the UPE phase, being ${\cal B}$ the baryon load parameter. The transparency in regions with ${\cal B} \sim 50$ and Lorentz factors of $\Gamma \lesssim 5$ explain the SXFs and HXFs \citep[see][for numerical simulations]{2018ApJ...852...53R}. The emission is visible at intermediate angles between the binary plane and the rotation axis \citep[see, e.g.,][]{2021MNRAS.504.5301R}. We notice that low Lorentz factors $\Gamma \lesssim 5$ are indeed inferred from the time-resolved analysis of the X-ray data, which rule out any ultrarelativistic bulk motion (e.g., massive jets) of the emitter \citep[see][for details]{2018ApJ...852...53R}.  

We expect SXFs and/or HXFs to appear only in BdHN I since they are related to the transparency in the high-density regions of the $e^+e^-$ plasma, originated in the formation of the newborn Kerr BH (explained above in the UPE). However, the emission is not observable in every BdHN I because of the angular dependence of the emission, which becomes visible only for lines-of-sight close to the binary plane \citep{2018ApJ...852...53R}.

\subsection{Multiwavelength (X, optical, radio) afterglow}\label{sec:3afterglows}

In the BdHN scenario, the synchrotron radiation generated by relativistic electrons in the ejecta expanding in the magnetized medium provided by the $\nu$NS magnetic field, and powered by the $\nu$NS rotational energy, explains the afterglow emission in the X-rays, optical, and radio wavelengths \citep{2018ApJ...869..101R, 2019ApJ...874...39W, 2020ApJ...893..148R,2022ApJ...939...62R}.

Because the afterglow emission depends only on the existence of the $\nu$NS, the SN ejecta, and the synchrotron radiation from an isotropic distribution of pitch angles is isotropic, the afterglow synchrotron emission must be present in \textit{all} BdHNe. Indeed, the X-ray afterglow is observed in all the $380$ BdHN I identified in \citet{2021MNRAS.504.5301R}, and in all observed BdHN II and III, as shown in this article, which proves that the afterglow emission is spherically symmetric with excellent approximation. A further implication comes from the nature of the BdHN progenitor. Every gravitational collapse of a CO star with a sufficient short orbital period must necessarily lead to a $\nu$NS (see Conclusions). 

A semi-analytic theoretical treatment of the above synchrotron emission in BdHN can be found in \citet{2022ApJ...939...62R, 2022ApJ...936..190W}. The synchrotron luminosity follows a power-law behavior with the same power-law index in all energy bands. The fit of the multiwavelength afterglow data with the above model gives information on the SN ejecta expansion velocity, the $\nu$NS magnetic field, the energy and distribution of electrons in the ejecta, and the power injected by the $\nu$NS into the SN ejecta. This description of the GRB afterglow within the BdHN scenario differs from traditional GRB models, which consider that an ultra-relativistic jet with Lorentz factor $>100$ produces the prompt emission and then continues to expand, leading to the afterglow by the synchrotron emission from the accelerated electrons swept in.

In general, the X-ray emission has the contribution of the synchrotron emission and the $\nu$NS pulsar. The $\nu$NS pulsar luminosity is characterized by a plateau, followed by a power-law decay at times longer than the characteristic spin-down timescale. Thus, in the X-rays, the sum of the synchrotron and the pulsar emission can result in a power-law luminosity that is shallower than the power-law luminosity of pure synchrotron radiation. Therefore, from the energetics of the afterglow, and the fit of the X-ray light curve, it is possible to infer the evolution of the $\nu$NS rotation period and magnetic field strength \citep[see, e.g.][]{2018ApJ...869..101R, 2019ApJ...874...39W, 2020ApJ...893..148R, 2021MNRAS.504.5301R, 2022ApJ...939...62R, 2022ApJ...936..190W}.

\subsection{The classic SN emission powered by nickel decay}\label{sec:episodesclassicSN}

Finally, the emission is observed in the optical band powered by the energy release of nickel decay (into cobalt) in the SN ejecta. We refer to \citet{2021IJMPD..3030007R, 2021ARep...65.1026R, 2019Univ....5..110R} for recent reviews on the BdHN scenario of long GRBs and the related physical phenomena.

The nuclear energy released by the decay of nickel into cobalt within the SN ejecta powers the observed energy of the SN Ic emission. The SNe associated with GRBs are similar to each other irrespectively on the GRB energetics (see, e.g., \citealp{2017AdAst2017E...5C} and this article). The GRB-SN connection is one of the most relevant observational properties constraining GRB models. We introduce in this article additional observational features of the GRB-associated SNe and discuss how they constrain GRB models.

Therefore, within the BdHN model, the SN optical emission is always present and observable for $z<$1 with current telescopes or $z>$1 for future missions. Using the BdHN model, we have successfully predicted the time of occurrence and luminosity of the SN optical emission for the BdHN I, GRB 130427A \citep{2013GCN.14526....1R}, GRB 190114C \citep{2019GCN.23715....1R}, GRB 211023A \citep{2021GCN.31056....1A}, and GRB 221009A \citep{2022GCN.32780....1A}; for the BdHN II, GRB 180728A \citep{2018GCN.23066....1R}, and GRB 190829A \citep{2022ApJ...936..190W}; for the BdHN III, GRB 171205A \citep{2022ApJ...936..190W}.

Having given the details of the physical origin of each episode and the information about the time-resolved spectral analysis, we now turn to specific examples of two BdHNe I; GRB 180720B in Section~\ref{sec:6}, and GRB 190114C in Section~\ref{sec:7}, one BdHN II; GRB 190829A in Section~\ref{sec:8}, and one BdHN III; GRB 171205A in Section~\ref{sec:9}. 

\begin{table*}
\centering
\caption{The episodes and afterglows of GRB 180720B. This table reports the name, the underlying astrophysical process, the duration (s), the best-fit spectrum, and the isotropic energy (erg) for each event in GRB 180720B. GRB 180720B has a redshift $z=0.654$ and $T^{\rm total}_{90}=29.56$ s (corrected in the rest frame). The NS-rise in GRB 180720B is not observable because of the formation of the BH.}
\label{tab:Summary}
\small\addtolength{\tabcolsep}{2pt}
\begin{tabular}{l|c|c|c|c|l}
\hline
Episode &Event & duration(s) & Spectrum & $E_{\rm iso}$ (erg)&Physical phenomena\\
\hline
\hline
\rowcolor{white}0&{\bf SN-rise}&{ --}&{--}&--& CO$_{\rm core}$ collapse\\
\hline 
\rowcolor{white}I&\textcolor{black}{\textbf{$\nu$NS--rise }}& \textcolor{black}{}& \textcolor{black}{}& \textcolor{black}{}&\textcolor{black}{$\nu$NS accretion}\\
\rowcolor{white}&\textcolor{black}{$\nu$NS-rise I}& \textcolor{black}{4.84}& \textcolor{black}{Band}& \textcolor{black}{$(1.53\pm 0.09) \times 10^{53}$}&\textcolor{black}{}\\
\rowcolor{white}&
\cellcolor{white}\textcolor{black}{$\nu$NS-rise II}&\textcolor{black}{3.02} &\textcolor{black}{CPL}& \textcolor{black}{$(1.13\pm0.04) \times 10^{53}$}&\textcolor{black}{}\\
\hline 
\rowcolor{white}II&{\bf NS-rise}&{ Not observable}&{Not observable}&Not observable&Companion NS accretion\\
\hline 
\rowcolor{white}III&\textcolor{black}{\textbf{BH-rise (overcritical)}}& \textcolor{black}{}& \textcolor{black}{}& \textcolor{black}{}&\textcolor{black}{BH QED}\\
\rowcolor{white}&\textcolor{black}{UPE I}& \textcolor{black}{1.21}& \textcolor{black}{CPL+BB}& \textcolor{black}{$(6.37\pm 0.48) \times 10^{52}$}&\textcolor{black}{}\\

\rowcolor{white}&\textcolor{black}{UPE II}& \textcolor{black}{1.82}& \textcolor{black}{CPL+BB}& \textcolor{black}{$(1.60\pm 0.10) \times 10^{53}$}&\textcolor{black}{}\\
\hline
\rowcolor{white}IV&\textcolor{black}{\textbf{BH-rise (undercritical)}}& \textcolor{black}{}& \textcolor{black}{}& \textcolor{black}{}&\textcolor{black}{BH CED}\\

\rowcolor{white}&\textcolor{black}{Jetted GeV emission}& \textcolor{black}{600}& \textcolor{black}{PL}& \textcolor{black}{$(2.2\pm 0.2) \times 10^{52}$}&\textcolor{black}{}\\
\hline
\rowcolor{white}IV&\textcolor{black}{\textbf{BH-echoes}}& \textcolor{black}{}& \textcolor{black}{}& \textcolor{black}{}&\textcolor{black}{BH disk accretion}\\
\rowcolor{white}&\textcolor{black}{Cavity}& \textcolor{black}{3.02}& \textcolor{black}{CPL}& \textcolor{black}{$(4.32\pm 0.19) \times 10^{52}$}&\textcolor{black}{}\\
\rowcolor{white}&\textcolor{black}{HXF}& \textcolor{black}{6.03}& \textcolor{black}{CPL+BB}& \textcolor{black}{$(3.93\pm 0.33) \times 10^{52}$}&\textcolor{black}{}\\

\rowcolor{white}&\textcolor{black}{SXF}& \textcolor{black}{15.12}& \textcolor{black}{PL}& \textcolor{black}{$(2.89\pm 042) \times 10^{52}$}&\textcolor{black}{}\\
\hline
\cellcolor{white}VI&\cellcolor{white}\textcolor{black}{\textbf{The Afterglows}}&&&&\parbox{2.0cm}{$\nu$NS synchrotron+pulsar emission}\\
\rowcolor{white}&\textcolor{black}{X-ray}& \textcolor{black}{$ 10^7$}& \textcolor{black}{PL}& \textcolor{black}{$(2.61\pm 1.01) \times 10^{52}$}&\\
\rowcolor{white}&\textcolor{black}{TeV} & \textcolor{black}{$\sim 3\times 10^3$}& \textcolor{black}{PL}& \textcolor{black}{$(2.40\pm 1.80) \times 10^{50}$}&\\

\rowcolor{white}&\textcolor{black}{Optical}& \textcolor{black}{$\sim 3\times 10^5$}& \textcolor{black}{PL}& \textcolor{black}{$(6.10\pm 1.00) \times 10^{50}$}&\\

\rowcolor{white}&\textcolor{black}{Radio}& \textcolor{black}{$\sim 2.21 \times 10^6$}& \textcolor{black}{PL}& \textcolor{black}{$(2.21\pm 0.24) \times 10^{46}$}&\\
\hline 
\rowcolor{white} VII&\textcolor{black}{ \textbf{SN Ic \& HN}} & \textcolor{black}{No data}&  \textcolor{black}{No data}& \textcolor{black}{No data}&\textcolor{black}{Nickel decay}\\
\hline 
\end{tabular}
\end{table*}

\begin{figure*}
\centering
\includegraphics[width=0.95\hsize,clip]{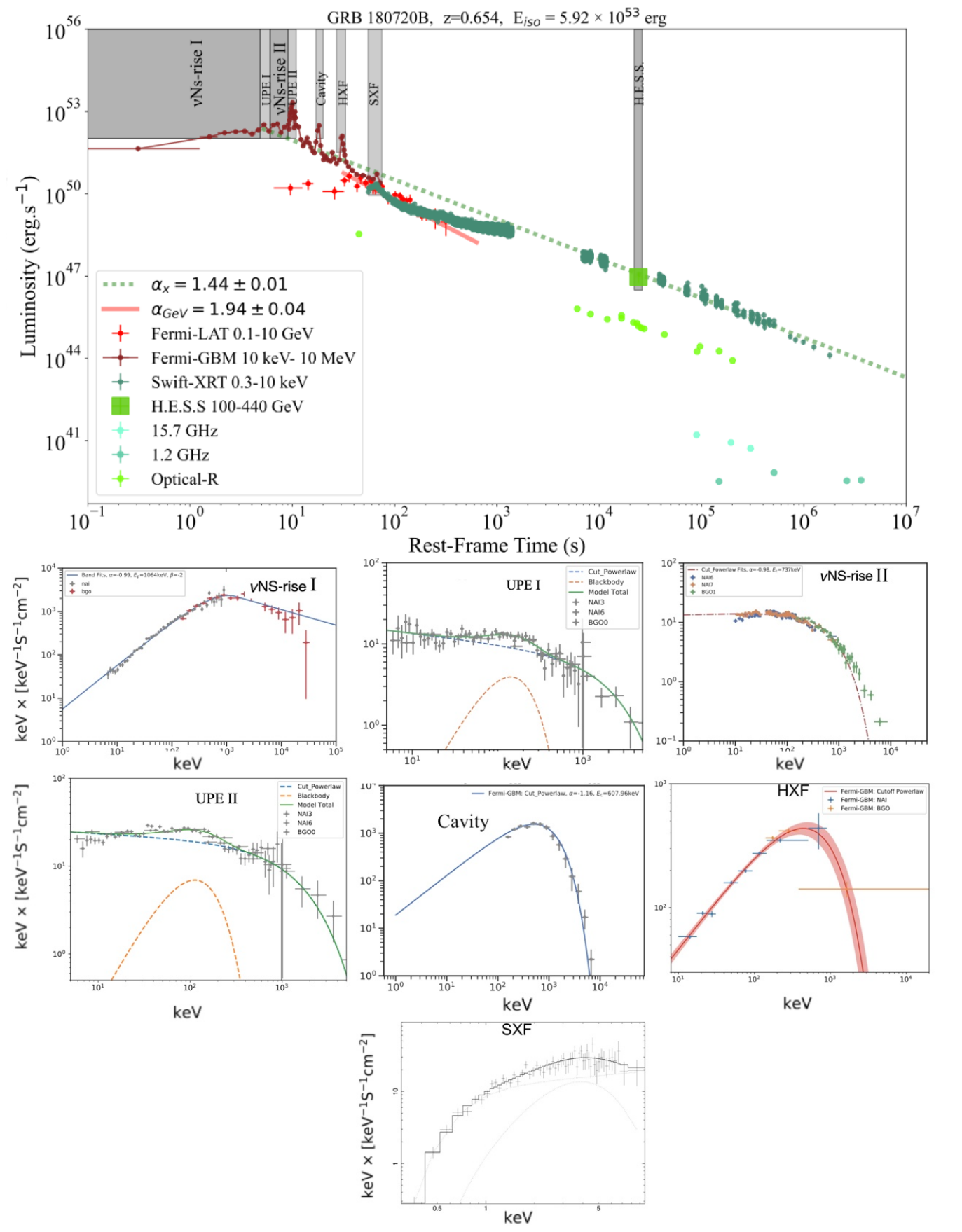} 
\caption{Luminosity light-curve of GRB 180720B and spectra related to the different Episodes identified in GRB 180720B. The energetics of the Episodes are given in Section \ref{sec:6} and Table~\ref{tab:Summary}. See also \citet{2022EPJC...82..778R} for the analysis of the UPE phase.}\label{fig:BdHNeI:180720B} 
\end{figure*}

\section{GRB 180720B as an example of BdHN I}\label{sec:6}

GRB 180720B was detected by Fermi-GBM \citep{2018GCN.22981....1R}, CALET Gamma-ray Burst Monitor \citep{2018GCN.23042....1C}, Swift-BAT \citep{2018GCN.22973....1S}, Fermi-LAT \citep{2018GCN.22980....1B} and Konus-Wind \citep{2018GCN.23011....1F}, in the gamma-rays. The High Energy Stereoscopic System (H.E.S.S.) also observed this source in the $100$--$440$~GeV bandwidth \citealp{2019Natur.575..464A}). In the X-rays, the \textit{Swift}-XRT started to observe the GRB afterglow from $91$ s after the Fermi-GBM trigger \citep{2018GCN.22973....1S}, MAXI/GSC at $296$ s \citep{2018GCN.22993....1N} and NuStar from $243$ ks to $318$ ks \citep{2018GCN.23041....1B}. In the optical and near-infrared, the $1.5$-m Kanata telescope observed the source at $78$ s from the GRB trigger time \citep{2018GCN.22977....1S}. Complementary observations in the optical, infrared, and radio telescopes are also reported in \citet{2018GCN.22976....1M,2018GCN.22977....1S,2018GCN.22983....1I,2018GCN.22985....1K,2018GCN.22988....1C,2018GCN.23017....1W,2018GCN.23020....1S,2018GCN.23021....1C,2018GCN.23023....1L,2018GCN.23024....1J,2018GCN.23033....1Z,2018GCN.23037....1S,2018GCN.23040....1I,2019Natur.575..464A}. With the redshift, $z = 0.654$, identified by the Fe II and Ni II lines in the optical observations by the VLT/X-shooter telescope \citep{2018GCN.22996....1V}, the GRB 180720B isotropic energy is $E_{\rm iso}=5.92 \times 10^{53}$~erg \citep{2018GCN.23019....1R, 2019Natur.575..464A, 2019ApJ...885...29F}. 

GRB 180720B possesses different episodes relating to specific astrophysical processes identified in the time-resolved spectral analysis of GRB 180720B \citep[see  ][ Table~\ref{tab:Summary} and Fig.~\ref{fig:BdHNeI:180720B}]{2021arXiv210309158M,2022ApJ...939...62R,2022EPJC...82..778R}. 

We summarize in Table~\ref{tab:Summary} the name of each episode, their physical event, the duration, the spectrum, $E_{\rm iso}$, and the physical phenomena originating each event. Similarly, in Fig.~\ref{fig:BdHNeI:180720B}, we represent the luminosity in wavelengths ranging from radio to TeV and show the spectra corresponding to each physical process. 

\textit{The $\nu$NS-rise I}. The radiation originating from the fallback of the SN ejecta onto the $\nu$NS \citep{2019ApJ...871...14B,2022PhRvD.106h3002B}. The first evidence of this episode in GRB 180720B, referred to as the \textit{$\nu$NS-rise}, extends from $t_{\rm rf}=0$~s to $~t_{\rm rf}=4.84$~s time interval, with isotropic energy of $E_{\rm iso}=(1.53 \pm 0.09) \times 10^{53}$~erg. A Band model best fits its spectrum with $E_p=1064$ keV, $\alpha =-0.99$, and $\beta = -2.00$.

\textit{The UPE I}. This episode pinpoints the first emission originating from the BH (BH-rise). The UPE I of GRB 180720B occurs from $~t_{\rm rf}=4.84$~s to $~t_{\rm rf}=6.05$~s. Its measured isotropic energy is $E^{\rm MeV}_{\rm UPE I}=(6.37\pm0.48) \times 10^{52}$~erg, and its spectrum is best fitted by a CPL+BB model (index $\alpha=-1.13$, cutoff energy $E_{\rm c}=2220.569$~keV, and blackbody (BB) temperature $k T = 50.31$~keV in the observer frame). 

\textit{The $\nu$NS-rise II}. It spans from $t_{\rm rf}=6.05$~s to $t_{\rm rf}=9.07$~s. The isotropic energy of this phase is $E^{\rm MeV}_{\nu \rm NS}=(1.13\pm0.04) \times 10^{53}$~erg, and its spectrum is best fitted by a CPL model ($\alpha=-0.98$, and  $E_{\rm c}=737$~keV, in the observer frame). 

\textit{The UPE II}. It is evidenced by the first significant observed GeV photon at $~t_{\rm rf}=7.06$~s. The UPE phase is also continued during this phase (UPE II), which lasts from $~t_{\rm rf}=9.07$~s to $~t_{\rm rf}=10.89$~s, with isotropic energy of $E_{\rm UPE II}^{\rm MeV}=(1.6 \pm 0.95) \times 10^{53}$~erg. A CPL+BB model with the following model parameters of $\alpha= -1.06^{+0.01}_{-0.01}$, $E_{\rm c}=1502.5^{+88.6}_{-87.5}$~keV and $kT= 39.8^{+1.6}_{-1.6}$~keV best fits the spectrum. 

\textit{The Cavity}. This emission extends from $t_{\rm rf}=16.94$~s to $~t_{\rm rf}=19.96$~s, with an isotropic energy of $E_{\rm CV}^{\rm MeV}=(4.32 \pm 0.19) \times 10^{52}$~erg, characterized by a CPL spectrum ($\alpha=-1.16$, $E_c = 607.96$~keV) with an energy of $\sim 10^{52}$ erg and a luminosity of $\sim 10^{51}$ erg s$^{-1}$. 

\textit{The HXF and SXF}. The HXF of GRB 180720B extends from $t_{\rm rf}= 28.95$~s to $t_{\rm rf}= 34.98$~s, with $L_{\rm HXF,iso}^{\rm MeV}=(7.8 \pm 0.07) \times 10^{51}$~erg~s$^{-1}$. Its spectrum is best fitted by a CPL model with $E_{\rm c}=(5.5_{-0.7}^{+0.8}) \times 10^2$~keV, $\alpha = -1.198 \pm 0.031$. The SXF occurs from $t_{\rm rf}= 55$~s to $t_{\rm rf}= 75$~s, with $L_{\rm SXF,iso}^{\rm X}=1.45\times 10^{50}$~erg s$^{-1}$. Its spectrum is best fitted by a PL+BB model with $\alpha = -1.79 \pm 0.23$, and $k T=0.99 \pm 0.13$~keV; see Table \ref{tab:Summary}. Because the interaction of expanding $e^+e^-$ with the SN ejecta produces the cavity, the HXF, and the SXF, their energetics are similar to the UPE phase \citep[see][and references therein]{2021MNRAS.504.5301R}.

\textit{The GeV emission}. The 0.1--10 GeV emission of GRB 180720B observed by Fermi-LAT starts at $t_{\rm rf}=7.01$~s. The highest photon energy corresponding to this GRB is $4.9$ GeV, which was detected 137 seconds after the Fermi—GBM trigger \citep{2020A&A...636A..55R}. The luminosity rises up to $t_{\rm rf} \sim 40$~s. After $t_{\rm rf} \sim 40$~s the GeV luminosity follows a temporal decaying luminosity of $L_{\rm GeV}= 4.6 \times 10^{53}~ t^{-1.94\pm0.0.13}$~erg~s$^{-1}$. It has a total isotropic energy of $E_{\rm iso,GeV}= (2.2\pm 0.2) \times 10^{52}$~erg.

\textit{The radio, optical, and X-ray afterglows}. The X-ray afterglow luminosity observed by Swift-XRT starts at $t_{\rm rf}=52$~s with a time decaying luminosity of $L_{\rm X} =2.5\times 10^{53} t^{-1.44\pm 0.01}$~erg~s$^{-1}$ and its isotropic energy is $E_{\rm iso, X} =2.61\times 10^{52}$~erg. The X-ray afterglow is accompanied by the radio, optical, and TeV afterglows with isotropic energies of $E_{\rm iso, radio} =2.21\times 10^{46} $~erg, $E_{\rm iso, opt} =6.1\times 10^{50}$~erg, and $E_{\rm iso, TeV} =2.4\times 10^{50} $~erg, respectively. 

In \citet{2022ApJ...939...62R}, the above afterglows of GRB 180720B have been explained within the synchrotron scenario described in Section \ref{sec:3afterglows}.  The X-ray afterglow of GRB 180720B exhibits two distinct power-laws, the first at times $10^2$--$10^3$ s and the second at times $>10^4$ s (there is a data gap at $10^3$--$10^4$ s). The X-ray luminosity in the time interval $10^2$--$10^3$ s exhibits a shallower power-law than the pure synchrotron luminosity, as evidenced by comparing it with the power-laws of the optical and radio synchrotron at times $>10^4$ s. The above is explained by the contribution of the $\nu$NS magnetic-braking radiation (see Section \ref{sec:3afterglows}). Around $10^2$ s, the critical synchrotron radiation energy falls below the keV range, so the X-rays synchrotron luminosity decays exponentially afterward. At lower energies, the power-law behavior remains. The subsequent dominance of the pulsar emission in the observed X-ray emission has allowed us to infer the strength of the magnetic field dipole and quadrupole and the rotation period of the $\nu$NS. We refer to \citet{2022ApJ...939...62R} for more details. There is a technical difficulty in detecting the early (from the GRB trigger up to a few tens of seconds) X-ray afterglow by Swift-XRT. Only recently, thanks to the cosmological time dilation effect, it has been possible to pinpoint this $\nu$NS emission in its early phase using high-$z$ sources (Bianco et al., submitted). The extrapolation of the X-ray afterglow power-law behavior, backward in time from $10^4$ s, indicates our theoretical prediction at early times, confirmed in the few cases where observations have allowed us to do it.

\textit{The optical SN}. As a BdHN I source, GRB 180720B was expected to have an associated SN emission, with an optical peak at $21.8 \pm 4.3$ day after the trigger \citep{2018GCN.23019....1R}. Unfortunately, no telescope observed the source at those times to confirm the SN appearance. 

In conclusion, the total energy released by the GRB 180720B is $E_{\rm tot}=6.5 \times 10^{53}$~erg of which  $3.57 \times 10^{53}$~erg is due to the BH with mass with a lower limit of $M= 2.4 M_\odot$ and initial spin with an upper limit of $\alpha=0.6$. The remaining $2.93\times 10^{53}$~erg is due to the accreting $\nu$NS with the period of $1$ ms.

\section{GRB 190114C as an example of BdHN I}\label{sec:7}

\begin{figure*}
\centering
\includegraphics[width=0.90\hsize,clip]{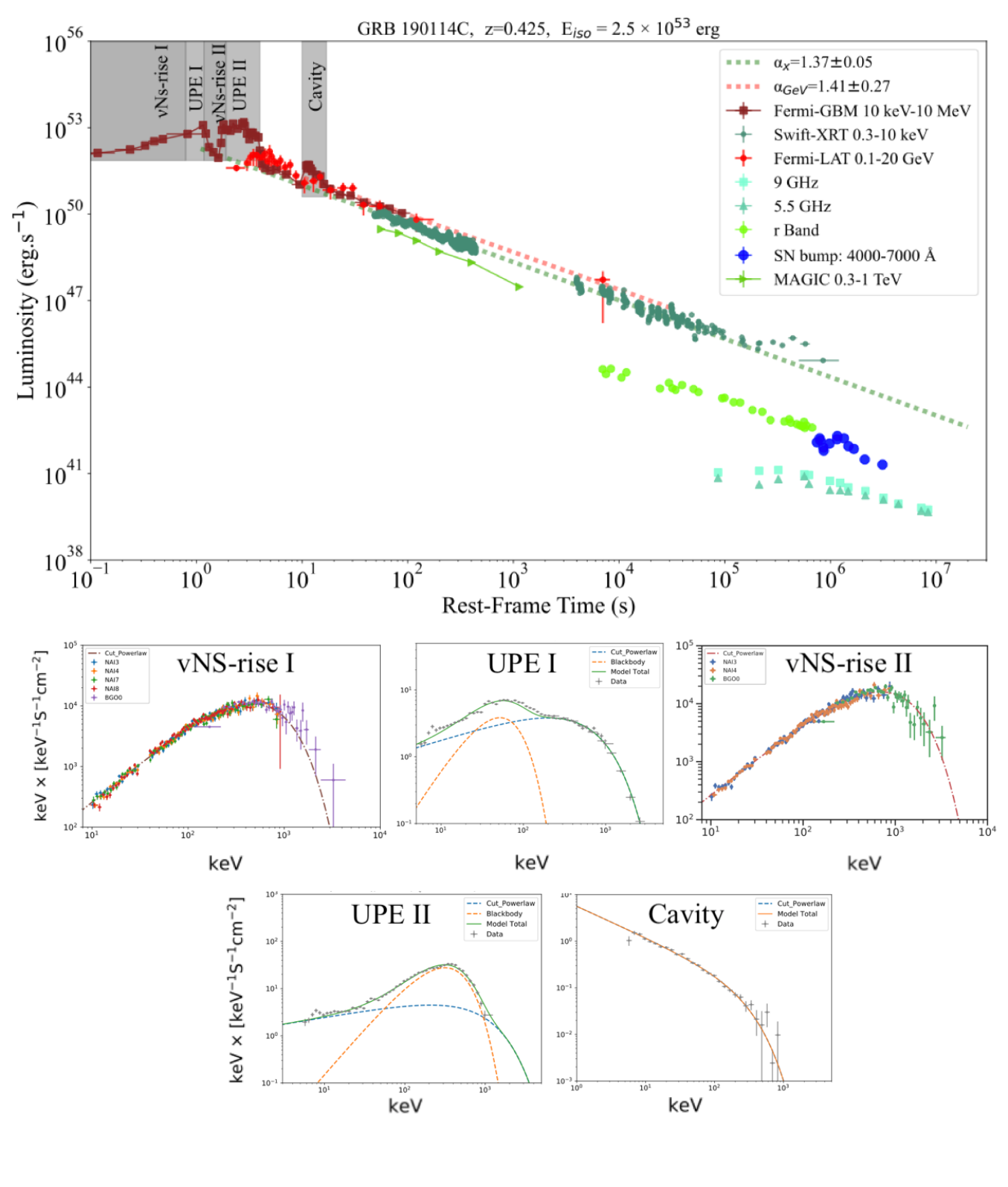} 
\caption{BdHNe I: GRB 190114C.
Luminosity light-curves obtained from \textit{Fermi}-GBM, in $10$~keV--$10$~MeV, \textit{Fermi}-LAT in $0.1$~GeV--$10$~GeV, \textit{Swift}-BAT in $15$~keV--$50$~keV, \textit{Swift}-XRT in $3$~keV--$10$~keV and optical R-band. The late X-ray afterglow luminosity of BdHN I GRB 190114C observed by \textit{Swift}-XRT is best fit by a temporal decaying power law of $L_{\rm X}= (2.5\pm 0.4)\times 10^{53}~t^{1.44\pm 0.01}$~erg s$^{-1}$. The light curve of \textit{Fermi}-LAT in is fitted by temporal decaying power law of $L_{\rm GeV}= (4.6\pm 2.9)\times 10^{53}~t^{-1.94\pm 0.04}$~erg.
The prediction of the associated SN by \citep{2019GCN.23715....1R} has been successfully observed by \citep{2019GCN.23983....1M} and has made GRB 190114C as a prototype of BdHN I \citep{2021PhRvD.104f3043M} to study the properties of GRB-SN sources. The rest-frame visual absolute magnitude of the SN associated with GRB 190114C is $\sim -18$ mag \citet{2019GCN.23983....1M}, which is $\sim 1$ mag less than famous SN 1998bw \citep{2001ApJ...555..900P}. This fainter brightness can be due to the extinction of this event \citep{2019GCN.23710....1K}. 
The energetic of the Episodes are given in section \ref{sec:7} and Table~\ref{tab:Summary1}.}\label{fig:BdHNeI:190114C} 
\end{figure*} 

\begin{table*}
\centering
\caption{The episodes and afterglows of GRB 190114C. This table reports the name, the underlying astrophysical process, the duration (s), the best-fit spectrum, and the isotropic energy (erg) for each event in GRB 190114C. GRB 190114C has a redshift $z=0.424$ and $T^{\rm total}_{90}=81.4$ s (corrected in the rest frame).}
\label{tab:Summary1}
\small\addtolength{\tabcolsep}{2pt}
\begin{tabular}{l|c|c|c|c|l}
\hline
Episode &Event & duration(s) & Spectrum & $E_{\rm iso}$ (erg)&Physical phenomena\\
\hline
\hline
\rowcolor{white}0&{\bf SN-rise}&{ --}&{--}&--& CO$_{\rm core}$ collapse\\
\hline 
\rowcolor{white}I&\textcolor{black}{\textbf{$\nu$NS-rise }}& \textcolor{black}{}& \textcolor{black}{}& \textcolor{black}{}&\textcolor{black}{$\nu$NS accretion}\\
\rowcolor{white}&\textcolor{black}{$\nu$NS-rise I}& \textcolor{black}{0.79}& \textcolor{black}{CPL}& \textcolor{black}{$(3.52\pm 0.15) \times 10^{52}$}&\textcolor{black}{}\\
\rowcolor{white}&
\cellcolor{white}\textcolor{black}{$\nu$NS-rise II}&\textcolor{black}{0.84} &\textcolor{black}{CPL}& \textcolor{black}{$(3.75\pm0.11) \times 10^{52}$}&\textcolor{black}{}\\
\hline 
\rowcolor{white}II&{\bf NS-rise}&{ Not observable}&{Not observable}&Not observable&Companion NS accretion\\
\hline 
\rowcolor{white}III&\textcolor{black}{\textbf{BH-rise (overcritical)}}& \textcolor{black}{}& \textcolor{black}{}& \textcolor{black}{}&\textcolor{black}{BH QED}\\
\rowcolor{white}&\textcolor{black}{UPE I}& \textcolor{black}{0.39}& \textcolor{black}{CPL+BB}& \textcolor{black}{$(1.00\pm 0.11) \times 10^{53}$}&\textcolor{black}{}\\

\rowcolor{white}&\textcolor{black}{UPE II}& \textcolor{black}{2.09}& \textcolor{black}{CPL+BB}& \textcolor{black}{$(1.47\pm 0.20) \times 10^{53}$}&\textcolor{black}{}\\
\hline
\rowcolor{white}IV&\textcolor{black}{\textbf{BH-rise (undercritical)}}& \textcolor{black}{}& \textcolor{black}{}& \textcolor{black}{}&\textcolor{black}{BH CED}\\

\rowcolor{white}&\textcolor{black}{Jetted GeV emission}& \textcolor{black}{600}& \textcolor{black}{PL}& \textcolor{black}{$(1.8\pm 1.3) \times 10^{53}$}&\textcolor{black}{}\\
\hline
\rowcolor{white}IV&\textcolor{black}{\textbf{BH-echoes}}& \textcolor{black}{}& \textcolor{black}{}& \textcolor{black}{}&\textcolor{black}{BH disk accretion}\\
\rowcolor{white}&\textcolor{black}{Cavity}& \textcolor{black}{13.1}& \textcolor{black}{CPL}& \textcolor{black}{$(2.49\pm 0.12) \times 10^{52}$}&\textcolor{black}{}\\
\rowcolor{white}&\textcolor{black}{HXF}& \textcolor{black}{--}& \textcolor{black}{--}& \textcolor{black}{--}&\textcolor{black}{}\\

\rowcolor{white}&\textcolor{black}{SXF}& \textcolor{black}{--}& \textcolor{black}{--}& \textcolor{black}{--}&\textcolor{black}{}\\
\hline
\cellcolor{white}VI&\cellcolor{white}\textcolor{black}{\textbf{The Afterglows}}&&&&\parbox{2.0cm}{$\nu$NS synchrotron+pulsar-like emission}\\
\rowcolor{white}&\textcolor{black}{X-ray}& \textcolor{black}{$\sim  10^7$}& \textcolor{black}{PL}& \textcolor{black}{$(3.20\pm 1.28) \times 10^{52}$}&\\
\rowcolor{white}&\textcolor{black}{TeV} & \textcolor{black}{$\sim 3\times 10^3$}& \textcolor{black}{PL}& \textcolor{black}{$(4.00\pm 1.80) \times 10^{51}$}&\\

\rowcolor{white}&\textcolor{black}{Optical}& \textcolor{black}{$\sim 3\times 10^5$}& \textcolor{black}{PL}& \textcolor{black}{$(7.10\pm 1.20) \times 10^{50}$}&\\

\rowcolor{white}&\textcolor{black}{Radio}& \textcolor{black}{$\sim 2 \times 10^6$}& \textcolor{black}{PL}& \textcolor{black}{$(3.31\pm 0.34) \times 10^{46}$}&\\
\hline 
\rowcolor{white} VII&\textcolor{black}{ \textbf{SN Ic \& HN}} & \textcolor{black}{$\sim10^7$}&  \textcolor{black}{BB}& \textcolor{black}{$3\times 10^{49}$}&\textcolor{black}{Nickel decay}\\
&$M_{\rm ej} = (6.0 \pm 4.0)~ M_\odot$ && & &  \\
&$M_{\rm Ni} = (0.4 \pm 0.2) ~M_\odot$ && & &   \\
&$E_K = (2.5 \pm 1.8)\times 10^{52}$~erg && & &  \\
\hline 
\end{tabular}
\end{table*}

\begin{figure*}
\centering
\includegraphics[angle=90, width=0.84\hsize]{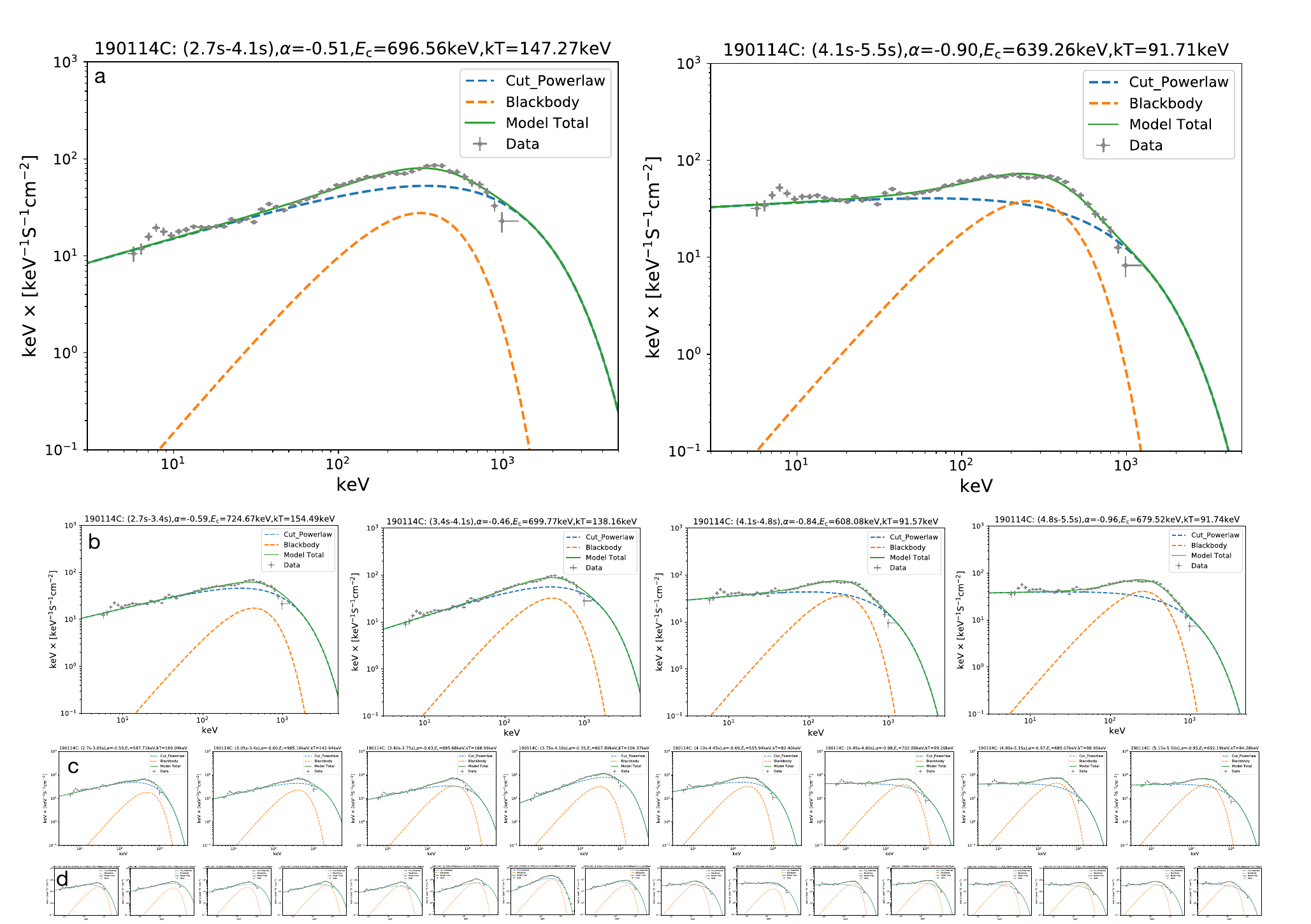}
\caption{Time-resolved spectral analysis of UPE II phase of GRB 190114C from $t=2.7$~s ($t_{\rm rf}=1.9$~s) to $t=5.5$~s ($t_{\rm rf}=3.9$~s). The self-similar spectral structure is present when (a) the time interval is divided into two parts, (b) four parts, (c) eight parts, and (d) sixteen parts, respectively. The plot is adapted from \citet{2019arXiv190404162R} with the authors' permission.}\label{alltogether}
\end{figure*}

GRB 190114C was first detected by the \textit{Fermi}-GBM \citep{2019GCN.23707....1H}, and the \textit{Neil Gehrels Swift} Burst Alert Telescope (BAT) \citep{2019GCN.23688....1G}.  The highest-energy GeV photon detected by \textit{Fermi}-LAT (with a boresight angle of $68$~degrees) is a $22.9$~GeV event which is observed $15$~s after the GBM trigger \citep{2019GCN.23709....1K}. Nordic Optical Telescope (NOT)  announced  the redshift of $z=0.424$ \citep{2019GCN.23695....1S} which leads the isotropic energy of $E_{\rm iso}= (2.48 \pm 0.22) \times 10^{53}$~erg. The late-time $0.3$--$10$ keV light curve observed by \textit{Swift} X-ray Telescope (XRT) revealed a temporal power-law decay \citep{2019GCN.23706....1D}. Given the above observations, at time 15:29:54 GMT on January 15, 2019, we identified by \citet{2019GCN.23715....1R} this GRB as a BdHN I and predicted that an optical SN should appear in the same location of the GRB within $18.8\pm 3.7$~days, which indeed was confirmed by \citet{2019GCN.23983....1M}. This successful prediction and the following detection of TeV radiation by MAGIC \citep{2019GCN.23701....1M} have made GRB 190114C a prototype in which all the BdHN phases have been observed \citep[][]{2019arXiv190404162R}. 

The GRB 190114C reveals different episodes of specific astrophysical processes identified in the time-resolved spectral analysis; see Table~\ref{tab:Summary1} and Fig.~\ref{fig:BdHNeI:190114C}.

We summarize in Table~\ref{tab:Summary1} the name of each episode, their physical event, the duration, the spectrum, $E_{\rm iso}$, and the physical phenomena originating each event. Similarly, in Fig.~\ref{fig:BdHNeI:190114C}, we represent the luminosity in wavelengths ranging from radio to TeV and show the spectra corresponding to each physical process. 

\textit{The $\nu$NS-rise I}. With an isotropic energy of $E_{\rm iso}=(3.52 \pm 0.15) \times 10^{52}$~erg, it extends from $t_{\rm rf}=0$~s to $~t_{\rm rf}=0.79$~s time interval. Its spectrum is best fitted by a CPL model with $E_c=710^{+21.3}_{-26.1}$.

\textit{The UPE I}. It starts from $t_{\rm rf}=0.79$~s and ends at $~t_{\rm rf}=1.18$~s. Its spectrum is best fitted by a cutoff power-law plus blackbody (CPL+BB) with the parameters of power-law index $\alpha= -0.62^{+0.03}_{-0.03}$, cut-off energy $E_c=524.7^{+20.1}_{-20.1}$, temperature, $kT= 18.4^{+0.5}_{-0.5}$ keV,  with isotropic energy of $E_{\rm iso}=(1.00 \pm 0.11) \times 10^{53}$~erg.

\textit{The $\nu$NS-rise II}. With an isotropic energy of $E_{\rm iso}=(3.75 \pm 0.11) \times 10^{52}$~erg, it spans from $t_{\rm rf}=1.18$~s to $t_{\rm rf}=1.9$~s time interval. Its spectrum is best fitted by a CPL model with $E_c=770^{+22.4}_{-21.8}$. 

\textit{The UPE II}. It is signed by a CPL+BB spectrum with power-law index $\alpha= -0.71^{+0.02}_{-0.02}$,  cut-off energy $E_c=717.6^{+25.4}_{-25.4}$, temperature, $kT= 111.64^{+2.5}_{-2.5}$ keV, and a self-similar structure deduced from an appropriate time-resolved analysis \citep{2021PhRvD.104f3043M}; see Fig.~\ref{alltogether}. With an isotropic energy of $E_{\rm iso}=(1.47 \pm 0.20) \times 10^{53}$~erg, it starts from $t_{\rm rf}=1.9$~s, and ends at $~t_{\rm rf}=3.99$~s. The following mass and spin parameter of the newborn BH have been inferred, $M=4.5 M_\odot$, and $\alpha= 0.54$, respectively \citep[see, e.g.][for details]{2021PhRvD.104f3043M}.

\textit{The Cavity}. It extends from $t_{\rm rf}=11$~s to $t_{\rm rf}=17$~s. Its spectrum is best fitted by a CPL model with the photon index $\alpha= -1.67$ and the cutoff energy $E_c = 251$~keV. The enclosure of the companion NS and the accreted material (i.e., $\approx 10^{57}$ baryons) inside the BH horizon creates a large cavity of $\approx 10^{11}$~cm around it. The density distribution around the newborn BH has been inferred in \citet{2019ApJ...883..191R}, and the data have confirmed the spatial extension of the cavity (see Table~\ref{tab:Summary1} and Fig.~\ref{fig:BdHNeI:190114C}).

\textit{The HXF and SXF}. It is demonstrated in \citet{2018ApJ...869..151R} and \citet{2021MNRAS.504.5301R} that the HXF and SXF are observable when the BdHNe viewing angle is closed to the equatorial plane of the binary progenitors. GRB 190114C is a BdHNI observed with a viewing angle orthogonal to the orbital plane of the GRB binary \citep{2021MNRAS.504.5301R}. Therefore, the HXF and SXF of GRB 190114C are not observable from the polar axis. 

\textit{The GeV emission}. The onset of the GeV radiation is also signed by the first GeV photon in the range $0.1$--$100$~GeV observed by \textit{Fermi}-LAT. The total energy emitted by this source in the above GeV range is $E_{\rm GeV}=(1.8 \pm 0.9) \times 10^{53}$~erg \citep{2021MNRAS.504.5301R}, comparable to the energy observed by the GBM.

\textit{The radio, optical, and X-ray afterglows}. The X-ray afterglow luminosity observed by Swift-XRT starts at $t_{\rm rf}=52$~s with a temporal decaying luminosity of $L_{\rm X} =5.14\times 10^{52} t^{-1.37\pm 0.05}$~erg~s$^{-1}$ and its equivalent isotropic energy is $E_{\rm iso, X} =3.2\times 10^{52}$~erg. The X-ray afterglow of GRB 190114C is accompanied by the radio, optical, and TeV afterglows with isotropic energies of $E_{\rm iso, radio} =3.31\times 10^{46} $~erg, $E_{\rm iso, opt} =7.1\times 10^{50}$~erg, and $E_{\rm iso, TeV} =4.0\times 10^{51}$~erg, respectively. These afterglows originated from synchrotron radiation powered by the interaction of the $\nu$NS, with an initial period of $P_0=1$~ms, and the SN ejecta \citet{2022ApJ...939...62R} (see Table~\ref{tab:Summary1} and Fig.~\ref{fig:BdHNeI:190114C}).

\textit{The optical SN}. The optical signal of SN 2019jrj, a typical GRB-associated SN Ic (see Figs. \ref{fig:iso-tpeak} and \ref{fig:redlpeak}), peaks at $\sim 10^6$~s (see also Fig.~\ref{fig:BdHNeI:190114C}). Deducing certain physical properties of SN 2019jrj is difficult due to the relatively low quality of the light curve and spectra \citep[see, e.g.,][]{2022AandA...659A..39M}. Therefore, we use the average values reported in \citet{2017AdAst2017E...5C} obtained using the nickel radioactive-heating model for the bolometric SN light curve \citep{1982ApJ...253..785A}. The corresponding total SN ejected mass, nickel mass, and SN kinetic energy are, respectively, $M_{\rm ej} = 6.0 \pm 4.0~ M_\odot$, $M_{\rm Ni} = 0.4 \pm 0.2 ~M_\odot$, and $E_K = (2.5 \pm 1.8)\times 10^{52}$~erg \citep{2017AdAst2017E...5C}. 

In conclusion, the total energy released by the GRB 180720 is $E_{\rm tot}=3.8 \times 10^{53}$~erg of which  1) $2.7 \times 10^{53}$~erg is due to the BH with mass with a lower limit of $\rm M= 4.53 M_\odot$ and initial spin with an upper limit of $\rm \alpha=0.54$, 2) $1.1\times 10^{53}$~erg is due to the accreting $\nu$NS with the period of 1 ms, and  3) $3\times 10^{49}$~erg is due to the optical SN emission corresponding to the HN ejecta with the kinetic energy of $2.5\times 10^{52}$~erg. 

\section{GRB 190829A as an example of BdHN II}\label{sec:8}

\begin{figure}
\centering
\includegraphics[width=1\hsize,clip]{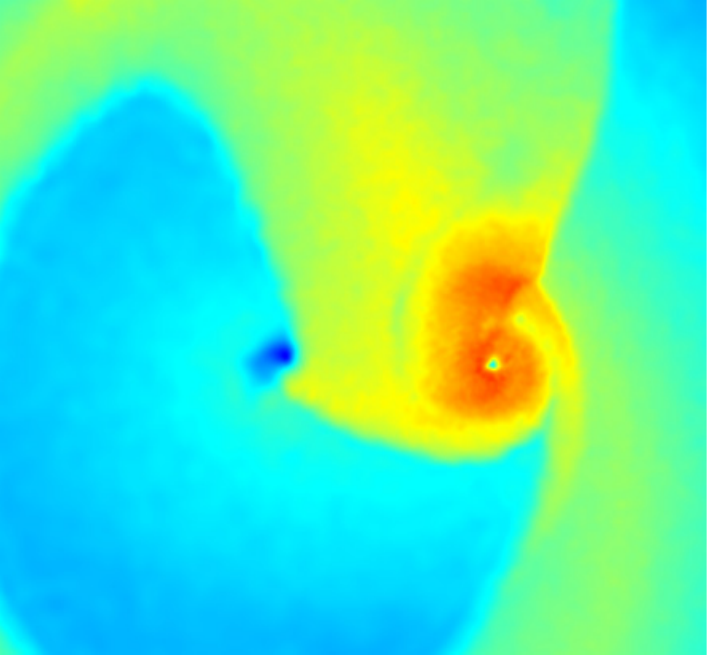}
\caption{Ongoing accretion process of SN ejecta onto the $\nu$NS and the NS companion, simulated in \citet{2019ApJ...871...14B}. The $\nu$NS is located at the center of the dark blue spot accumulating material around it. And at the center of the green spot, the NS companion is also accreting SN ejecta. Also, we notice that a portion of the SN ejecta is flowing back towards $\nu$NS due to the distortion of SN ejecta caused by the companion NS.}
\label{fig:accretion}
\end{figure}

\begin{table*}
\centering
\caption{The episodes of GRB 190829A. The episodes of accretion onto the companion star and the $\nu$NS are triggered by SN explosion. According to the BdHN terminology, they can be classified as sub-episodes of SN-rise. Times are measured in the source rest frame. }
\label{tab:190829A}
\small\addtolength{\tabcolsep}{2pt}
\begin{tabular}{l|c|c|c|c|l}
\hline
Episode &Event & duration(s) & Spectrum & $E_{\rm iso}$ (erg)&Physical phenomena\\
\hline
\hline
\rowcolor{white}0&{\bf SN-rise}&{ --}&{--}&--& CO$_{\rm core}$ collapse\\
\hline 
\rowcolor{white}I&\textcolor{black}{\textbf{$\nu$NS-rise }}& \textcolor{black}{16.23}& \textcolor{black}{Band}& \textcolor{black}{$(3.5\pm 0.5) \times 10^{50}$}&\textcolor{black}{$\nu$NS accretion}\\

\hline 
\rowcolor{white}II&{\bf NS-rise}&{ 8.16}&{CPL}&$(4.2\pm 1.0) \times 10^{49}$&Companion NS accretion\\
\hline 
\rowcolor{white}III&\textcolor{black}{\textbf{BH-rise (overcritical)}}& \textcolor{black}{--}& \textcolor{black}{--}& \textcolor{black}{--}&\textcolor{black}{BH QED}\\

\hline
\rowcolor{white}IV&\textcolor{black}{\textbf{BH-rise (undercritical)}}& \textcolor{black}{--}& \textcolor{black}{--}& \textcolor{black}{--}&\textcolor{black}{BH CED}\\

\hline
\rowcolor{white}IV&\textcolor{black}{\textbf{BH-echoes}}& \textcolor{black}{--}& \textcolor{black}{--}& \textcolor{black}{--}&\textcolor{black}{BH disk accretion}\\

\hline
\cellcolor{white}VI&\cellcolor{white}\textcolor{black}{\textbf{The Afterglows}}&&&&\parbox{2.0cm}{$\nu$NS synchrotron+pulsar--like emission}\\
\rowcolor{white}&\textcolor{black}{X-ray}& \textcolor{black}{$>10^7$}& \textcolor{black}{PL}& \textcolor{black}{$>4 \times 10^{50}$}&\\
\rowcolor{white}&\textcolor{black}{TeV} & \textcolor{black}{$\sim 2\times 10^5$}& \textcolor{black}{PL}& \textcolor{black}{$>3 \times 10^{49}$}&\\

\rowcolor{white}&\textcolor{black}{Optical}& \textcolor{black}{$>10^7$}& \textcolor{black}{PL}& \textcolor{black}{$>4 \times 10^{48}$}&\\

\rowcolor{white}&\textcolor{black}{Radio}& \textcolor{black}{$>10^7$}& \textcolor{black}{PL}& \textcolor{black}{$>  10^{44}$}&\\
\hline 
\rowcolor{white} VII&\textcolor{black}{ \textbf{SN Ic \& HN}} & \textcolor{black}{$\sim10^7$}&  \textcolor{black}{BB}& \textcolor{black}{$> 3\times 10^{49}$}&\textcolor{black}{Nickel decay}\\
&$M_{\rm ej} = 5.67 \pm 0.72~ M_\odot$& && &   \\
&$M_{\rm Ni} = 0.5 \pm 0.1 ~M_\odot$& && &    \\
&$E_K = (1.35 \pm 0.51)\times 10^{52}$~erg& && &   \\
\hline 
\end{tabular}
\end{table*}

\begin{figure*}[ht]
\centering
\includegraphics[width=0.850\hsize,clip]{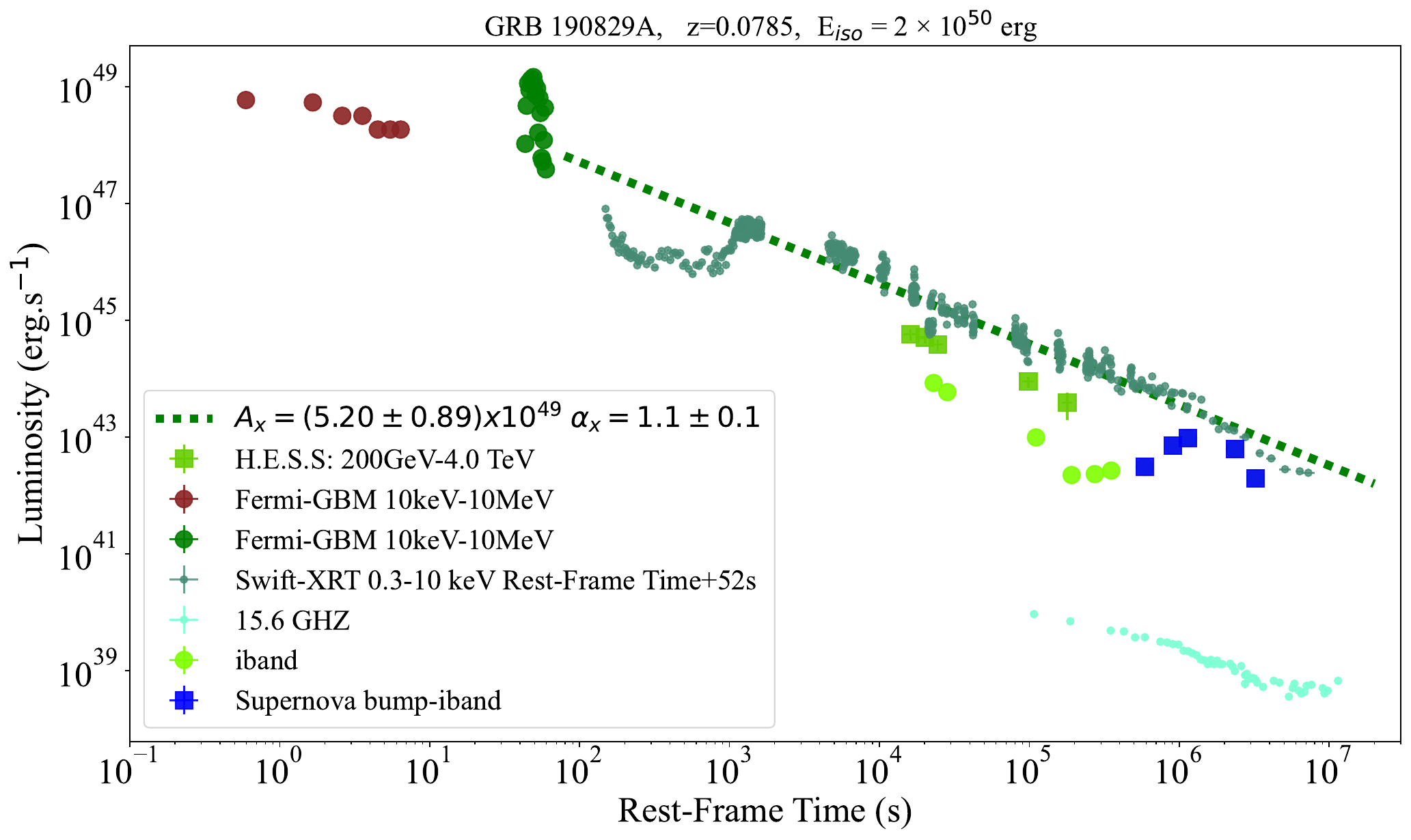} 
\caption{BdHN II: GRB 190829A. Luminosity light-curves obtained from H.E.S.S. in $200$~GeV -- $4$~TeV, \textit{Fermi}-GBM in $10$~keV--$10$~MeV, \textit{Swift}-BAT in $15$--$50$~keV, \textit{Swift}-XRT in $3$--$10$~keV and i band and radio band. An SN component at $\sim 10^6$ is indicated as the blue color. The power-law fitting of the X-ray, shown as a green dotted line, gives a power-law index of $-1.1$. The $T_0$ is taken from the trigger of \textit{Fermi}-GBM to which the initial time of other telescopes is aligned.}\label{fig:BdHNeII} 
\end{figure*} 

GRB 190829A triggered the \textit{Fermi}-GBM at 19:55:53 UT on 2019 August 29 \citep{2019GCN.25551....1F}. \textit{Swift}-BAT was triggered $51$~s later. The \textit{Swift}-XRT started observing $148.3$~s later after the Fermi trigger \citep{2019GCN.25552....1D}. \textit{Swift}-UVOT \citep{2019GCN.25552....1D}, Half Meter Telescope (HMT) \citep{2019GCN.25555....1X}, Nordic Optical Telescope (NOT) \citep{2019GCN.25563....1H} and GTC \citet{2021AandA...646A..50H} detected the redshift of $z=0.0785 \pm 0.005$, as one of the nearest GRBs. The flattening of the optical light-curve observed  by \citep{2019GCN.25623....1P,2019GCN.25651....1B,2019GCN.25657....1P}  provided the initial evidence for the optical SN emergence. Finally, the confirmation of an associated Type Ic-BL SN named SN 2019oyw came from the spectroscopic observation performed by \citet{2019GCN.25677....1D}. 

The SN-rise is not observed for this source. Two pulses are observed in the Fermi-GBM and The \textit{Neil Gehrels Swift}--BAT light curves \citep{2022ApJ...936..190W}. The initial pulse rises at time $-0.70$~s, peaks at $1.02$~s, and declines at time $7.46$~s. After a time delay of $35.65$~s, the second, more luminous pulse begins at $43.11$~s, peaks at $47.89$~s, and declines at $59.34$~s. All the times are indicated in the rest frame. A cutoff power-law function best fits the first pulse. It has isotropic energy $4.25\pm1.02 \times 10^{49}$~erg~s$^{-1}$ and averaged luminosity $4.84 \pm 1.16 \times 10^{48}$~erg~s$^{-1}$. The second pulse shows a Band function spectrum. It is nearly one order of magnitude more energetic than the first pulse, with isotropic energy $3.56 \pm 0.50 \times 10^{50}$~erg and an averaged luminosity is ($2.05 \pm 0.29) \times 10^{49}$~erg~s$^{-1}$; see Table \ref{tab:190829A} for the summary of the GRB 190829A episodes.

The accretion onto the NS companion and the enhanced fallback accretion onto the $\nu$NS are responsible for the above two pulses. The comparison of GRB 190829A, especially the time separation between two pulses ($\sim 50$~s), with several CO-NS binaries simulated in \citet{2019ApJ...871...14B} and \citet{2022PhRvD.106h3002B}, indicates as the possible progenitor of this GRB a binary composed of a CO star and an NS with an orbital period in the range $20$--$40$ min. 

Figure~\ref{fig:accretion} shows the visualization of the three-dimensional numerical simulation that shows the $\nu$NS and the NS companion surrounded by high-density material and undergoing their corresponding accretion processes.

The first peak corresponds to the SN ejecta accretion onto the companion NS; see details in \citet{2022ApJ...936..190W}. A part of the ejecta is altered by the companion NS and flows back to the $\nu$NS, leading to a second fallback accretion episode onto the $\nu$NS, leading to the second peak. 

At the time $>100$~s, the afterglow started and was observed by Swift-XRT for the soft X-ray band, GTC for the optical band, and AMI-LA for the radio band, as shown in figure \ref{fig:BdHNeII}. The X-ray afterglow from $\sim 1000$~s follows a power-law decay with an index of $\sim -1.1$. A single power-law function best fits its spectrum with a photon index $\sim -2.15$. The optical and radio light curves share similar power-law behavior. The total energy released till $10^7$~s is $\sim 4 \times 10^{50}$~erg. We attribute this energy to the rotational energy of the $\nu$NS, which leads to an initial period of 8 ms \citep{2022ApJ...936..190W}. 

In addition, the ejected mass by the CO core collapse, $M_{\rm ej} = 5.67 \pm 0.72 M_\odot$, contributes in three different ways: 1) in spinning up the $\nu$NS, which then releases dipole and/or multipole radiation; 2) in the accretion on the NS; and 3) in the kinetic energy, $E_K = (1.35 \pm 0.51)\times 10^{52}$~erg, of the remaining SN ejecta moving with mildly relativistic velocities. All these three components contribute to the overall energetics, which reaches its peak emission within the first 100~s. 

The optical emission of the SN Ic 2019oyw, due to a nickel mass of $M_{\rm Ni} = (0.5 \pm 0.1)~M_\odot$, commonly occurs around $\sim10^6$~s with the emission of $10^{49}$~erg; see Figs.~\ref{fig:redlpeak}--\ref{fig:iso-tpeak}, \citep[see][for an in-depth discussion of the SN observation and calculations]{2017AdAst2017E...5C,2021AandA...646A..50H}. 

The remaining kinetic energy of expansion of the ejecta leads to establishing the HN associated with GRB 190829A with the total energy of $1.35 \times 10^{52}~erg$ kinetic energy+all the radiation energy). 

In addition to being a very close GRB at $z=0.0785$, which has allowed a specially significant data analysis of GRB 190829A, one of the remarkable peculiarities of this source, has been the discovery of the TeV emission very similar to the case of GRB 180720B and GRB 190114C. In {all these} systems, the TeV emission behavior closely follows the $\sim 10\%$ level of the X-ray {afterglow power-law} emission. This is the most significant since {being a BdHN II, no BH is present in this source, which suggest linking the TeV radiation to} the $\nu$NS activity. {However, the explanation of the TeV emission within the BdHN model needs still further research that we are currently pursuing (see Section \ref{sec:newphysics4}).} We can now conclude that the total energy of BdHN 190829A, observed in the keV, sub-MeV, TeV, optical, and radio bands, is $E_{\rm tot}> 8.56 10^{50}$~erg. 

\section{GRB 171205A as an example of BdHN III}\label{sec:9}

\begin{figure*}[ht]
\centering
\includegraphics[width=0.80\hsize,clip]{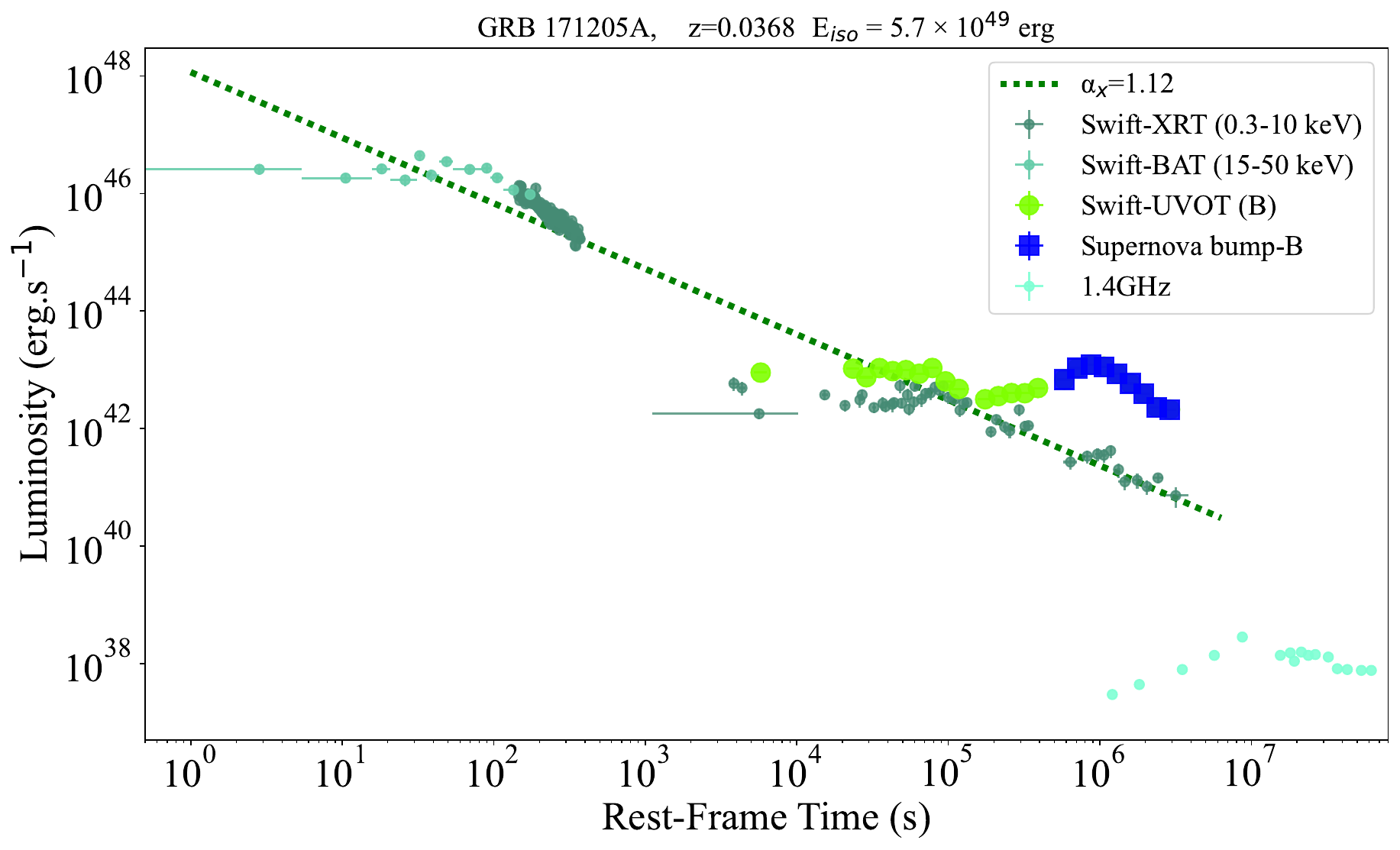} 
\caption{BdHN III: GRB 171205A. Luminosity light-curves obtained from \textit{Swift}-BAT in $15$--$50$~keV, \textit{Swift}-XRT in $3$--$10$~keV and \textit{Swift}-UVOT in V and B band. After $t_{\rm rf} \sim 10^5$~s it follows a decaying power-law with index $\alpha_X=1.12\pm 0.08$ and amplitude of $A_{\rm X}= (1.1\pm 0.8) \times 10^{48}$~erg s$^{-1}$. The optical and radio data were taken from \citet{2018AandA...619A..66D} and \citet{2021ApJ...907...60M}, and the X-ray data were retrieved from the Swift-XRT repository. The blue color indicates an SN bump.}\label{fig:BdHNeIII:171205A} 
\end{figure*}

\begin{table*}
\centering
\caption{The episodes of GRB 171205A. The first episode of prompt emission contains energy from the fallback accretion onto the $\nu$NS and the emission from the heated SN ejecta; the latter contributes the most energy. The optical afterglow emission is dominated by the cooling of  fast-moving ejecta and the supernova nickel radioactive decay. The synchrotron emission mainly contributes to the X-ray and radio bands.}\label{tab:171205A}
\small\addtolength{\tabcolsep}{2pt}
\begin{tabular}{l|c|c|c|c|l}
\hline
Episode &Event & duration(s) & Spectrum & $E_{\rm iso}$ (erg)&Physical phenomena\\
\hline
\hline
\rowcolor{white}0&{\bf SN--rise}&{ --}&{--}&--& CO$_{\rm core}$ collapse\\
\hline 
\rowcolor{white}I&\textcolor{black}{\textbf{$\nu$NS-rise }}& \textcolor{black}{182.5}& \textcolor{black}{CPL}& \textcolor{black}{$(1.7\pm 0.4) \times 10^{49}$}&\textcolor{black}{$\nu$NS accretion}\\

\hline 
\rowcolor{white}II&{\bf NS-rise}&{ --}&{--}&--&Companion NS accretion\\
\hline 
\rowcolor{white}III&\textcolor{black}{\textbf{BH-rise (overcritical)}}& \textcolor{black}{--}& \textcolor{black}{--}& \textcolor{black}{--}&\textcolor{black}{BH QED}\\

\hline
\rowcolor{white}IV&\textcolor{black}{\textbf{BH-rise (undercritical)}}& \textcolor{black}{--}& \textcolor{black}{--}& \textcolor{black}{--}&\textcolor{black}{BH CED}\\

\hline
\rowcolor{white}IV&\textcolor{black}{\textbf{BH-echoes}}& \textcolor{black}{--}& \textcolor{black}{--}& \textcolor{black}{--}&\textcolor{black}{BH disk accretion}\\

\hline
\cellcolor{white}VI&\cellcolor{white}\textcolor{black}{\textbf{The Afterglows}}&&&&\parbox{2.0cm}{$\nu$NS synchrotron+pulsar emission}\\
\rowcolor{white}&\textcolor{black}{X-ray}& \textcolor{black}{$>10^8$}& \textcolor{black}{PL}& \textcolor{black}{$>1 \times 10^{48}$}&\\

\rowcolor{white}&\textcolor{black}{Optical}& \textcolor{black}{$>10^8$}& \textcolor{black}{PL}& \textcolor{black}{$>2 \times 10^{47}$}&\\

\rowcolor{white}&\textcolor{black}{Radio}& \textcolor{black}{$>10^8$}& \textcolor{black}{PL}& \textcolor{black}{$>  10^{44}$}&\\
\hline 
\rowcolor{white} VII&\textcolor{black}{ \textbf{SN Ic \& HN}} & \textcolor{black}{$\sim10^7$}&  \textcolor{black}{BB}& \textcolor{black}{$> 3\times 10^{49}$}&\textcolor{black}{Nickel decay}\\
&$\rm M_{\rm ej} = 4.9 \pm 0.9~ M_\odot$& && &   \\
&$\rm M_{\rm Ni} = 0.18 \pm 0.01 ~M_\odot$& && &    \\
&$\rm E_K = (2.4 \pm 0.9)\times 10^{52}$~erg& && &   \\
\hline 
\end{tabular}
\end{table*}

At 07:20:43 UT, GRB 171205A (\textit{Swift} trigger 794972) with $T_{\rm 90}$ of $189.4 \pm 35.0$~s and $z= 0.0368$ \citep{2017GCN.22180....1I}, was triggered and located by the \textit{Swift} Burst Alert Telescope (BAT) \citep{2017GCN.22177....1D,2017GCN.22184....1B}. The \textit{Swift}--XRT \citep{2017GCN.22183....1K} and \textit{Swift}--UVOT (GCN22181) started the observation after $134$~s and $154$~s, respectively, from the BAT trigger.

The prompt emission maintains its luminosity of $10^{46} - 10^{47}$~erg~s$^{-1}$ for $\sim 100$~s then drops following a power-law, see figure \ref{fig:BdHNeIII:171205A}. Its spectrum is best fitted by a cutoff power-law function with peak energy $148.55\pm 121.97$~keV and low-energy power-law index $-1.10\pm0.35$. The total isotropic energy within the $T_{90}$ of BAT gives $E_{\rm iso} = (1.71\pm 0.35) \times 10^{49}$~erg, see \citet{2022ApJ...936..190W} and table \ref{tab:171205A} for details.

As we discussed in the previous section for BdHN II, there are three episodes of accretion, and the last two are unique features of BdHNe. In the case of GRB 171205A, the progenitor system is a single CO star or a CO-NS binary with negligible interaction between the binary components because of a large orbital separation. Hence, only the first fallback accretion onto the $\nu$NS is expected. We also discussed that a large part of the energy from the accretion propagates inside the SN ejecta and accelerates its outermost layer, which has a steep density gradient, to a mild-relativistic speed of Lorentz factor $<10$. The fast-moving material produces the luminosity of $<10^{47}$~erg~s$^{-1}$ for some minutes, which is often missed by \textit{Fermi}-GBM or \textit{Swift}-BAT. But for GRB 171205A, one of the nearest GRBs at redshift $z=0.0368$, this weak signal is resolvable and detected by \textit{Swift}-BAT, shown as the initial hundreds of seconds of prompt emission. This physical picture is similar to the hot cocoon, which is produced by a narrow jet passing through the shells of the progenitor \citep[see, e.g.][]{2001ApJ...556L..37M,2002MNRAS.337.1349R,2004ApJ...608..365Z,2017ApJ...834...28N}. The difference comes from the outflow in our picture having a clear accretion origin onto the $\nu$NS, which emits radiation at a wider opening angle. The heated SN ejecta emits thermal emissions, a temperature of $\sim 80$~eV is observed by \textit{Swift}-XRT in the initial $\sim 400$~s (see Fig.~2 in \citealp{2022ApJ...936..190W}), then cools to optical bands observed by \textit{Swift}-UVOT, VLT/X-shooter, and GTC/OSIRIS. 

Different from more luminous GRBs, the emission from the accelerated fast-moving material has an obvious impact on the observation of the weak GRB 171205A. Before the transparency time $\sim 10^{5}$~s of the fast-moving material of mass $\sim 10^{-2} M_{\odot}$, the X-ray and optical light-curves form a long plateau phase (see Fig. \ref{fig:BdHNeIII:171205A}). The growing transparent part of the fast-moving material dominates the X-ray flux through the synchrotron mechanism, and the thermal radiation from the rest opaque part dominates the optical flux. After $\sim 10^{5}$~s, the X-ray light curve decays as a typical power-law of power-law index $\sim -1$, and optical emission starts to be taken over by the emission from the radioactive decay of SN ejecta. The $1000$ days radio observation by uGMRT \citep{2021ApJ...907...60M} shows the radio flux rises till $\sim 10^7$ then decays as a power-law, and no jet break signature was observed, indicating the outflow has a wide opening angle. Like GRB 190829A, the same synchrotron simulation for the fast-moving material was applied on GRB 171205A \citep{2022ApJ...936..190W}. To fit the power-law decay behavior of the X-ray and radio afterglow, an $\nu$NS with an initial magnetic field of $\sim 3 \times 10^{13}$~G and a spin period of $58$~ms is required (see Fig.~5 in \citealp{2022ApJ...936..190W}).

The optical signal of SN 2017iuk, a typical GRB-associated Ic SN (see Figs. \ref{fig:redlpeak}, \ref{fig:iso-tpeak} and \ref{fig:BdHNeIII:171205A}), peaks at $\sim 10^6$~s. Using the nickel radioactive-heating model for the bolometric SN light curve, the estimated total SN ejected mass, nickel mass, and total SN kinetic energy are $M_{\rm ej} = 4.9 \pm 0.9 ~M_\odot$, and $M_{\rm Ni} = 0.18 \pm 0.01 ~M_\odot$, $E_K = (2.4 \pm 0.9)\times 10^{52}$~erg, respectively, comparable to the average value deduced for the GRB-SN sample \citep{1982ApJ...253..785A,2017AdAst2017E...5C,2019Natur.565..324I}.

\section{New physics regimes in Hypernovae and GRBs physics}\label{sec:10}

The above description of the richness of physical phenomena triggered by the SN in the BdHN brings us to new physics in the explanation of long GRBs and which deserves to be highlighted. Below, we summarize new physics regimes made possible by understanding long GRBs and HNe.

\subsection{Evidence from triaxiality in the $\nu$NS early evolution}\label{sec:newphysics1} 

The $\nu$NS-rise and the afterglow emission are powered by the rotational energy of the $\nu$NS. The recent analysis of the $\nu$NS parameters and energetics in GRB 180720B and GRB 190114C \citep{2022PhRvD.106h3004R}, has shown that the $\nu$NS at the beginning of the $\nu$NS-rise, is characterized by a rotation period at the verge of the bifurcation point of the Maclaurin sequence of equilibrium spheroids into the Jacobi ellipsoidal sequence. The presence of the highly spinning $\nu$NS deserves deeper attention in the core collapse of the CO star.

Therefore, the $\nu$NS might have evolved from a triaxial Jacobi-like ellipsoid into the axially symmetric Maclaurin spheroid by emission of gravitational waves, as anticipated in early models of pulsars  \citep{1969ApJ...157.1395O, 1969ApJ...158L..71F, RWlincei1970}, and theoretically verified by \citet{1970ApJ...161..571C, 1974ApJ...187..609M}. The triaxial configuration lives for a short time, i.e., about less than a second, due to the copious emission of gravitational waves, before the GRB emission. The gravitational-wave emission could be, in principle, detected for sources located at distances closer than $100$ Mpc \citep[see][for details]{2022PhRvD.106h3004R}. This appears to be the only emission of gravitational waves associated with the long GRB in the BdHN scenario: the core-collapse leading to the $\nu$NS radiates poor gravitational waves ($\sim 10^{-7} M_\odot c^2 \sim 10^{47}$ erg; see \citealp{2002A&A...393..523D, 2011LRR....14....1F}). In addition, given the stringent limits on the ultrarelativistic jetted emission, both in the GeV radiation and in the X-ray afterglow, previous gravitational waves estimates \citep[e.g.][]{2021PhRvD.104j4002L} do not apply \citep{2022ApJ...929...56R}.

\subsection{QED radiation process in the UPE}\label{sec:newphysics2}

At every expansion and transparency of the $e^+e^-$ plasma, the energy radiated by the plasma is paid by the Kerr BH that reduces its mass and angular momentum by amounts $\Delta M$ and $\Delta J$, respectively (see Section \ref{sec:5} for details). The lower value of the BH spin leads to a lower value of the induced electric field, which implies that a new self-expansion and transparency can occur with a lower $e^+e^-$ plasma energy \citep{2021PhRvD.104f3043M, 2022EPJC...82..778R}. The QED process and the approach to transparency are analogous. Still, the plasma parameters are different, which explains the hierarchical structure and similarity of the spectra in the time-resolved analysis of the UPE.

\subsection{Classic electrodynamics radiation in the GeV emission}\label{sec:newphysics3}

At the end of the UPE phase, the induced electric field is still sufficiently high to power the GeV emission of the GRB, which is emitted in the polar regions above and below the BH within an angle $\approx 60^\circ$ from the polar axis. The radiation power, timescale, and the energy stored in the electric field to accelerate the electrons confabulate to power luminosities of the order of $10^{51}$ erg s$^{-1}$ in the GeV domain for magnetic fields $B_0 \sim 10^{11}$ G \citep{2019ApJ...886...82R, 2021A&A...649A..75M, 2022ApJ...929...56R}. The acceleration and radiation process occurs thanks to the magnetic dominance, $B^2-E^2 > 0$, and the existence of regions where the component of the electric field parallel to the magnetic field is non-zero, i.e., $\mathbf{E}\cdot \mathbf{B} \neq 0$. As for the UPE, the rotational energy of the BH, the reservoir, powers this radiation process. The extension of this approach to AGNs (e.g., M87*; see \citealp{2021A&A...649A..75M}).

\subsection{Additional knowledge from the Physics frontier: the TeV emission }\label{sec:newphysics4}

As we {have shown in the above} sections, the SN has triggered not only the path to the new physical processes and understandings of phenomena in the BdHN, but there is also a focus on the part of GRB radiation that is not yet theoretically understood and has only recently begun in earth band experiments: TeV radiation. In particular, what is most impressive is the presence of the TeV radiation in the prompt phase of BdHN I GRB 190114C \citep{2021MNRAS.504.5301R} as well as in the afterglow of a BdHN I, GRB 180720B \citep{2022ApJ...939...62R,2022EPJC...82..778R}, and in the afterglow of {a BdHN II,} GRB 190829A \citep{2022ApJ...936..190W}. 

The first crucial information possibly contributing to the understanding of these processes is the fact that the energy flux of the TeV is $10\%$--$60\%$ of the energy flux of the afterglow. The second essential information is that {TeV emission has been observed in} the case of {the} BdHN II, GRB 190829A, {hence,} without a BH \citep{2022ApJ...936..190W}. These two observations lead to privilege an energy emission of the TeV radiation {linked to the rapidly spinning $\nu$NS emission}. {All the above} has driven us to predict the TeV luminosity of GRB 221009A \citep{2022GCN.32802....1A}. 

{Finally, new perspectives emerge from the knowledge on the seven Episodes of BdHNe presented in this article for long GRBs, for the analysis of short GRBs previously studied, e.g., GRB 140619B \citep{2015ApJ...808..190R}, GRB 090510 \citep{2016ApJ...831..178R}, GRB 081024B and GRB 140402A \citep{2017ApJ...844...83A}.}

\section{conclusions}\label{sec:11}

A new era in physics and astrophysics started in 1996 when the 
Beppo SAX satellite promoted the extension of the observational 
techniques from the gamma-rays, the domain where GRBs were initially discovered, to the X-rays, optical, and radio observations. Further extensions to GeV, TeV, and VHE emissions observations were soon implemented. Three main discoveries were made possible at the time: a) 
the presence in long GRBs of an afterglow with long-lasting X-ray emission \citep{Costa1997}. As we here show, these afterglows have contributed significantly to the long GRB understanding; b) the cosmological nature of the GRBs, implying energies up to $10^{54}$~erg \citep{1997Natur.387..878M}; 
and c) the outstanding spatial and temporal coincidence between the 
Type Ic SN 1998bw, with optical emission of $10^{49}$ erg \citep{1998Natur.395..670G}, and the long GRB 980425 of $10^{48}$~erg  \citep{2007ESASP.622..561R}. This article is rooted in explaining this outstanding coincidence and illustrates, as well, the exponential growth of knowledge in physics and astrophysics made 
possible by an equally impressive growth of new technologies.

We have recalled in the Introduction the earlier description of long GRBs as originating from a single BH and an ultrarelativistic jet, the ``collapsar'' model. The lengthy and gradual evolution to a binary progenitor follows the pioneering work of \citet{1999ApJ...526..152F}. A further change of perspective happened with the introduction of the concept of induced gravitational collapse \citep[IGC][]{2012ApJ...758L...7R}. There the idea was advanced that BHs in long GRBs were not primordial but could be created by reaching the critical mass of an already existing accreting NS during the evolution of the binary progenitor. We have also recalled how motivated by a multiyear inquiry of long GRBs, and we finally proposed the BdHN model with a $\sim 10 M_\odot$ CO core and $\sim 2 M_\odot$ NS binary companion as progenitors for long  GRBs. The CO core collapse triggers the GRB event.

We have also recalled how the BdHN approach has gained relevance because of the observed spatial and temporal coincidences of long GRBs with type Ic SNe. Most SN Ic progenitors assume ultra-stripped binaries based on a multi-year effort evolution analysis. This fact has been a guiding factor in further developing our BdHN model, which naturally leads to comprehending the occurrence of the SN Ic in coincidence with a family of long GRBs, presented in this article.

Section~\ref{sec:2} recalls that the BdHN model assumes that the gravitational collapse of the CO core necessarily leads to an SN with $7$--$8 M_\odot$ ejecta and a millisecond spinning $\nu$NS of $1.5 M_\odot$, at its center. Both theoretical arguments and observational evidence for these assumptions are later justified in the article. Still, in Section~\ref{sec:2}, we recall that ultra-stripped binaries comprise $0.1$--$1\%$ of the total SNe; so the BdHN I population could be explained by a small subpopulation of $0.01$--$0.1\%$ of them \citep[see, e.g.,][]{2015PhRvL.115w1102F}. It is interesting to explore if that branch could only occur under specific conditions in the last evolution stages of the binary evolution after the common-envelope phase. {The description of the multiwavelength phenomenology of long GRBs with the BdHN model predicts the formation of CO-NS binaries with orbital periods from hours to days (BdHN II and III) to minutes (BdHN I), with a crucial role of the angular momentum. These binaries could be eventually observed in the Galaxy or nearby galaxies by sensitive facilities, e.g., like the James Webb Space Telescope (JWST). In addition, thanks to the cosmological time dilation, we have identified in BdHN at high redshift (e.g., GRB 220101A at $z=4.2$, GRB 090423 at $z=8.2$, GRB 090429B at $z = 9.4$) crucial information of the $\nu$NS-rise emission in Swift-XRT data \citep{2023arXiv230605855B}. The JWST is also gaining information on the galaxies hosting high redshift GRBs like the above ones. We have advanced that such $\nu$NS-rise emission identified in high redshift sources could be observed coincident with GWs in nearby sources by a new satellite overcoming the $43$ s gap between the GRB trigger and the Swift-XRT observations \citep[see][for details]{2023arXiv230605855B}.}

This article addresses the identification of the separatrix properties of the CO core's gravitational collapse occurring in CO-NS binaries and leading, alternatively, to a single SN Ic or a similar SN Ic and a variety of long GRBs. It is shown that the most general BdHN, in addition to a standard Ic SN, leads to 1) an HN $10^3$ times more energetic than a typical SN Ic, 2) to long GRBs, much more energetic than the SN Ic, in the range of $10^{49}$--$10^{54}$~erg, 3) these long GRBs being subdivided in BdHN I, BdHN II, and BdHN III.

From observations and theoretical analysis, we illustrate in Section~\ref{sec:2} the BdHN I with energies between $10^{52}$~erg and $10^{54}$~erg, the only BdHNe where the IGC process forms a BH, BdHN II with energies between $10^{50}$~erg and $10^{52}$~erg, and BdHN III with energies below $10^{50}$~erg. For each BdHN type, we have identified the typical CO-NS orbital period and the $\nu$NS spin: the former ranges from $\sim 4$--$5$ min in BdHN I, $\sim 20$ min in BdHN II, and to a few hours in BdHN III. The $\nu$NS spin ranges between $1$ and $100$ ms. A long-lasting X-ray afterglow is associated with \textit{each} GRB and is present in \textit{all} BdHN types. Specific examples are given in Section~\ref{sec:6}--\ref{sec:9}.

Already on these results, the important conclusion can be
inferred that BdHNe are intrinsically dominated by a large amount of rotational energy; see text and references therein:
\begin{enumerate}
    \item The $\nu$NS spin inferred from the energetics of the X-ray afterglows has an initial dimensionless angular momentum $a/M = c J/(G M^2)$, being $J$ and $M$ the $\nu$NS angular momentum and mass, of $\sim 0.5$ for BdHN I down to $\sim 10^{-3}$ in BdHN III. We have given an example of how the fast spinning $\nu$NS in GRB 180720B initially follows a Jacobi ellipsoid sequence \citep{2022PhRvD.106h3004R}, an absolute first in relativistic astrophysics.
    \item The BH is formed only in BdHN I by the IGC process due to the accretion of SN ejecta onto the companion NS. Also, in this case, an initial dimensionless parameter $\sim 0.5$ of the BH has been inferred from the two BHs in BdHNI,  GRB 180720B (see Section~\ref{sec:6}) and GRB 19014C (see Section~\ref{sec:7}). 
    \item As recalled above, the CO core gravitational collapse originates the entire energetics of the BdHN. Traditionally, the initial rotational energy of the CO core is assumed to be zero. Possibly the largest change of paradigm introduced by the BdHN model has been to point out that the zero angular momentum traditionally assumed in the description of the collapse of the CO core is untenable. In the BdHN model, the CO core has to be close to corotation with the binary NS companion: this implies, for a binary companion NS of $\sim 4$ min orbital period, a CO core with $a/m \sim 1$, assuming a radius $\sim 10^{10}$~cm and a mass $\sim 10 M_\odot$. All efforts should be directed at gaining observational evidence for this corotation and developing an SN explosion model consistent with this assumption.
\end{enumerate}

In Section~\ref{sec:3}, we have recalled relativistic transformations to evaluate the time measurement and the bolometric luminosities in the rest frame of the source. In
Section~\ref{sec:4}, we present a selected sample of 24 spectroscopically confirmed SN Ic and their associated long GRBs (see Table \ref{tab:24-grb-sn-spectroscopic}). The main outcome is that all observed SNe Ic have peak luminosities around an average value of $9.45\times 10^{42}$ erg s$^{-1}$ independently of the source redshift (see Fig. \ref{fig:redlpeak}). The time of occurrence of the peak optical luminosity, measured from the GRB trigger, peaks at an average value of $1.16\times 10^6$ s (see Fig. \ref{fig:redtpeak}), again independently of the redshift of the source.

The properties of the associated GRBs for the selected three BdHNe classes are correspondingly summarized: 1) Figure \ref{fig:iso-lpeak} shows that the luminosity of the SN Ic has roughly the same value, BdHNe $E_{\rm iso}$ ranges from $10^{48}$ erg to $10^{54}$ erg. 2) Figure \ref{fig:iso-tpeak} shows that the time of occurrence of the peak luminosity of the SN Ic is also independent of the energetics of the associated BdHN. 3) The HN energy is $10^3$ times larger than the common SN Ic. This decoupling between the GRBs and the Ic SN was highlighted in a pioneering work of \citet{2004ApJ...609..952Z} where this problem was announced, which we quantify and explain.

In Section~\ref{sec:5}, we indicate the BdHN approach in addressing using quantum and classical field theories, the conceptual description of a selected number of
Episodes are then subjected to observational scrutiny via a time-resolved spectral analysis in the rest frame of the BdHN. The case is presented for the necessity of introducing and verifying new physical laws: either in extrapolating well-known physical laws already studied on Earth-bound experiments \citep[see, e. g.,][]{2010PhR...487....1R}, now extended to new more extreme regimes encountered for the first time in extragalactic sources. This is the case of the classical electrodynamics processes extended to overcritical fields. Equally important has been to review the introduction of new physical laws in the quantum electrodynamics regimes to probe the process of rotational energy extraction from non-stationary and non-asymptotically flat Kerr solution as explaining the high energy GeV emission of GRBs. Particularly important has been the observational verification of the energy extraction process from a Kerr BH embedded in a fully ionized low-density plasma with a non-flat asymptotic solution given by a magnetic field aligned with the rotation axis of the Kerr solution. These new approaches, previously published in specific cases, are here directly applied in interpreting \textit{all} seven Episodes of the most general BdHN, which are here briefly recalled, and details are presented in Section~\ref{sec:5}.

\textit{Section~\ref{sec:SNrise}: the SN-rise.} {We introduce, in this Episode (0), the first appearance of the CO core collapse and the SN explosion. This episode has been possibly observed in three BdHNe, i.e., GRB 160625B \citep{2021MNRAS.504.5301R}, GRB 221009A and GRB 220101A (Ruffini et al., in preparation), and needs further examples to verify its spectrum unambiguously. What makes this Episode's observation particularly difficult is its intrinsically low luminosity, with total energy $\sim 10^{52}$~erg, that in all three above BdHN I precedes, by a time interval between $1$ s and $100$~s, the $\nu$NS-rise and the UPE, the two first Episodes of the prompt radiation of energy $10^{53}$--$10^{54}$~erg \citep[see also][]{2021MNRAS.504.5301R}.}

\textit{Section~\ref{sec:nuns}: the $\nu$NS-rise.} This Episode is identifiable by CPL spectra and its time of occurrence, manifesting the early presence of rapidly spinning $\nu$NS. Their periods range from $\sim 1$ ms in BdHN I to $\sim 100$ ms in BdHN III. The $\nu$NS-rise occurs in all BdHN types. It is followed by the synchrotron emission emitted by the $\nu$NS interacting with the SN ejecta and leading to the three components afterglow: in the X-ray, in the optical and the radio, further examined in Section~\ref{sec:3afterglows}. One of the main results reached in the analysis of the $\nu$NS-rise in the two BdHN I, GRB 180720B and GRB 190114C, has been the first observations of an initial triaxial Jacobi ellipsoid evolving in a Maclaurin spheroid, with possible emission of gravitational waves. It is interesting that the presence of afterglows in \textit{all} GRBs (observed in $380$ BdHN I and all BdHN II and III) necessarily also implies the presence of $\nu$NS in \textit{all} GRBs.

\textit{Section~\ref{sec:episodesUPE}: the UPE phase.} The SN accretion onto the binary NS companion, soon after the first observation of the $\nu$NS-rise, leads to the formation by the IGC process of a rapidly spinning Kerr BH whose presence is highlighted by the emission of the Ultra-relativistic Prompt Emission (UPE). In this section, we present an extended introduction of the theoretical works developed to extend to overcritical fields, i.e., $E \geq E_c = m_e^2 c^3/(e \hbar)\approx 1.32\times 10^{16}$ V cm$^{-1}$, to the multiyear theoretical works on vacuum polarization. This treatment is now finally reaching its observational verification in the GRBs. The overcritical field is due to an effective charge given by $Q_{\rm eff}=2 B_0 J G/c^3,$ being $B_0$ the magnetic field and $J$ the angular momentum of the Kerr BH. These verifications on two selected BdHNe I, GRB 190114C, and GRB 180720B, have allowed explaining the existence of detailed new spectral features with the presence of self-similarities and structures on ever-decreasing time intervals to $10^{-9}$~s. The UPE phase has allowed to test observationally and verify the
validity of the Christodoulou-Ruffini\citep{1970PhRvL..25.1596C, 1971PhRvD...4.3552C}-Hawking \citep{1971PhRvL..26.1344H} mass-energy formula. This has allowed us to estimate the initial mass-energy of the Kerr BH and the associated magnetic field, $B_0$, in both BdHNe I examined.

\textit{Section~\ref{sec:episodesGEV}: High-energy jetted (GeV) emission}. In this section, we study the high energy GeV emission originating from the classical electrodynamics process that transitions from the overcritical field, characterizing the UPE phase, to an undercritical field. The theoretical analysis of the emission originated from a Kerr BH in the presence of a magnetic field of $\sim 10^{10}$--$10^{11}$ G has allowed inferring an emission of the GeV radiation within a cone of half-opening of $\approx$60$^{\circ}$ \citep{2022ApJ...929...56R}. This has been confirmed by the statistical analysis of the $54$ BdHNe observed by Fermi-LAT. Only $25$ emit the GeV radiation, and the remaining $29$, confirming not observable given the beamed radiation \citep{2021MNRAS.504.5301R}. Equally important has been the specific temporal power-law behavior of the GeV luminosity, well evidenced in Section~\ref{sec:6} dedicated to GRB 180720B and Section~\ref{sec:7} dedicated to GRB 190114C.

\textit{Section~\ref{sec:episodesCav}: The BH echoes}. The cavity radiation, explained by the collision and partial reflection of the expanding $e^+e^-$ with the cavity's wall, originated from the BH formation \citep{2019ApJ...883..191R}, is evidenced for GRB 180720B in Section~\ref{sec:6} and GRB 190114C in Section~\ref{sec:7}. The HXFs and SXFs, previously explained by the interaction of the expanding $e^+e^-$ with the surrounding accretion matter, are observable in sources with observation angle in the equatorial plane of the BdHN \citep{2018ApJ...869..151R}. These processes are identified in Section~\ref{sec:6} for GRB 180720B. Neither HXF nor SXF is present in GRB 190114C, given the viewing angle orthogonal to the plane of orbit. 

\textit{Section~\ref{sec:3afterglows}: Multiwavelength (X, optical, radio) afterglow}. In this section, the afterglow's multi-wavelength X-ray, optical, and radio emissions are recalled with references to their theoretical treatments. We here recall that the afterglows are observed in \textit{all} BdHN types, implying a large angle emission perfectly explained in terms of the synchrotron radiation emission process originating in a millisecond period of spinning $\nu$NS as described in the following Sections~\ref{sec:6} and \ref{sec:7}. The afterglow is observed in \textit{all} BdHNe, implying that all these GRBs originate from a CO-NS binary. 

\textit{Section~\ref{sec:episodesclassicSN}: The classic SN emission powered by nickel decay}. In this section, we address the optical SN emission due to the nickel decay well expressed by the theoretical work of Nadyozhin  \citep[see, e.g., Nadyozhin's lectures,][]{2011-Nadyozhin1,2011-Nadyozhin2} and Arnett \citep{1982ApJ...253..785A}. The crucial point is to recall that  SN Ic is present in all BdHN types and observable with current telescopes for $z \lesssim 1$. New telescopes, e.g., the James Webb Space Telescope, should probe the presence of an SN, which is predicted to exist also for higher $z$ values, following the BdHN model. We refer to Table \ref{tab:24-grb-sn-spectroscopic} for a summary.

We turn then to the two examples of BdHNe I. In Section~\ref{sec:6}, we have summarized the results of GRB 180720B, and in Section~\ref{sec:7} of GRB 190114C. In
Section~\ref{sec:8}, we give the example of a BdHN II, GRB 190829A, and finally, in Section~\ref{sec:9}, the case of a BdHN III, GRB 171205A.

For each source, we have given: 1) the complete references to the observational papers we have utilized to perform the theoretical and the time-resolved spectral analysis; 2) a Figure summarising the luminosities for each Episode as a function of the rest-frame time and concerning the specific instruments and bandwidths. The same figure shows the specific examples of the spectra of each Episode; and 3) again, for each source, we present a Table summarizing the names of the observed Episodes: for each, we give the name of the event, the duration, the spectrum the corresponding $E_{\rm iso}$ and the underlying physical phenomena. A specific time-resolved spectral analysis of the UPE phase is exemplified in the case of
GRB 190114C. In addition to the complete material for the description of two BdHN I, one BdHN II, and one BdHN III, we would like to mention that preliminary results have already been obtained for the UPE phase of two additional BdHN I, namely GRB 160626B and GRB 160509A \citep{2023ApJ...945...10L}. There, one can find the detailed UPE analysis for GRB 160625B in Table 2, Fig. 4, as well as Fig. 5, and for GRB 160509 in Table 4, Fig. 8, and Fig. 9. We are currently working on the identification of the other six Episodes present in both sources.

Following the above, we identify the primary energy source of all BdHN, independently of their type. The most remarkable property which has allowed us to understand the nature of GRBs has been the possibility
to retrace back from the extraordinary observed spectral data,  the specific energy sources, and their fundamental new physics. This has been made possible by the guidance of the  BdHN model. We refer to the Figures and the Tables in the text and the references to the data acquisitions we have performed.

In Section~\ref{sec:10}, we briefly highlight the three main topics in which the analysis of the BdHN has promoted new research perspectives with the discovery of new physical laws and the verification of existing laws in new regimes made possible by the unique GRBs and HNe observations. The study of rotating figures of equilibrium represents one of the topics of research in which the best intellectuals have addressed their attention for over two centuries: from the self-gravitating Maclaurin spheroids to the discovery of the triaxial Jacobi ellipsoids. Now, for the first time, we have given evidence that triaxial ellipsoids can play a fundamental role in relativistic astrophysics and be the most prominent source of gravitational waves \citep{2022PhRvD.106h3004R}. Furthermore, far from being a conclusion, this is just the beginning of a new era in relativistic astrophysics leading to a new understanding of the physics of gravitational collapse of the creation of new physical systems by gravitational fission, to a new physics of SN explosion based on quantum and classical electrodynamics process coupled to the rotational energy extraction.

Similarly, we have indicated the perspectives of classic and quantum electrodynamics energy extraction processes from rotating NS and Kerr BHs. The examples of the UPE phase and the GeV emission are here recalled just as interesting prototypes to be further extended. But far from being self-exhaustive, the GRB observations still present new challenges by the observations of vast amounts of TeV radiation up to luminosities of $10^{52}$ erg s$^{-1}$. At the same time, these emissions have recently been observed, in very low fluxes, in Earth-based accelerators, e.g., at CERN. Possibly, the most exciting new perspective is that there is evidence that this most energetic emission \textit{does not} originate from a rotating BH, as already shown in this article. 

{It looks equally promising the interpretation of previous results on short GRBs using the knowledge acquired from the BdHN seven Episodes, e.g., in GRB 140619B \citep{2015ApJ...808..190R}, GRB 090510 \citep{2016ApJ...831..178R}, GRB 081024B and GRB 140402A \citep{2017ApJ...844...83A}.}

Having said all the above, we can return to  explain the enormous energetic difference between an SN Ic and the associated HN and long GRB through the occurrence of seven specific episodes in the most general BdHN leading to these concluding remarks: 
\begin{enumerate}
    \item The associated SN Ic bolometric energy of $10^{49}$~erg originates from the nuclear physics process leading to the decay of a common amount of a fraction of $0.2~M_\odot$ to $0.5~M_\odot$ of
       nickel \citep[see e. g.][]{2011-Nadyozhin1,2011-Nadyozhin2,1982ApJ...253..785A}, remarkably similar in all BdHN sources. The same explanation regarding nuclear physics applies to explain the common time of occurrence of the SN peak emission, identified as soon as the relativistic corrections are implemented.
  \item The HNe in BdHN have kinetic energies of $10^{52}$~erg originating from the kinetic energy of $7$--$8~M_\odot$ ejecta, expanding mildly-relativistically, observed in all BdHN types.
  \item Both the above kinetic energy and the formation of a highly spinning millisecond $\nu$NS at the SN center should find an explanation in a CO core-collapse, duly considering the contribution of the rotational energy, again observed in all BdHN types.
\end{enumerate}

Turning now to the GRBs: 

\begin{enumerate}
    \item The X-ray, optical, and radio emission of the afterglow, present in all BdHN types, ranging from a few $10^{52}$~erg in BdHN I (GRB 190114C) down to $10^{49}$~erg in BdHN III (GRB 171205A), are powered by the synchrotron emission originating from the rotational energy of the $\nu$NS
    interacting with the SN ejecta. The  $\nu$NS initial rotation period is $1$--$100$ ms.
    \item The MeV and GeV emissions observed in the prompt radiation phase, present only in BdHN I, ranging $10^{52}$--$10^{54}$~erg, are powered by quantum and classical electrodynamics process originating from the extractable rotational energy of a Kerr BH, embedded in a fully ionized low-density plasma. The Kerr solution is neither stationary nor asymptotically flat, but in the presence of a magnetic field, $B_0$, aligned with its rotation axis and fulfilling the Christodoulou-Ruffini\citep{1970PhRvL..25.1596C, 1971PhRvD...4.3552C}-Hawking \citep{1971PhRvL..26.1344H} mass-energy formula. For the latest developments, see Rueda and Ruffini (submitted).
    \item
    Only the MeV emission in the prompt radiation of BdHN II, of $\sim 10^{52}$~erg (see Table \ref{tab:190829A}), originates from the accretion of the SN ejecta into the slowing rotating binary NS companion.
\end{enumerate}

We can then conclude, generally, that SNe Ic associated with long GRBs originate from CO-NS binary progenitors. 

We advance the hypothesis that most CO-NS binaries, with a binary period longer than a few hours, lead only to SN Ic, without any associated pulsar, GRB, or HN. This point can be easily tested observationally. A CO core, with an initial $a/M \sim 1$, endowed with an initial magnetic field of $\sim 10^3$~G, and density of $\sim 10^4$ g cm$^{-3}$, can indeed lead, in the process of gravitational collapse, to a triaxial ellipsoid. The consequent fission, Roche lobe bifurcation, can lead to a fast spinning $\nu$NS and a most powerful explosion. Like in the UPE phase, this process is expected to be driven by a quantum electrodynamical process originating from an overcritical ``effective charge''. This overcritical field can complete the comprehension of the GRB-SN connection and lead to a new understanding of some of the current open issues. This will undoubtedly sign a good starting point for approaching the yet unsolved problem of the SN explosion, mainly examined in the absence of rotation. But this brings us to a different topic: the multi-century works on the rotating equilibrium configurations, as recalled above, developed by Elie Cartan, Bernhard Riemann, James Hopwood Jeans and summarized in a series of articles by Subrahmanyan Chandrasekhar \citep{1969efe..book.....C}, also in collaboration with Enrico Fermi \citep{1953ApJ...118..116C}. These works are finally reaching the test of astrophysical observations in relativistic astrophysics, which is only partly the article's topic today.

\acknowledgements
{It has been a pleasure to collaborate with a knowledgeable anonymous referee who has promoted an in-depth and constructive dialogue with us.} We are developing the BdHN model in a series of papers and are in continuous contact with many scientists and collaborators. Particularly important have been the discussions with Roy Patrick Kerr and with editors and referees of the journals: they have constructively contributed to finalizing the presentation of our research. To all of them goes our heartfelt thanks. L.M.B. is supported by the Vicerrector\'ia de Investigaci\'on y Extensi\'on - Universidad Industrial de Santander Postdoctoral Fellowship Program No. 2023000107.


\end{CJK*}
\end{document}